\documentclass[%
 reprint,
 prd,
nofootinbib,
 amsmath,amssymb,
 aps,
]{revtex4-2}

\usepackage{graphicx}
\usepackage{dcolumn}
\usepackage{bm}
\usepackage{comment}
\usepackage{xcolor}
\usepackage{todonotes}
\usepackage{subfigure}
\usepackage{float}
\usepackage[normalem]{ulem}
\usepackage{amsmath}

\begin{document}

\preprint{APS/123-QED}

\title{The High W Challenge: Robust Neutrino Energy Estimators for LArTPCs}

\author{Christopher Thorpe$^1$, Elena Gramellini$^1$}
\address{$^1$The University of Manchester, Manchester M13 9PL, UK}

\date{\today}

\begin{abstract}
Accurate determination of the neutrino energy is central to precision oscillation measurements. In this work, we introduce the W$^2$-based estimator, a new neutrino energy estimator based on the measurement of the final-state hadronic invariant mass. This estimator is particularly designed to be employed in liquid-argon time-projection chambers exposed to broadband beams that span the challenging transition region between shallow inelastic scattering and deep inelastic scattering. The performance of the W$^2$-based estimator is compared to four other commonly used estimators. The impact of the estimator choice is evaluated by performing measurements of $\delta_{CP}$ and $\Delta m^2_{23}$ in a toy long-baseline oscillation analysis. We find that the W$^2$-based estimator shows the smallest bias as a function of true neutrino energy and it is particularly stable against the mismodelling of lepton scattering angle and momentum, missing energy, hadronic invariant mass and final state interactions. However, studies of the resolution of each estimator as a function of true neutrino energy show this is somewhat offset by worse energy resolution when perfect modeling of these quantities is assumed. This estimator is valid for events with at least one proton and any number of pions; an inclusive channel that complements the strength of more exclusive methods that optimize the energy resolution. By providing a detailed analysis of the strengths, weaknesses and domain of applicability of each estimator, this work informs the combined use of energy estimators in any future LArTPC-based oscillation analysis. 

\end{abstract}

\maketitle



\section{Introduction}
\label{sec:Introduction}

Neutrino oscillations are the only evidence of physics beyond the Standard Model (BSM) reproducible in the laboratory and provide a unique window into unresolved questions in neutrino physics, such as the ordering of neutrino masses and the potential for CP violation in the lepton sector~\cite{T2K:2019bcf}. 

In a simplified but pedagogically useful scenario, the oscillation probability between two neutrinos of different flavors can be expressed as:

\begin{equation}
P(\nu_\alpha \rightarrow \nu_\beta) = \sin^2(2\theta) \sin^2\left(1.27 \frac{\Delta m^2 \; L\, [\mathrm{km}]}{E_{\rm true}\, [\mathrm{GeV}]}\right),
\end{equation}

\noindent where $L$ is the baseline (source-to-detector distance), $E_{\rm true}$ is the true neutrino energy, $\theta$ is the mixing angle, and $\Delta m^2$ is the squared mass difference between the two neutrino mass eigenstates.  While the oscillation probability is a function of the true neutrino energy, $E_{\rm true}$, detectors can only measure visible final state quantities to estimate the neutrino energy, $E_{\rm est}$. The relationship between these quantities is non-trivial and depends on the neutrino–nucleus interaction processes involved even before accounting for detector response.

 Current and future accelerator-based neutrino experiments such as the Short-Baseline Neutrino (SBN) \cite{MicroBooNE:2015bmn,Machado:2019oxb} program and the Deep Underground Neutrino Experiment (DUNE)~\cite{DUNE:2015lol,DUNE:2016hlj} 
 are placed on-axis in broadband energy beams, which enable sampling of very wide energy ranges and probing of multiple oscillation maxima. This approach enhances sensitivity to oscillation parameters, including the CP-violating phase and the neutrino mass ordering. Its drawback, however, is that a wide neutrino energy spectrum inevitably samples a broad range of neutrino–nucleus interaction channels.  These span from charged-current quasi-elastic (CCQE) scattering at lower energies, through baryonic resonance production (RES) and meson-exchange current (MEC) processes, up to deep inelastic scattering (DIS) at higher energies. The overlapping nature of these regimes complicates both the modeling of interaction dynamics and the reconstruction of the incident neutrino energy, which is essential for precise oscillation analyses. While previous work has focused on exclusive channels~\cite{Furmanski:2016wqo}, calorimetry~\cite{Ankowski:2015jya,DeRomeri:2016qwo} or on advanced machine learning reconstruction techniques~\cite{Kopp:2024lch}, we are interested in exploring more inclusive channels, which, by definition, improve the sample statistics. We assume that ongoing progress in LArTPC reconstruction algorithms will eventually enable high-fidelity reconstruction of most events. Under these conditions, more inclusive analyses may prove to be not only statistically advantageous, but also more resilient to model and detector-related uncertainties.
 
 This paper explores a new approach to analyzing the final state kinematics, based on the calculation of the final state hadronic invariant mass. This method targets high hadron multiplicity events expected from the interactions of multi-GeV neutrinos in future LArTPCs. This method is then benchmarked against several established methods of estimating neutrino energy, introduced in Section~\ref{sec:EnergyEstimators}. 
 
To model the complex neutrino-nucleus interactions, we rely on several event generators, each of which embodies different theoretical assumptions to probe different kinematic scenarios as a function of the true energy. The event generators employed in this study are introduced in Section~\ref{sec:Simulation}, together with the adopted reconstruction assumptions.  We compare the performance of the different estimators by first calculating their bias and variance with respect to neutrino energy in Section~\ref{sec:BiasAndVariance}, and then with respect to other kinematic variables in Section~\ref{sec:VariableDependences}. Finally, we assess their impact on oscillation measurements through a series of tests presented in Sections~\ref{sec:ImpactOnOsc} and~\ref{sec:ImpactOnOsc2}.

\section{Simulation of Neutrino Events}
\label{sec:Simulation}

The evaluation of neutrino energy estimators necessarily relies on a set of working assumptions. As a case study, we adopt the neutrino flux provided by the DUNE Collaboration~\cite{DUNE:2015lol}. We do not include the effects of oscillation outside of the studies presented in Sections~\ref{sec:ImpactOnOsc} and ~\ref{sec:ImpactOnOsc2}. We restrict the true neutrino energies to the range 0.5–6~GeV, which is most relevant for oscillation measurements, and considerably more energetic than the beams used in existing LArTPC experiments~\cite{MicroBooNE:2015bmn}. Any comparable flux that covers the shallow to deep inelastic transition region would serve equally well for the purposes of this study. The neutrino interaction event generators are listed in Section~\ref{sec:EventGenerators}, and our idealized particle reconstruction is outlined in Section~\ref{sec:perfectReco}. These ingredients provide a controlled environment in which to test and compare the five different neutrino energy estimators.

\subsection{Event generators}
\label{sec:EventGenerators}

Simulations were conducted with four neutrino interaction event generators:
\begin{itemize}
    \item GENIE version 3.6.2~\cite{Andreopoulos:2009rq}, from which we use the \verb|G18_10a_02_11a| tune. The tune employs the Nieves model of quasielastic interactions~\cite{Nieves:2005rv} with the BBBA05 vector form factor parameterization~\cite{Bradford:2006yz} and including random phase approximation (RPA) corrections. The Beger-Sehgal for resonance production~\cite{Rein:1980wg,Berger:2007rq} and Valencia model of CC meson exchange current~\cite{Sobczyk:2020dkn,Nieves:2011pp,Gran:2013kda}. Resonances up to $W = 1.9$~GeV are simulated after which the generator transitions to the Bodek-Yang DIS model~\cite{Bodek:2004pc}. Hadronization is performed with the AGKY model~\cite{Yang:2009zx}. Final state interactions are simulated using the \verb|hA2018| model.

    \item NuWro 25.03.1~\cite{NuWro:UserGuide}. NuWro employs the LLwelyn-Smith model~\cite{LlewellynSmith:1971uhs} of quasielastic interactions augmented with RPA corrections, BBBA05 vector form factors~\cite{Bradford:2006yz} and the MINERvA parameterisation~\cite{MINERvA:2023avz} of the axial form factor. Meson exchange current is simulated with Valencia model~\cite{Sobczyk:2020dkn,Nieves:2011pp,Gran:2013kda}, and the Ghent hybrid model of single pion production~\cite{Yan:2024kkg}. Unlike GENIE, NuWro does not simulate higher resonances and instead switches to modeling the interaction as DIS when the invariant mass of the W boson-nucleon system exceeds 1.5~GeV. DIS interactions are simulated with the Bodek-Yang model~\cite{Bodek:2004pc} with hadronization routines outsourced to Pythia 6~\cite{Sjostrand:2006za}. Final state interactions are modeled with a cascade simulation developed in-house, the most complete description of which is found in~\cite{Golan:2012wx}.

    \item NEUT version 5.8.0~\cite{Hayato:2002sd,Hayato:2021heg}. This generator uses the Llwelyn-Smith model~\cite{LlewellynSmith:1971uhs} for CCQE interactions. Meson exchange current is simulated with the Valencia model~\cite{Sobczyk:2020dkn,Nieves:2011pp,Gran:2013kda}, and single pion production using the Berger-Sehgal model~\cite{Berger:2007rq}. Mutli-pion production events are modeled using a series of fits to hadron multiplicity data from bubble chambers~\cite{Derrick:1977zi,WA25:1988mir} before the model switches to deep inelastic scattering at $W = 2.0$~GeV, simulated with Pythia 5~\cite{Sjostrand:1993yb}.

    \item GiBUU 2025~\cite{Buss:2011mx}. This is distinct from the other three generators in that the neutrino-nucleus scattering cross sections are calculated through solution of nuclear transport equations, with single nucleon cross section calculations detailed in ~\cite{Buss:2011mx}, excluding DIS which is modeled using Pythia~\cite{Sjostrand:2006za}

\end{itemize}
We use the default settings for NuWro, NEUT and GiBUU. These generators provide varied interaction models and event topologies, allowing a thorough estimator performance analysis.
\subsection{Reconstruction Assumptions}
\label{sec:perfectReco}

In this work, we assume an idealized particle identification and reconstruction accuracy, in which the detection of final-state particles is taken to be 100\% efficient above the set of realistic detector thresholds reported in Table~\ref{tab:dune_thresholds}, and there is no misidentification of one particle type as another. 
The adopted thresholds are consistent with literature \cite{Friedland:2018vry} and motivated by expected capabilities for future LArTPCs \cite{DUNE:2015lol}.  In all calculations, we assume neutrons are not detected. We also ignore flavored hadrons such as kaons and hyperons as their impact on the estimated energy spectra will be negligible on account of their rarity and the detection performance of LArTPCs on these particles is not yet well understood.

\begin{table}[htbp]
  \caption{Detection thresholds  for perfect reconstruction, with values corresponding to the momentum of each particle.}
  \label{tab:dune_thresholds}
   \begin{tabular*}{\linewidth}{@{\extracolsep{\fill}}lccccc}
    \hline\hline
                  & $p$ & $\pi^{\pm}$ & $\pi^0$ & $\mu$ & $e$ \\ \hline
    Thresholds (GeV) & 0.3 & 0.1 & 0.0 & 0.1 & 0.05 \\ \hline\hline
  \end{tabular*}
\end{table}

Although these assumptions do not represent the true performance of any detector, they provide a useful starting point for understanding the energy reconstruction techniques under study.

Unless otherwise stated, we assume perfect kinematic reconstruction of particles above the thresholds in Table~\ref{tab:dune_thresholds}. When discussing the impact of detector resolution effects on neutrino energy estimation, we assume the momentum resolutions shown in Table.~\ref{tab:lartpc_resolutions}. These are estimated from the performance of the MicroBooNE detector on fully contained particles~\cite{MicroBooNE:2025pxk,MicroBooNE:2024bnl,MicroBooNE:2025phj,MicroBooNE:2022tdd}. Additionally, when considering the calorimetric method introduced in Section~\ref{sec:EDepEst}, we assume a resolution on the total energy of 20\%. This choice is justified by the recent MicroBooNE results in Ref.~\cite{MicroBooNE:2019efx,MicroBooNE:2022tdd}.

\begin{table}[htbp]
  \caption{Momentum resolutions for the measurements of different final state particles used to estimate neutrino energy. When applying the calorimetric method, we also include a separate calculation using an overall energy resolution of 20\%.}
  \label{tab:lartpc_resolutions}
   \begin{tabular*}{\linewidth}{@{\extracolsep{\fill}}lccccc}
    \hline\hline
                  & $p$ & $\pi^{\pm}$ & $\pi^0$ & $\mu$ & $e$ \\ \hline
    Resolution (\%) & 15 & 25 & 25 & 10 & 10 \\ \hline\hline
  \end{tabular*}
\end{table}
\section{Energy Estimators}
\label{sec:EnergyEstimators}
We compare five energy estimators: (i) one that assumes all interactions are CCQE-like (\textit{CCQE-like estimator}); (ii) a new approach that we wish to benchmark against existing methods, that improves on the CCQE-like estimator by accounting for multiproton knockout and pion-producing interactions through the hadronic invariant mass (\textit{$W^2$ estimator}); (iii) the sum of the lepton energy and proton kinetic energies (\textit{proton-based estimator}); (iv) the total energy of all final-state particles, intended to mimic calorimetric reconstruction (\textit{calorimetric method}); and (v) the Sobczyk–Furmanski method~\cite{Furmanski:2016wqo} (\textit{SF Estimator}). 

\subsection{CCQE-like Estimator}
\label{sec:CCQEEst}

This estimator assumes charged-current quasi-elastic (CCQE) kinematics to reconstruct neutrino energies, predominantly using the lepton momentum and deflection angle with respect to the beam direction. The estimated neutrino energy ($E^{\rm CCQE}_{\rm est}$) is defined as~\cite{Lalakulich:2012hs}:
\begin{equation}
    E^{\mathrm{CCQE}}_{\rm est} = \frac{m_p^2 - (m_n - E_b)^2 - m_l^2 + 2(m_n - E_b)E_l}{2(m_n - E_b - E_l + p_l \cos\theta_l)},
\end{equation}
where $m_p$, $m_n$, and $m_l$  are the masses of the proton, neutron, and scattered charged lepton, respectively, $E_b$ is a binding energy correction which we take to be 30~MeV for argon~\cite{MicroBooNE:2021ccs}, and $E_l$, $p_l$,and $\theta_l$  are the energy, momentum, and scattering angle of the charged lepton. We will study the performance of this estimator under three different conditions: applying this formula to every event regardless of topology, only applying this estimator to events with one visible proton and no visible pions (1p0$\pi$ events), and similar to~\cite{Ankowski:2015jya}, applying this formula inclusively but replacing $m_p$ with $m_{\Delta} = 1.232$~GeV when any mesons are observed. The first of these three approaches is intended to mimic the performance of water Cherenkov detectors in that they are unable to detect most final state hadrons, and therefore must apply this formula inclusively. The second choice is to only apply the estimator to the topology that matches the assumptions used to derive it. The use of the $\Delta$ mass in the third case is motivated by interpreting the event as a resonance producing interaction if pions are observed.

\subsection{W$^2$-based Estimator}
\label{sec:W2Est}

The W$^2$-based estimator improves on the CCQE-like method through utilization of the visible hadronic invariant mass:
\begin{equation}
    W^2_{\rm vis} = \left(\sum_i E_i\right)^2 - \left(\sum_i \textbf{p}_i\right)^2,
\end{equation}
where $E_i$ and $\textbf{p}_i$ are the total energy and 3-momentum of the $i$th visible final state particle respectively, and initial state mass is adjusted according to the number of visible final state protons ($n_p$). These modifications extend assumptions made by the CCQE-like method to account for the existence of meson producing and multi-nucleon knockout interactions. The expression for the estimated energy $\left(E^{W^2}_{\rm est}\right)$ is:
\begin{equation}
    E^{W^2}_{\rm est} = \frac{W_{\rm vis}^2 - n_p^2(m_n - E_b)^2 - m_l^2 + 2n_p(m_n - E_b)E_l}{2(n_p m_n - n_p E_b - E_l + p_l \cos\theta_l)}.
\end{equation}
This method is only applied to final states with at least one visible proton. 

\subsection{Proton-Based Estimator}
\label{sec:PeLEEEst}

Designed specifically for low-energy neutrino interactions, this estimator combines detailed track reconstruction with calorimetric measurements.  The expression for the estimated energy ($E^{\rm PB}_{\rm est}$) in this instance is given by the lepton total energy ($E_l$) plus the sum of the kinetic energies of all visible protons ($T_p$):
\begin{equation}
    E^{\rm PB}_{\rm est} = E_l + \sum_{\rm protons} T_p.
\end{equation}
Similar to the method used in the MicroBooNE analyses testing the MiniBooNE anomaly~\cite{MicroBooNE:2021wad,MicroBooNECollaboration:2024cpi}, this estimator is only applied to the exclusive topologies containing no visible mesons in the final state.

\subsection{Calorimetric Method}
\label{sec:EDepEst}

Some other tracking calorimeter detectors studying multi-GeV neutrinos, such as NOvA~\cite{NOvA:2007rmc} and MINERvA~\cite{MINERvA:2013zvz}, rely methods driven by calorimetry to measure the energy of the final state hadronic system~\cite{NOvA:2021eqi,MINERvA:2015ydy}, without attempting to identify individual particles. Furthermore, detectors focused on even very high energies such as IceCube~\cite{IceCube:2006tjp} detect exclusively DIS interactions, and are purely calorimetric (or measure quantities analogous to energy deposition). This method is distinguished from the others in this review in that high level reconstructed objects such as individual particle kinematics are not required; the energy measurement can be performed in a LArTPC by measuring the total charge collected by the readout that is not attributed to cosmic rays or radiological backgrounds. 

In this case the estimated energy ($E^{\rm SKE}_{\rm est}$) is modeled by summing the total lepton total energy ($E_l$), the kinetic energies of all visible protons ($T_p$), and total energies of all pions ($E_{\pi}$):
\begin{equation}
    E^{\rm SKE}_{\rm est} = E_l + \sum_{\rm protons} T_p + \sum_{\rm pions} E_{\pi}.
    \label{eq:Calo}
\end{equation}
Unlike the other methods explored in this paper, this method does not rely on high level reconstructed objects such as particle tracks or showers. 

\subsection{Sobczyk-Furmanski Estimator}
\label{sec:SFEst}

The Sobczyk-Furmanski (SF) method~\cite{Furmanski:2016wqo} analyses the energy balance by separating the kinematics into transverse (T) and longitudinal (L) components, to infer the additional energy supplied by the motion of the struck nucleon. The neutrino energy is expressed in terms of the mass of the parent nucleus, $M_A$, the mass of the daughter nucleus $M_{A-1}$, and the combined transverse/longitudinal momenta of the lepton-proton system, $p^{T,L} =p^{T,L}_{\rm prot} + p^{T,L}_{l}$. The estimated neutrino energy ($E^{\rm SF}_{\rm est}$) is then obtained from the following expressions:
\begin{gather}
    p^L = \frac{(M_A + p_{l}^{L} + p_{\rm prot}^{L} - E_l - E_{\rm prot})^2 - (p^T)^2 - M_{A-1}^2}{2(M_A + p_{l}^{L} + p_{\rm prot}^{L} - E_l - E_{\rm prot})}, \\ 
    E^{\rm SF}_{\rm est} = p_{l}^{L} + p_{\rm prot}^{L} - p^L.
\end{gather}
This estimator is only be applied to exclusive topologies containing a single visible proton and the charged lepton.
\subsection{Inclusivity Considerations}
\label{sec:incexcl}

An important factor in comparing different energy estimation techniques is the degree of inclusiveness of the event sample to which each method can be validly applied, as this directly impacts the available statistics. The CCQE-like and calorimetric estimators are valid for all hadronic final states, though the CCQE-like estimator is not expected for perform well on events that are not true CCQE, hence the two modified forms. The $W^2$ estimator both requires at least one visible proton, and the proton based estimator events with no visible pions. The SF estimator is the most exclusive, requiring a clean topology with exactly one visible proton and no pions.

\begin{figure}[h!]
    \centering
    \subfigure[Rates normalized to equal  arbitrary exposure.]{\includegraphics[width=\linewidth]{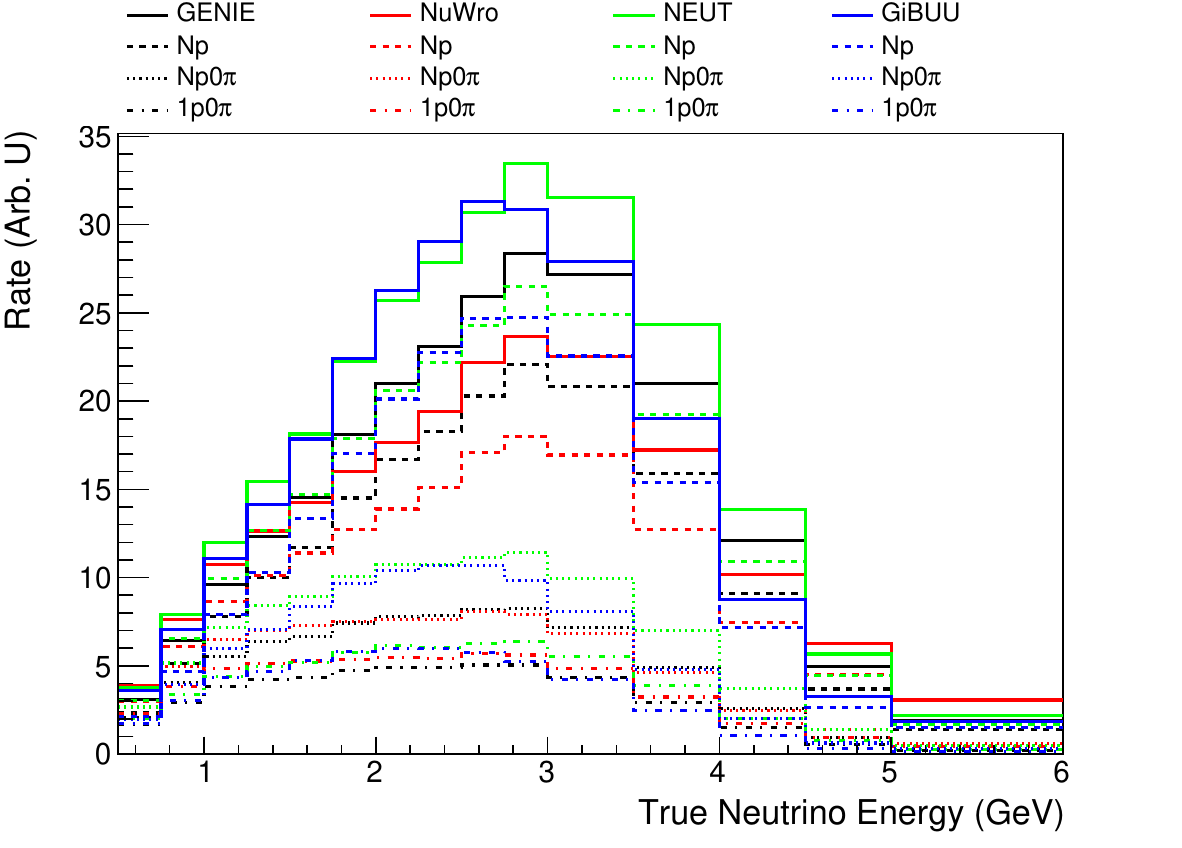}}
    \subfigure[Fractions of samples used by each exclusive channel.]{\includegraphics[width=\linewidth]{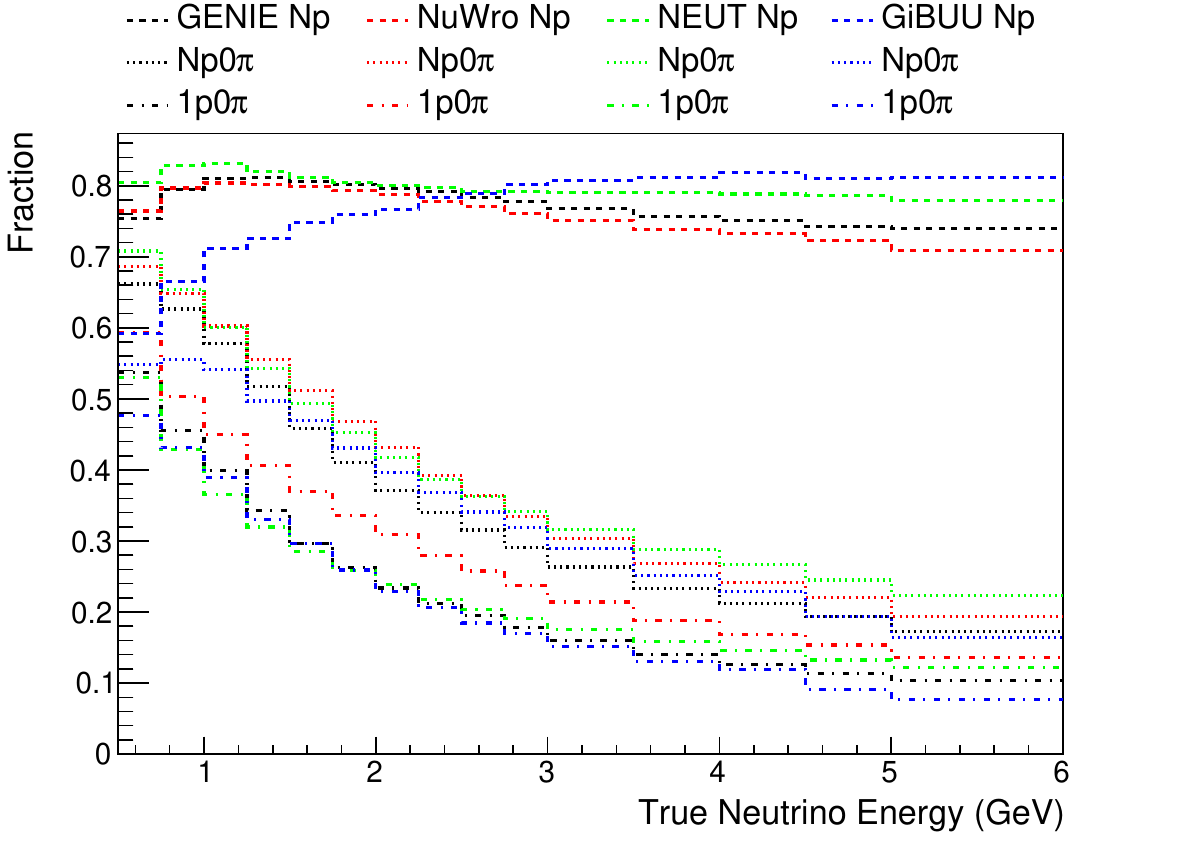}}
    \caption{Rates and fractions of $\nu_{\mu}$ interactions on argon that produce final states with at least one proton (Np), at least one proton and no pions (Np0$\pi$), and exactly one proton and no pions (1p0$\pi$). These are the subsets of events targeted by the $W^2$, proton based, and SF methods respectively. The CCQE-like and calorimetric estimators are fully inclusive.}
    \label{fig:Efficiencies}
\end{figure}

While the total predicted cross section is relatively consistent across generators, there are substantial variations in the predicted exclusive channels. This is illustrated in Fig.~\ref{fig:Efficiencies}, which shows the total rate normalized to arbitrary exposure for each exclusive channel, along with the fractions of all events producing final states with at least one proton and any number of pions (Np), at least one proton and no pions (Np0$\pi$), and exactly one proton and no pions (1p0$\pi$). This shows substantial loss of statistics in the multi-GeV region when excluding events with final state pions, where the second oscillation maximum is expected to appear at DUNE's baseline.

\section{Response Matrices}
\label{sec:ResponseMatrices}
\begin{figure*}
\centering     
    \subfigure[CCQE-like estimator.]{\includegraphics[width=0.33\linewidth]{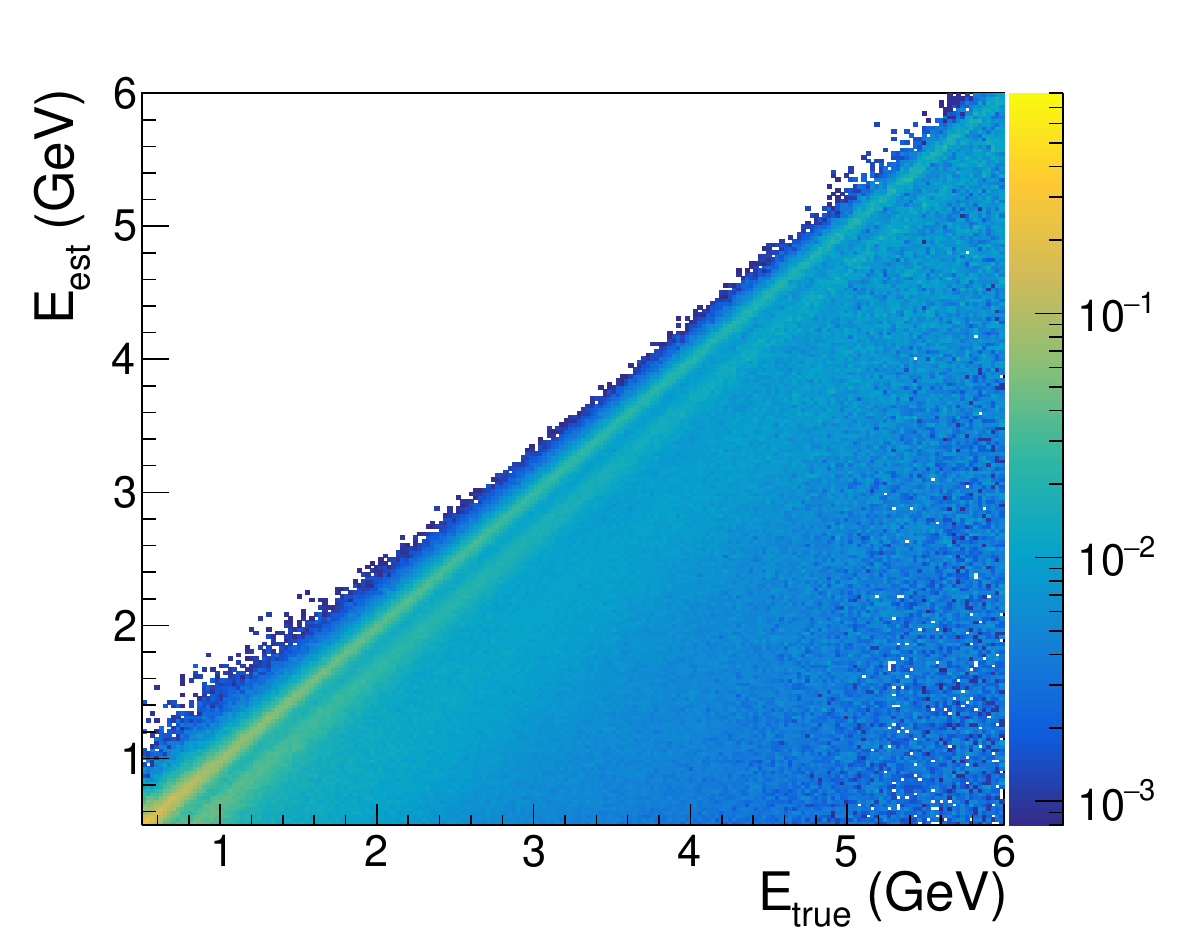}}%
    \subfigure[CCQE-like estimator only applied to single proton events with no pions.\label{fig:ConfusionMatricesMuonKinCCQE}]{\includegraphics[width=0.33\linewidth]{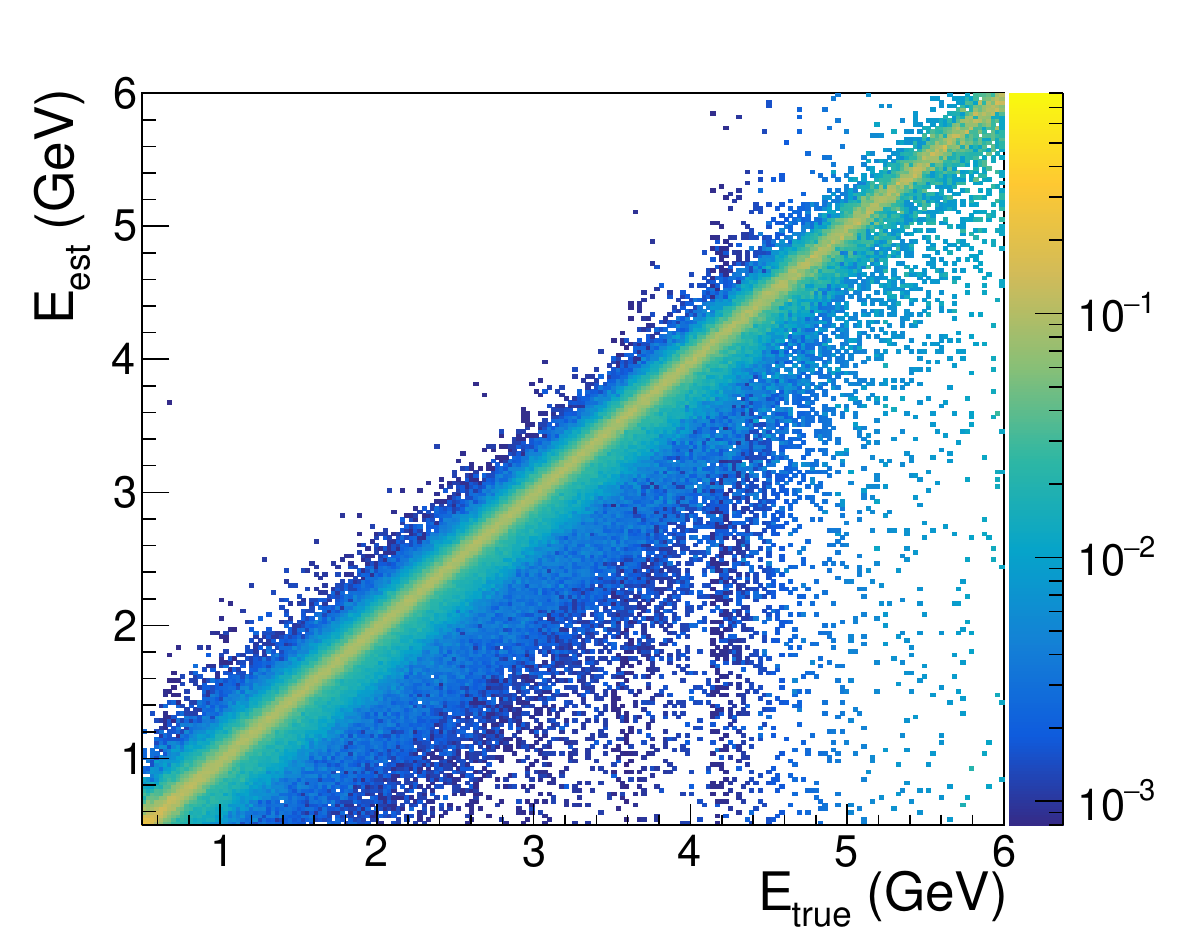}}%
    \subfigure[CCQE-like estimator with $\Delta$ mass correction in meson producing events.\label{fig:ConfusionMatricesMuonKinDelta}]{\includegraphics[width=0.33\linewidth]{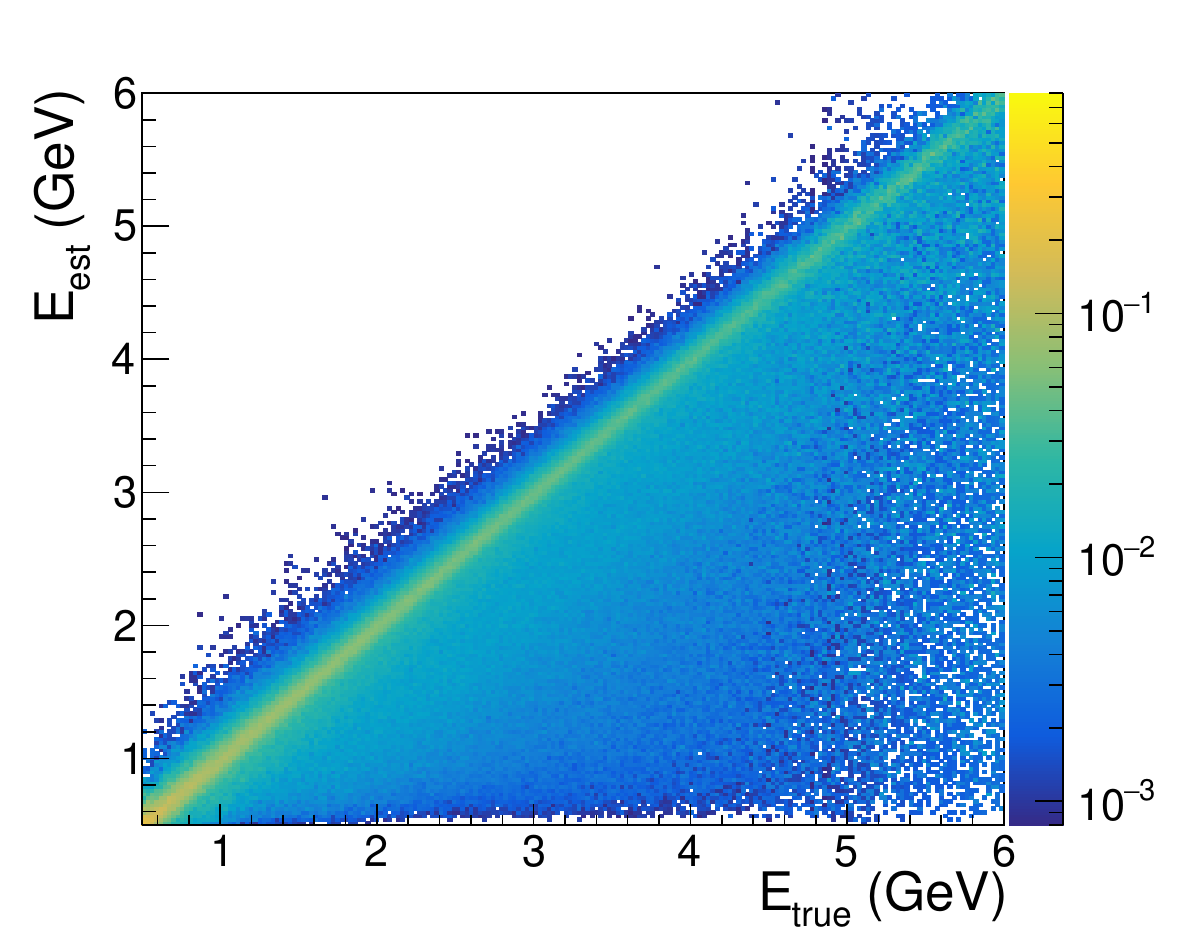}}
    \subfigure[$W^2$-based estimator.]{\includegraphics[width=0.33\linewidth]{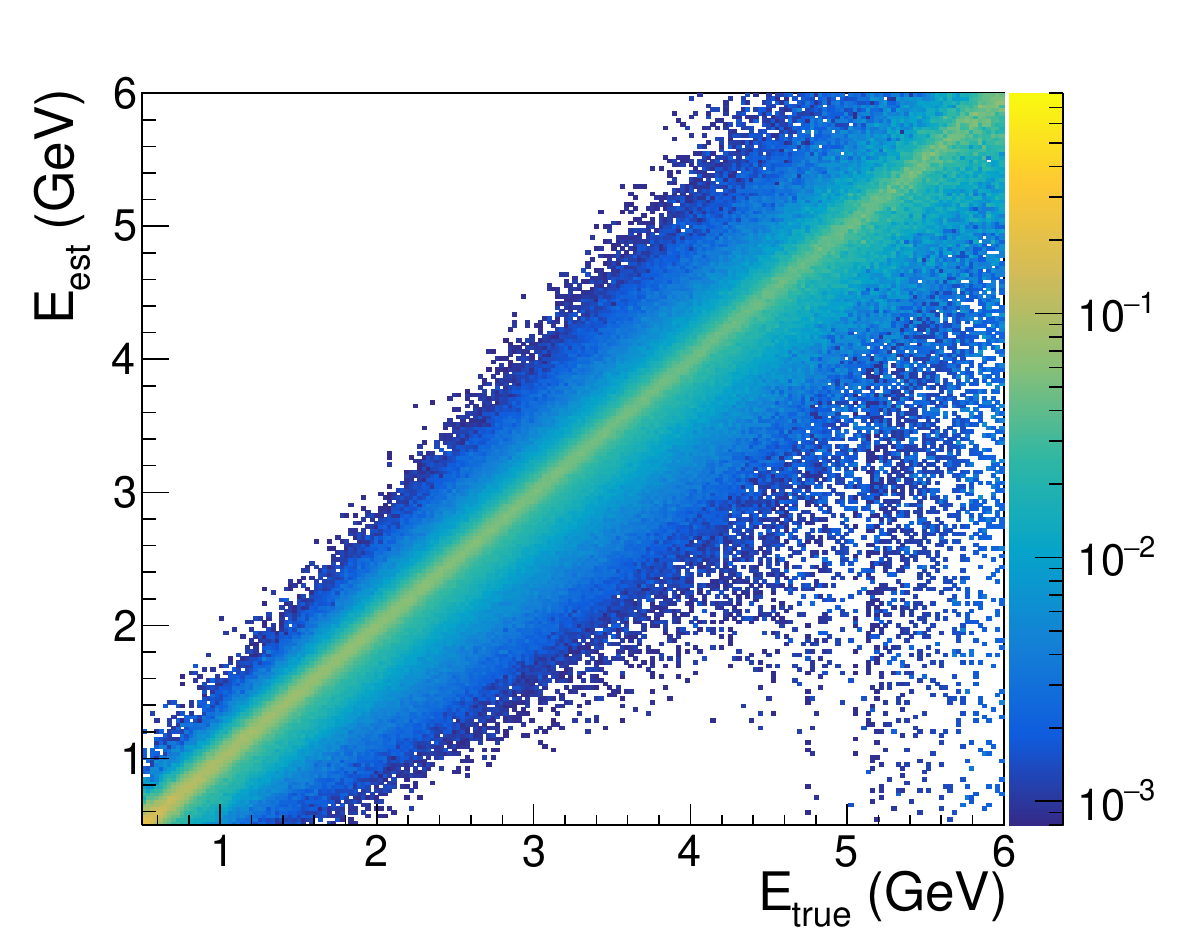}}%
    \subfigure[Proton-based estimator.]{\includegraphics[width=0.33\linewidth]{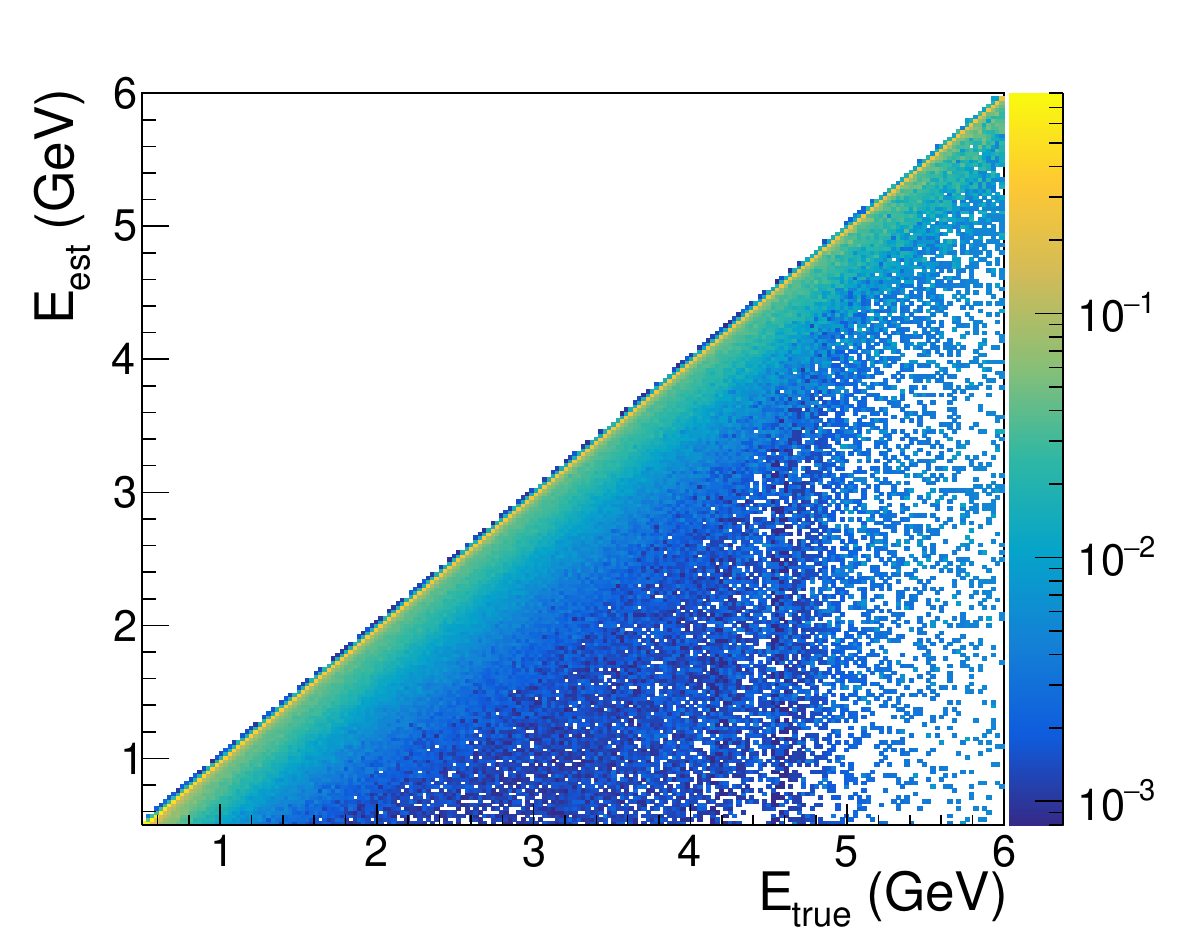}}
    \subfigure[SF estimator.]{\includegraphics[width=0.33\linewidth]{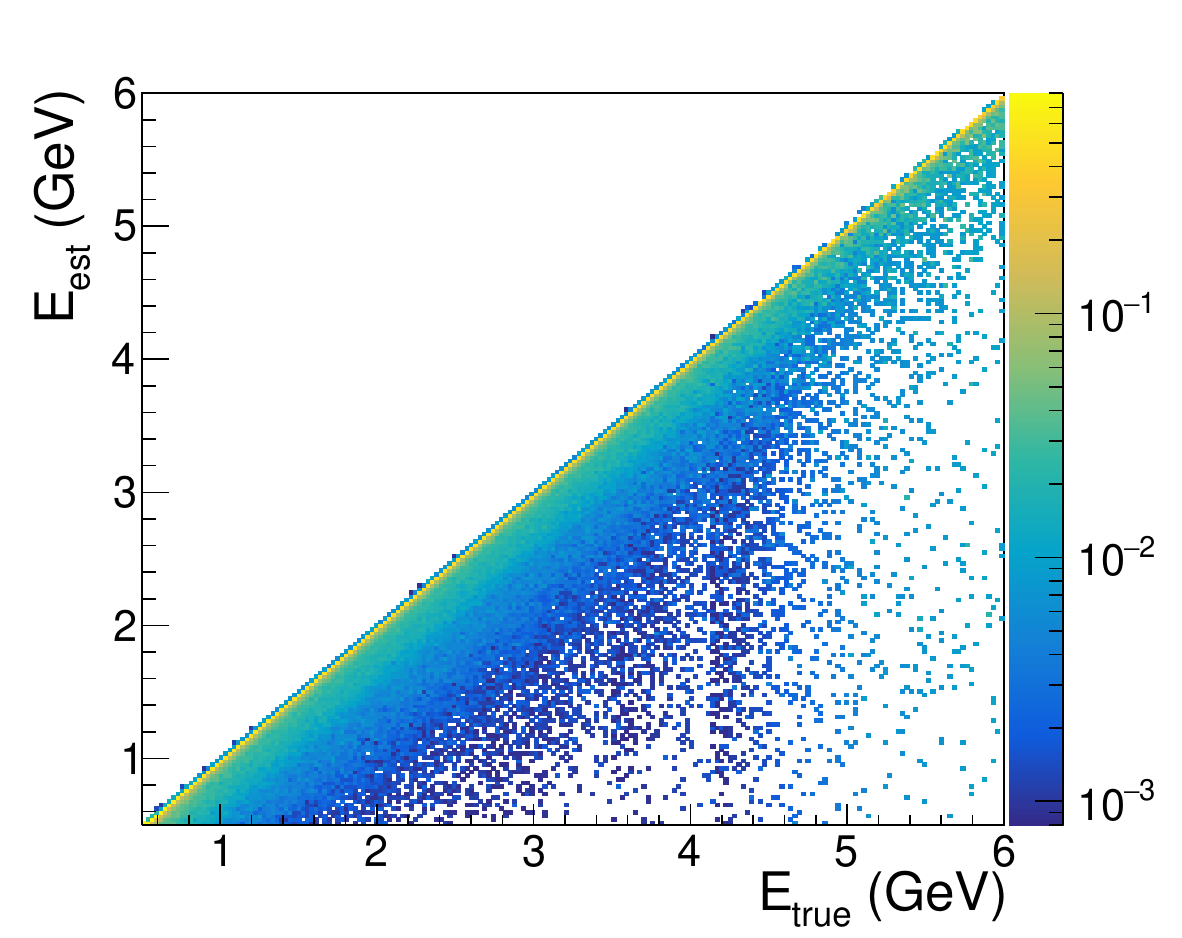}}%
    \subfigure[Calorimetric method.]{\includegraphics[width=0.33\linewidth]{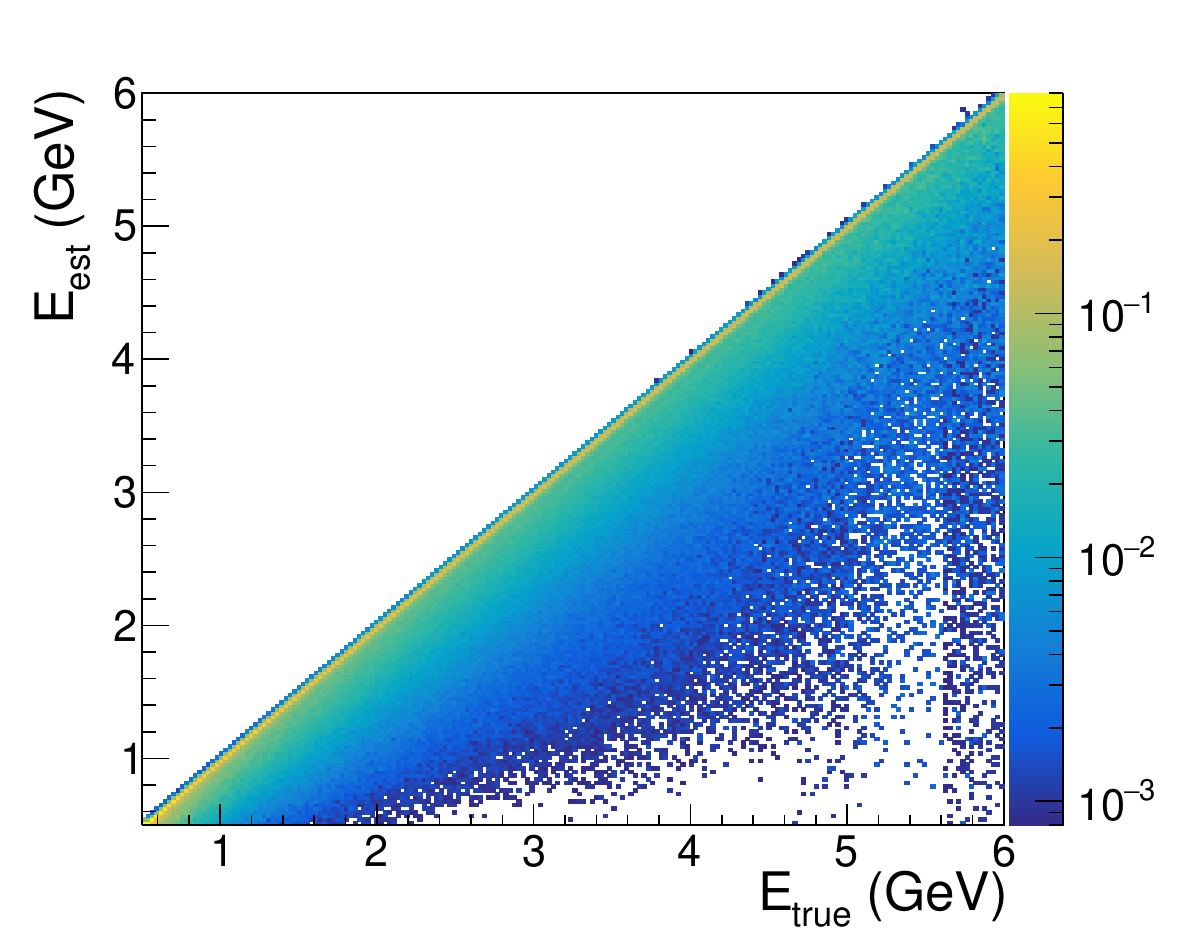}}
    \caption{Response matrices describing the smearing from true neutrino energy to estimated neutrino energy using each energy estimation.}
    \label{fig:ConfusionMatrices}
\end{figure*}

To understand the basic characteristics of each energy estimation method, we begin by calculating the corresponding response matrix. Defined as  
\begin{equation}
    R(E_{\rm est},E_{\rm true}) = \textrm{Prob}(E_{\rm est} | E_{\rm true}) ,
\end{equation}  
the response matrix represents the probability that a neutrino of true energy $E_{\rm true}$ is reconstructed with an estimated energy $E_{\rm est}$. These matrices are normalized such that $\int R(E_{\rm est},E_{\rm true}) dE_{\rm est} = 1$  regardless of the selection criteria applied to the underlying samples; the total number of events in each matrix thus reflects the degree of inclusiveness required by the corresponding energy estimator.
Fig.~\ref{fig:ConfusionMatrices} shows the response matrices for the five energy estimators, using events simulated with GENIE. Response matrices calculated with the other three generators can be found in the Supplemental Material~\cite{SupMat}.

 The matrix for the CCQE-like estimator exhibits a pronounced secondary band offset from the diagonal, primarily arising from multiproton knockout events and final states containing pions, processes neglected by this method. We can see the performance of this estimator improves considerably when only applied to events with single protons, shown in Fig.~\ref{fig:ConfusionMatricesMuonKinCCQE}, or when incorporating the $\Delta$ mass correction in Fig.~\ref{fig:ConfusionMatricesMuonKinDelta}. The $W^2$ estimator improves on these methods further, with similar performance to the single proton form over a far broader set of final states, and a narrower peak than the $\Delta$ corrected version.

The proton-based, SF, and calorimetric estimators all feature narrow diagonal peaks and a secondary population in the lower-right quadrant. This reflects that these methods essentially reconstruct the time component of the total final-state four-momentum, which equals the true neutrino energy in the absence of final state interactions (FSI), nucleon motion, or undetected particles. The off-diagonal population arises from events affected by FSI, which are challenging to model. The influence of FSI on the different estimators is discussed further in Section~\ref{sec:FSI_Focus}.

\begin{figure}
    \centering     
    \subfigure[Assuming perfect reconstruction.]{\label{fig:FractionalErrorNoSmear}\includegraphics[width=\linewidth]{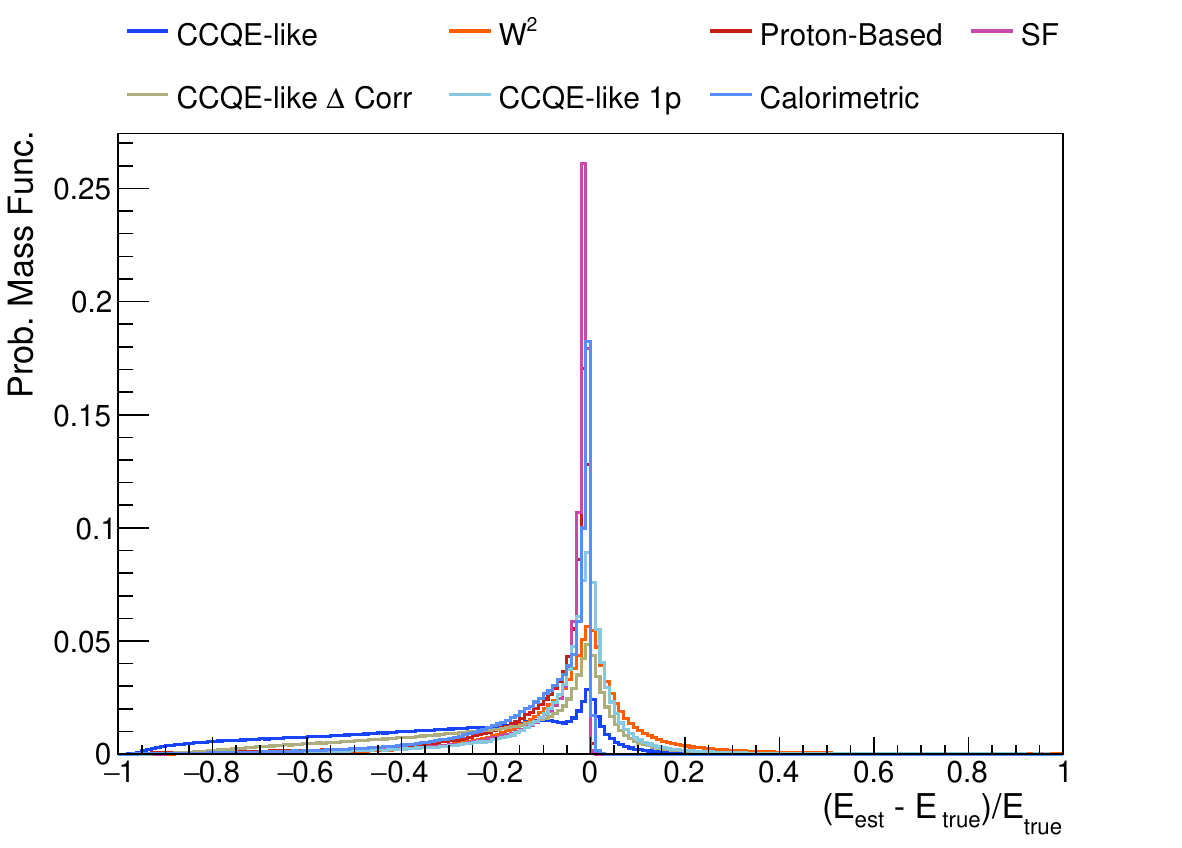}}
    \subfigure[Shape of fractional bias around zero.]{\label{fig:FractionalErrorNoSmearZoom}\includegraphics[width=\linewidth]{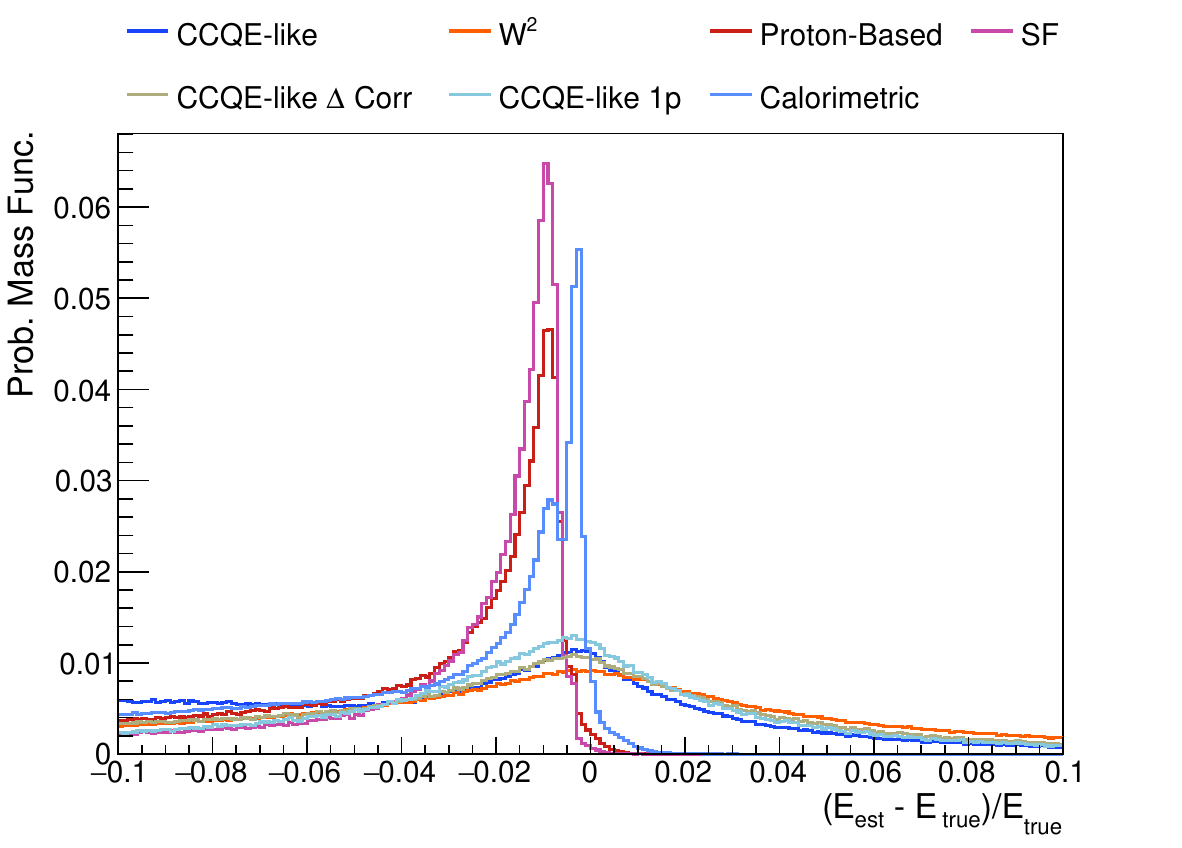}}
    \subfigure[After applying resolution effects.]{\label{fig:FractionalErrorSmear}\includegraphics[width=\linewidth]{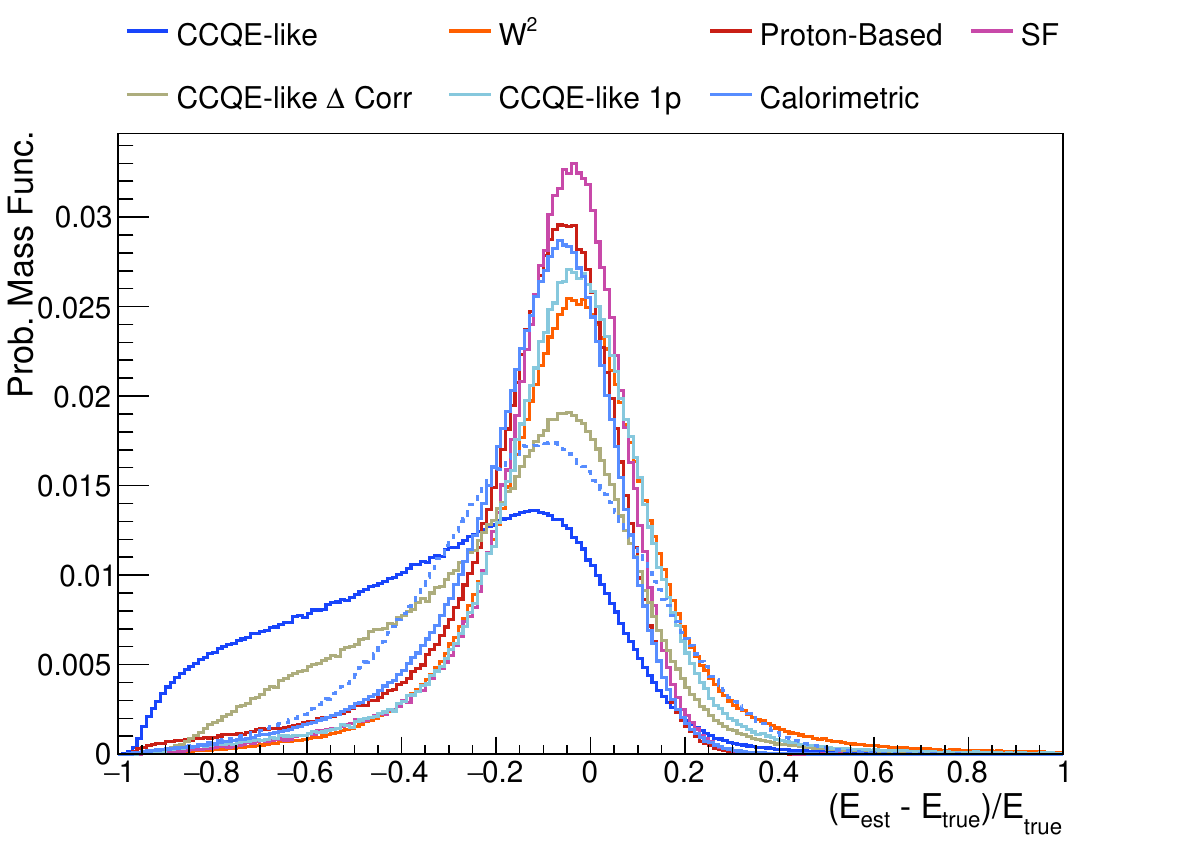}}
    \caption{The distribution of fractional error in neutrino energy using the different estimation methods, calculated using events simulated with the GENIE neutrino event generator. All distributions are normalized to 1. The dashed line is the calculation assuming overall 20\% smearing on the energy calculated with the calorimetric method. The curves labeled ``CCQE-like 1p'' and ``CCQE-like $\Delta$ Corr'' refer to the cases in which the CCQE-like estimator has been applied to 1p0$\pi$ events only and the CCQE-like estimator with the $\Delta$ mass correction respectively.}
    \label{fig:FractionalError}
\end{figure}

The fractional error, defined as $(E_{\rm est} - E_{\rm true})/E_{\rm true}$, further illustrates how the choice of estimator and the inclusion of resolution effects influence neutrino energy reconstruction.
Figure~\ref{fig:FractionalError} shows the fractional error distributions for the different estimators obtained using GENIE, assuming perfect kinematic resolutions in Fig.~\ref{fig:FractionalErrorNoSmear}, and after applying smearing effects in Fig.~\ref{fig:FractionalErrorSmear}. For the calorimetric method we separately calculate the resolution applying particle-by-particle smearing of the energies used in equation~\ref{eq:Calo} (solid pale blue line) and an overall 20\% effect (dashed pale blue line) to simulate effects impacting the quantity of charge detected by wire planes. The results for other generators are provided in the Supplementary Material~\cite{SupMat}. 

Assuming perfect kinematic reconstruction, the proton-based, SF, and calorimetric estimators exhibit the best performance. After application of resolution effects the best estimator is the SF method, benefiting from the better resolution on the momentum of protons compared with that of pions and the estimators that try to analyze them. The CCQE-like estimator shows a narrow peak near zero with a pronounced asymmetric tail, corresponding to CCQE and non-CCQE interactions, respectively, a feature that is greatly diminished when the estimator is restricted to only single proton events. A comparison between the CCQE-like and $W^2$ estimators highlights the improvement gained by incorporating the invariant mass into the energy reconstruction. Both produce similarly shaped peaks near zero, but the $W^2$ method notably lacks the long tail characteristic of the CCQE-like estimator.

\section{Bias, and Variance}
\label{sec:BiasAndVariance}
We further compare the effectiveness of different neutrino energy estimators by calculating the fractional bias $B(E_{\rm true})$ and fractional variance $V(E_{\rm true})$ of each estimator as a function of true neutrino energy. These are defined as:
\begin{align}
    B(E_{\rm true}) = \int R(E_{\rm est},E_{\rm true})  \left(\frac{E_{\rm est} - E_{\rm true}}{E_{\rm true}}\right) dE_{\rm est} \\
    V(E_{\rm true}) = \int R(E_{\rm est},E_{\rm true})  \left(\frac{E_{\rm est} - \overline{E_{\rm est}}}{\overline{E_{\rm est}}}\right)^2 dE_{\rm est}
\end{align}
where $\overline{E_{\rm est}} = \int R(E_{\rm est},E_{\rm true}) E_{\rm est}dE_{\rm est}$ is the mean estimated neutrino energy when the true neutrino energy is $E_{\rm true}$.

\subsection{Estimator Comparisons}
\begin{figure*}
    \centering     
    \subfigure[Fractional bias. We do not show the full range of bias from the CCQE-like and $\Delta$ corrected CCQE-like methods in order to display the other calculations in more detail.]{\label{fig:FractionalBias}\includegraphics[width=0.5\linewidth]{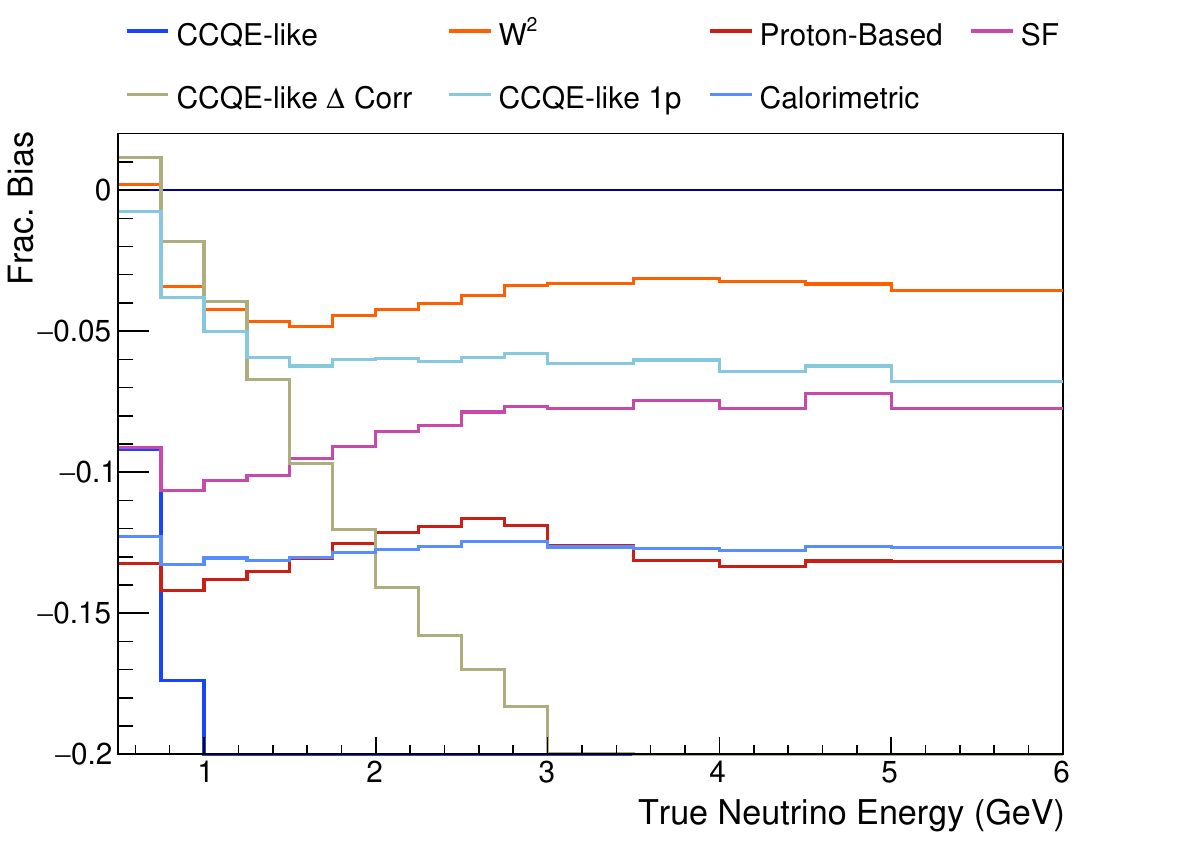}}%
    \subfigure[Fractional variance, not including resolution effects.]{\label{fig:FractionalVariance}\includegraphics[width=0.5\linewidth]{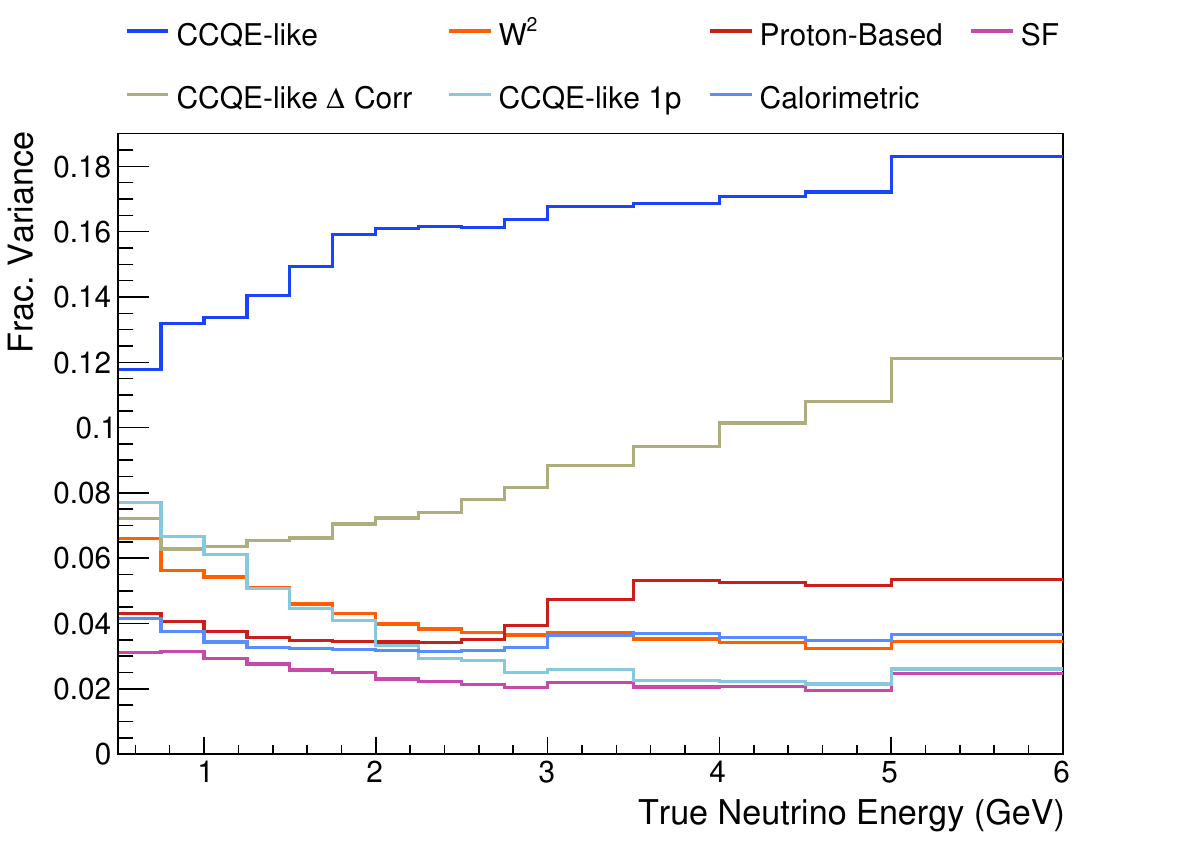}}
    \subfigure[Fractional variance, including resolution effects.]{\label{fig:FractionalVarianceSmeared}\includegraphics[width=0.5\linewidth]{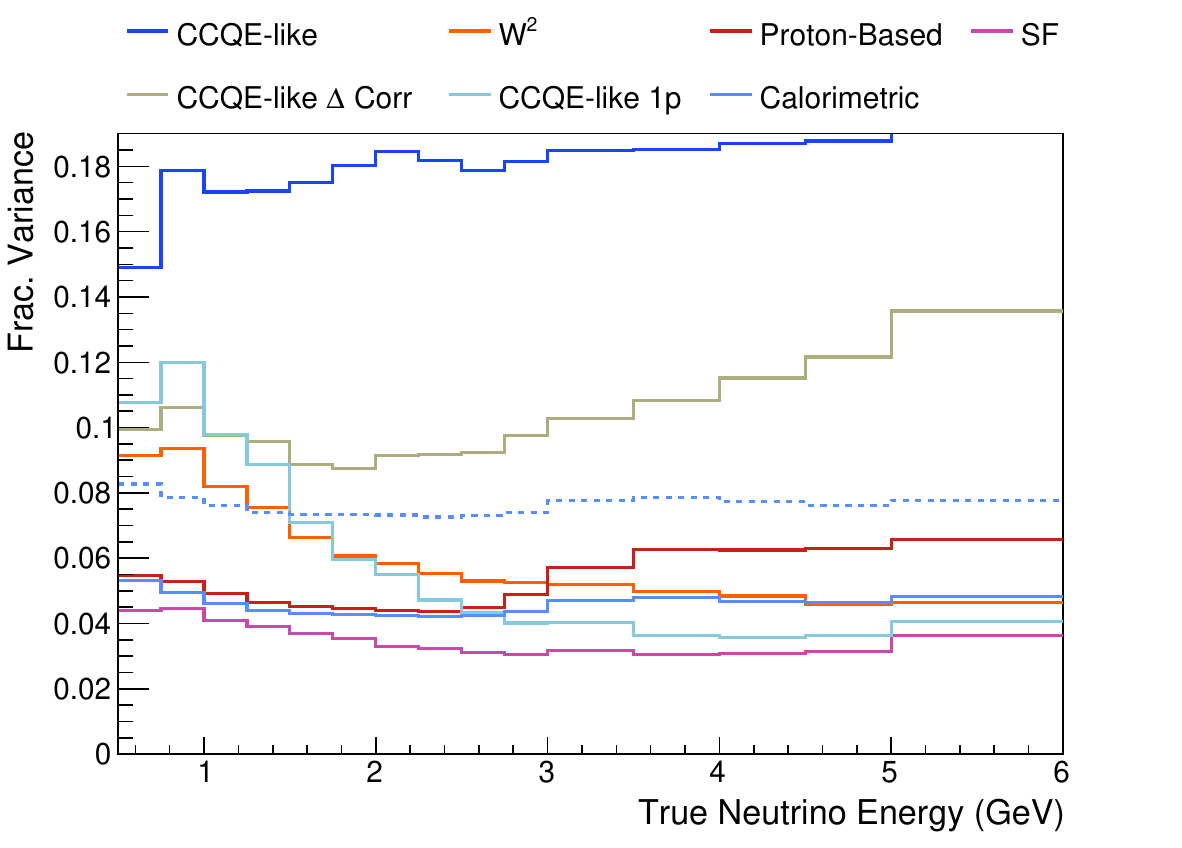}}%
    \subfigure[Increase in fractional variance from application of resolution effects.]{\label{fig:FractionalVarianceChange}\includegraphics[width=0.5\linewidth]{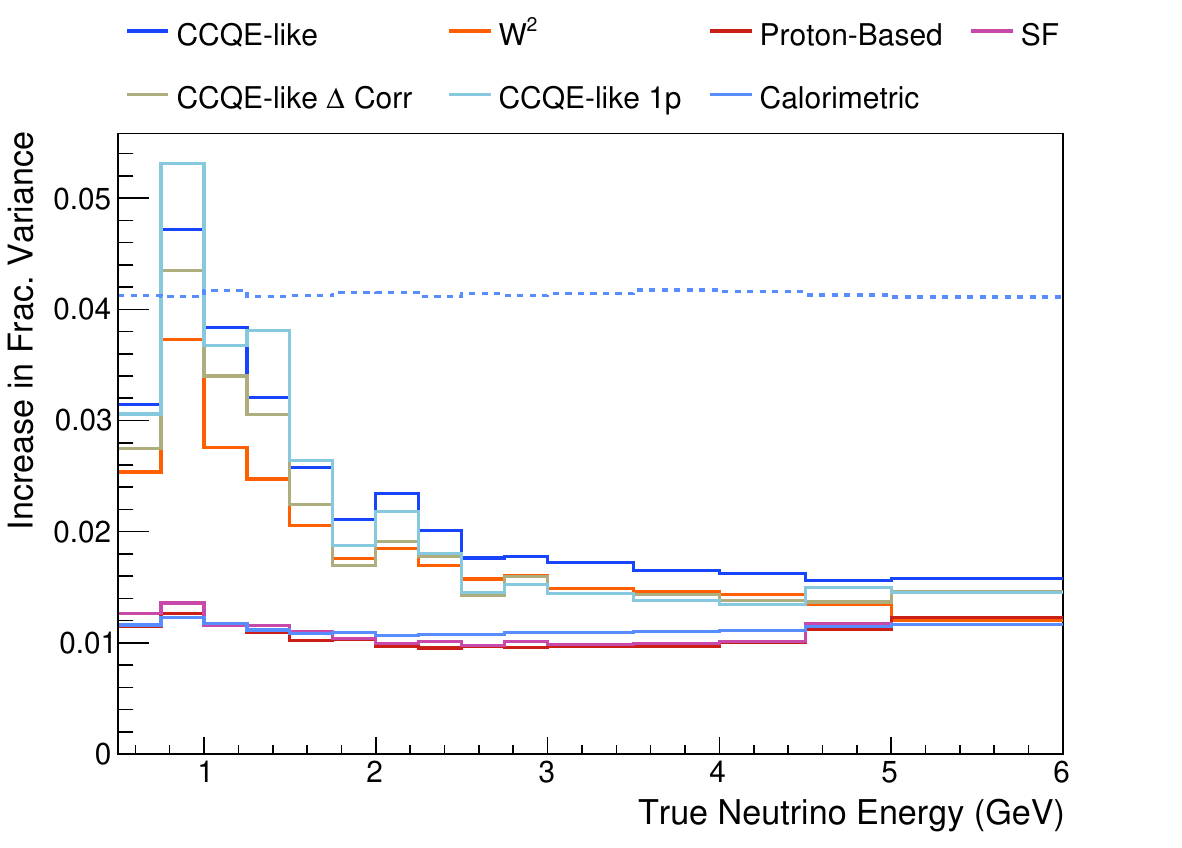}}
    \caption{Fractional bias and variance as a function of true neutrino energy. The dashed lines in panels (b) and (c) indicate the performance of the calorimetric method when applying the full 20\% smearing to the energy estimation.}
    \label{fig:BiasPlots}
\end{figure*}

Fig.~\ref{fig:BiasPlots} shows the fractional bias and variance obtained using GENIE. Among the estimators, the $W^2$ method exhibits the smallest overall bias. This behaviour can be understood by noting that the other estimators yield markedly asymmetric fractional error distributions, as illustrated in Fig.~\ref{fig:FractionalError}. Figs.~\ref{fig:FractionalVariance} and~\ref{fig:FractionalVarianceSmeared} shows the fractional variance in the energy estimators, before and after including the effects of detector resolution. In the case of the calorimetric method, we show the variance when applying both the individual particle resolutions from Table~\ref{tab:lartpc_resolutions}, and the overall 20\% smearing (dashed line). This calculation indicates the SF method offers the best variance across most of the energy spectrum. If individual particle resolutions are applied, the calorimetric method performs the best among the more inclusive methods, though its performance suffers considerably if the more conservative 20\% overall smearing is applied. This suggests experiments should consider whether to use this method if their calorimetric reconstruction performs well enough. After applying resolution effects, the performance of the SF method, proton based methods offer similar performance, followed by the $W^2$ method. 

Fig.~\ref{fig:FractionalVarianceChange} gives the increase in fractional variance for each estimator from applying the resolution effects. This indicates the the SF and proton-based methods are the least susceptible to these effects. The calorimetric method performs similarly, though this again only applies if the most optimistic model of calorimetric reconstruction in the detector is assumed. The $W^2$ and CCQE-like estimators are both more susceptible to resolution effects at low energies. In the case of the $W^2$ estimator this is because at low energies the hadronic system is often only a single proton and thus the energy estimation is purely driven by the kinematics, whereas the other methods will experience cancellations between the smearing effects to the lepton and hadronic kinematics. In the case of the CCQE-like method the resolution at higher energies is more dominated by the unrealistic assumptions this method makes about the neutrino interaction as opposed to the momentum resolution, and thus the momentum smearing is less impactful.

\subsection{Generator Comparisons}
\begin{figure*}
    \centering     
    \subfigure[CCQE-like.]{\includegraphics[width=0.33\linewidth]{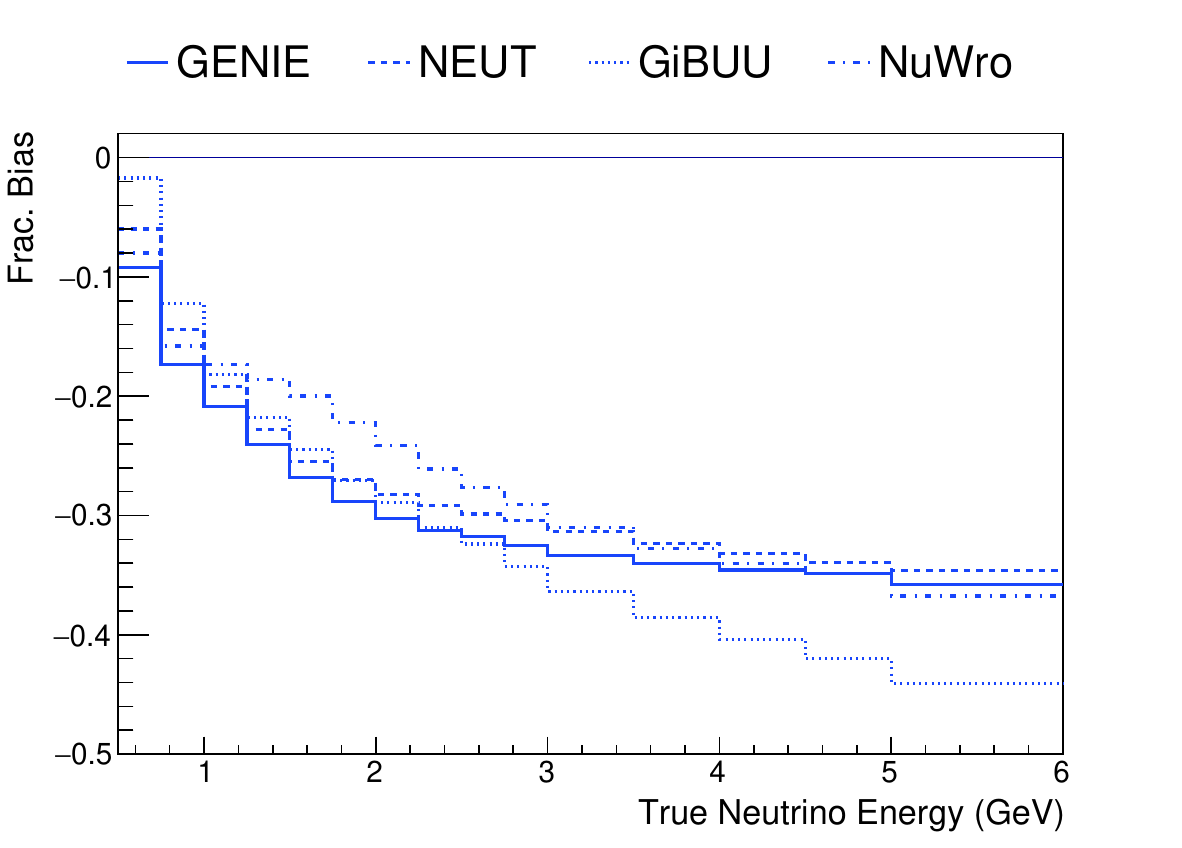}}%
    \subfigure[CCQE-like only applied to 1p0$\pi$ events.]{\includegraphics[width=0.33\linewidth]{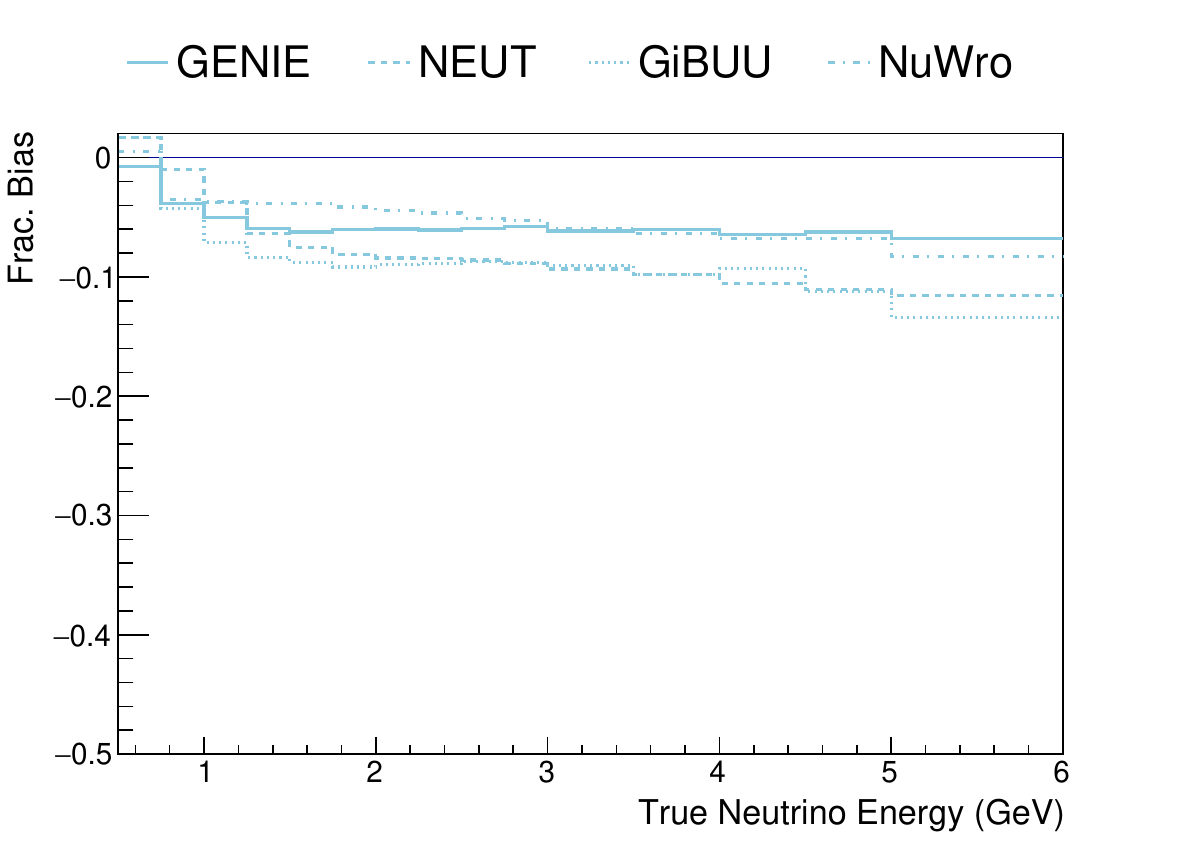}}%
    \subfigure[CCQE-like with $\Delta$ mass correction.]{\includegraphics[width=0.33\linewidth]{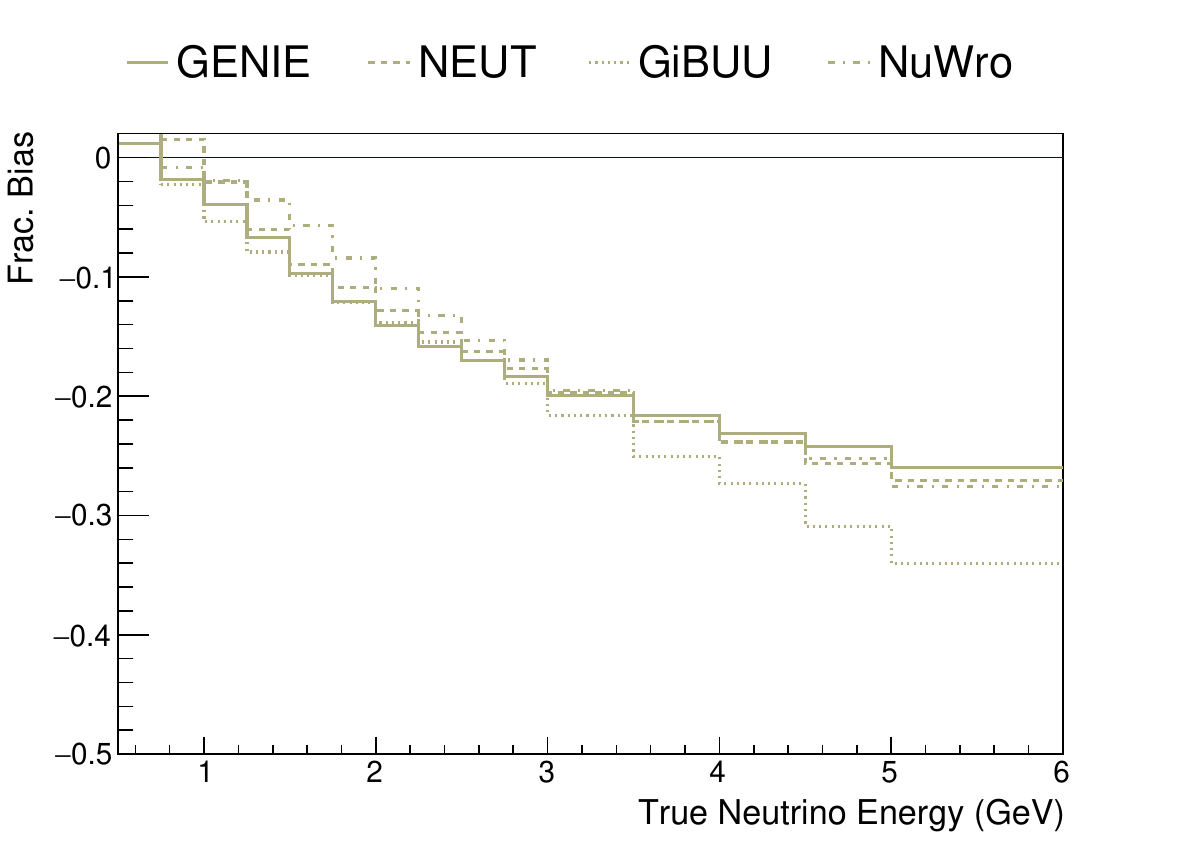}}
    \subfigure[$W^2$ method.]{\includegraphics[width=0.33\linewidth]{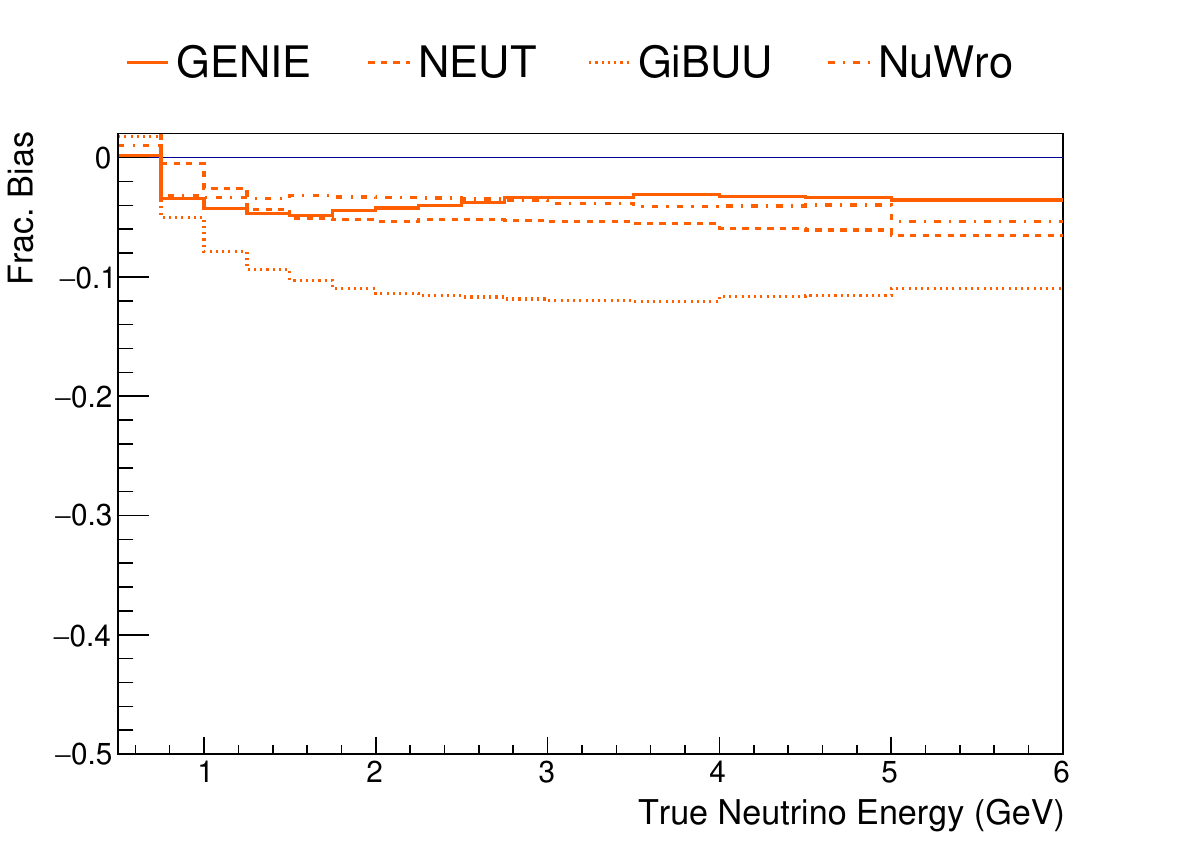}}%
    \subfigure[Proton based.]{\includegraphics[width=0.33\linewidth]{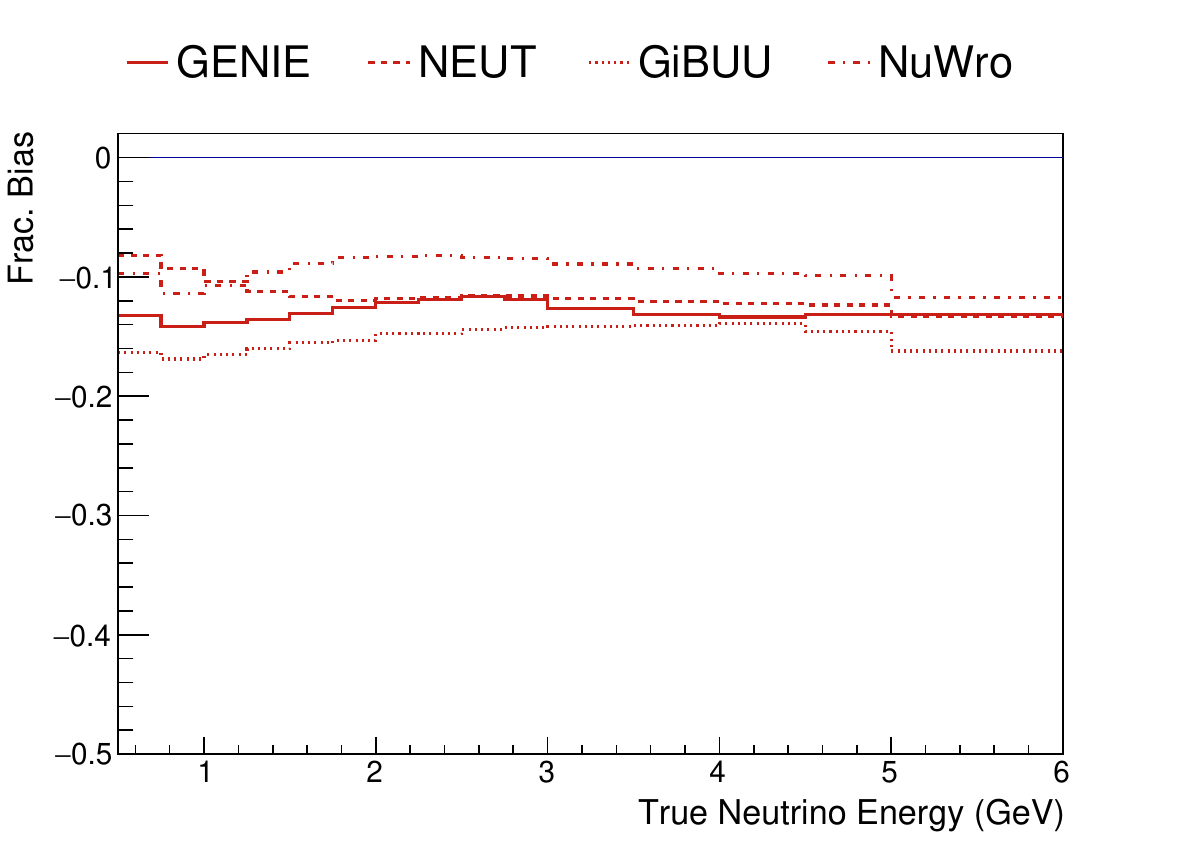}}
    \subfigure[SF method.]{\includegraphics[width=0.33\linewidth]{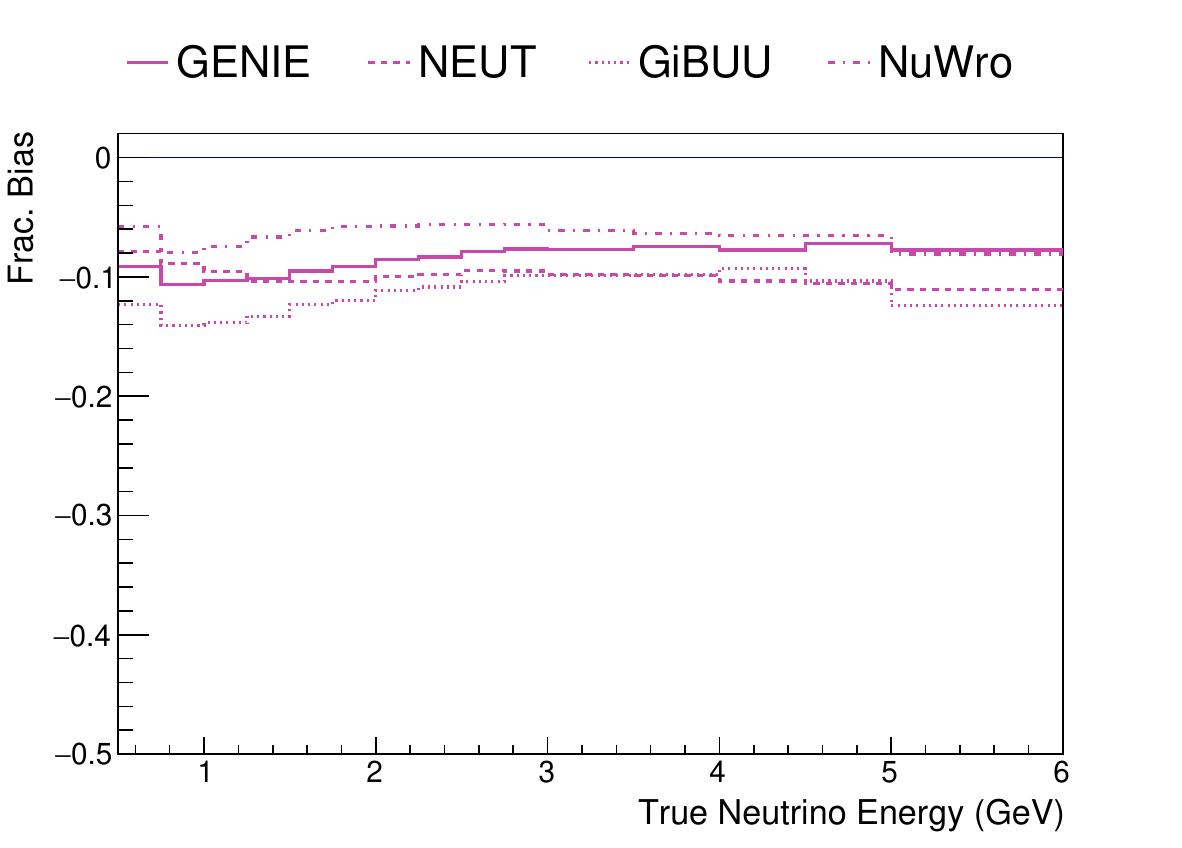}}%
    \subfigure[Calorimetric method.]{\includegraphics[width=0.33\linewidth]{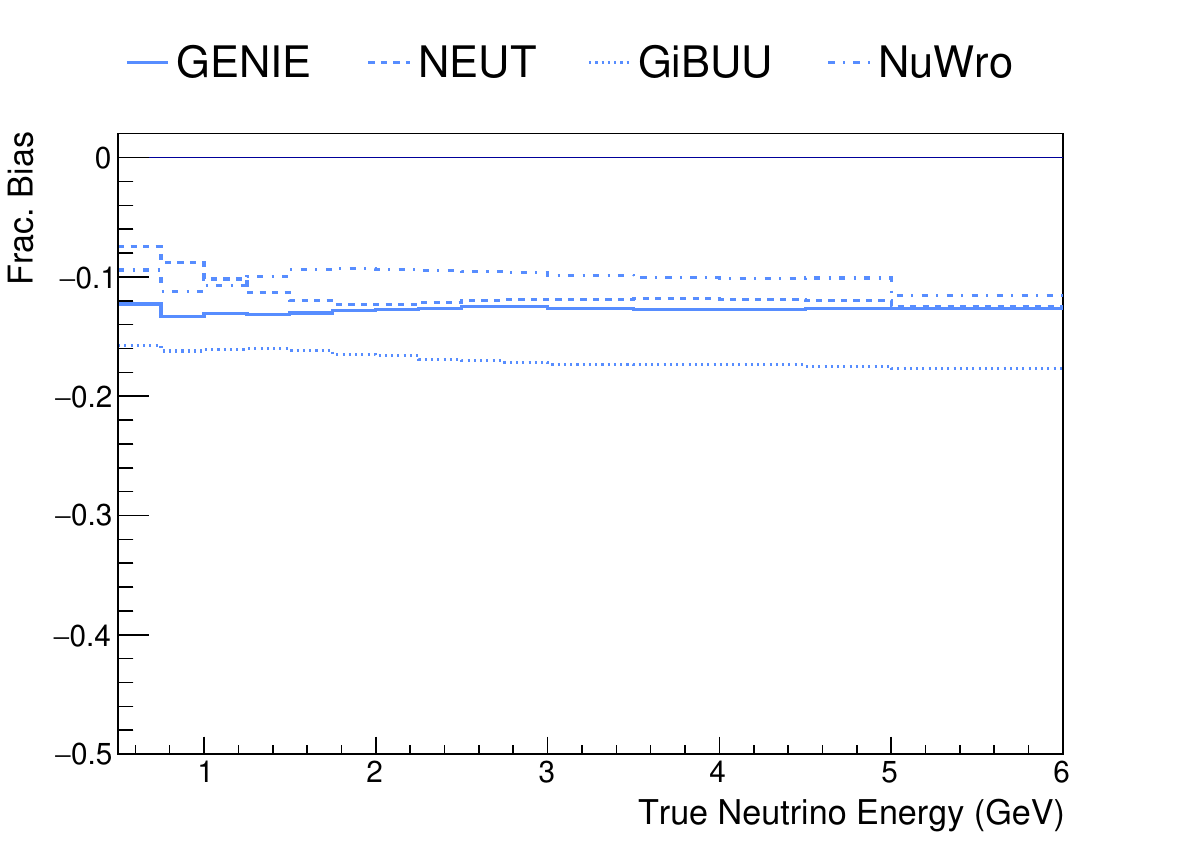}}
    \caption{Comparison of the biases in neutrino energy when calculated using different event generators, for each method of neutrino energy estimation.}
    \label{fig:BiasBands}    
\end{figure*}

Since realistic physics analyses typically employ multiple event generators to evaluate systematic uncertainties, we assess the robustness of each energy estimator with respect to the underlying nuclear model by varying the neutrino event generator and comparing the resulting bias and variance in the reconstructed neutrino energy. Fig.~\ref{fig:BiasBands} shows the bias as a function of true neutrino energy for each generator. The results for the variance are provided in the Supplementary Material~\cite{SupMat}. 

The benefits of the exclusive selection employed by the SF method and the modified versions of the CCQE-like estimator emerge here, as the spread between the biases predicted by each generator for these estimators are noticeably smaller than that of the other estimators. It is also interesting to note that the proton-based method does not perform substantially better than the inclusive methods in this test. Of the three inclusive methods, the $W^2$ and calorimetric methods offer very similar performance, while the CCQE-like method, with and without the $\Delta$ mass correction, shows a substantial divergence between GiBUU and the other generators above 2~GeV. The predictions from GiBUU disagree with the expectation from the other three generators for most of the estimators across much of the energy range studied, expected given GiBUU's different approach to nuclear transport modeling. 

\subsection{Focus on FSI}
\label{sec:FSI_Focus}

\begin{figure*}
    \centering
    \subfigure[GENIE.]{\includegraphics[width=0.47\linewidth]{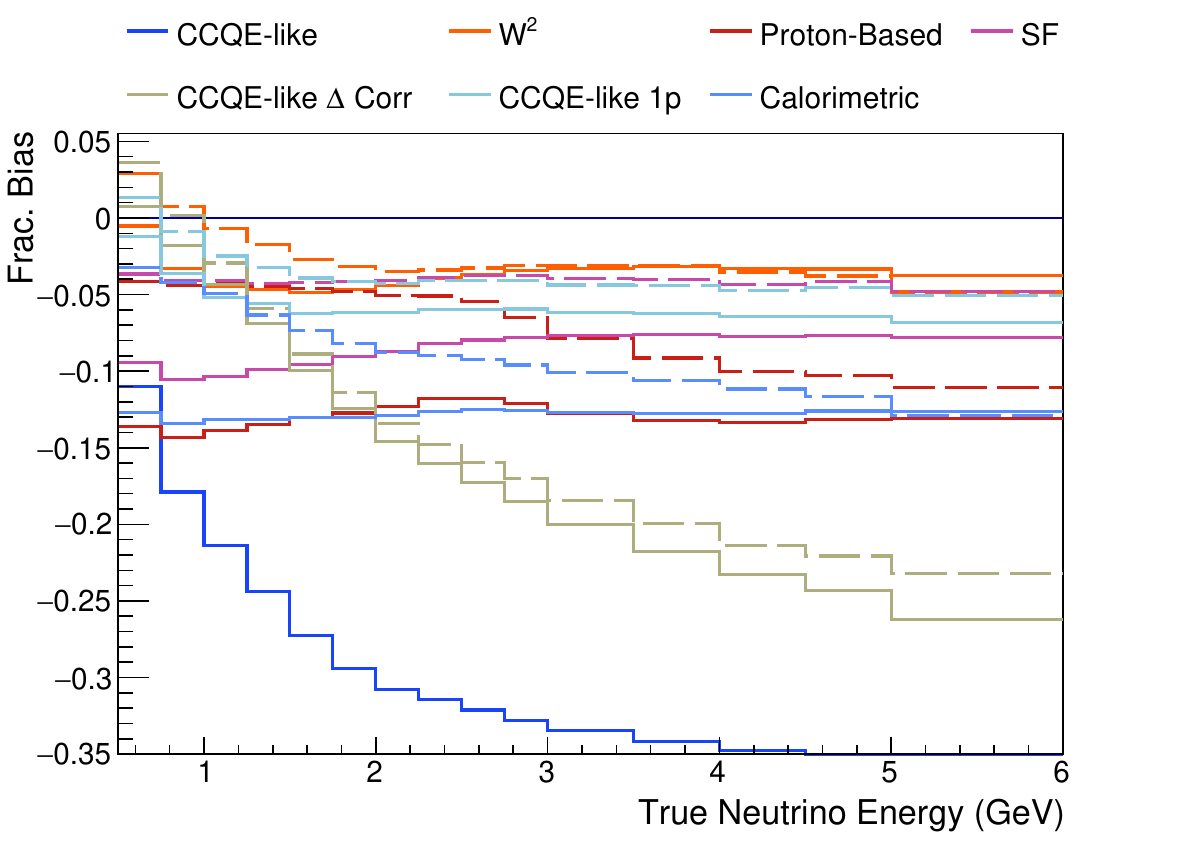}}%
    \hfill
    \subfigure[NuWro.]{\includegraphics[width=0.47\linewidth]{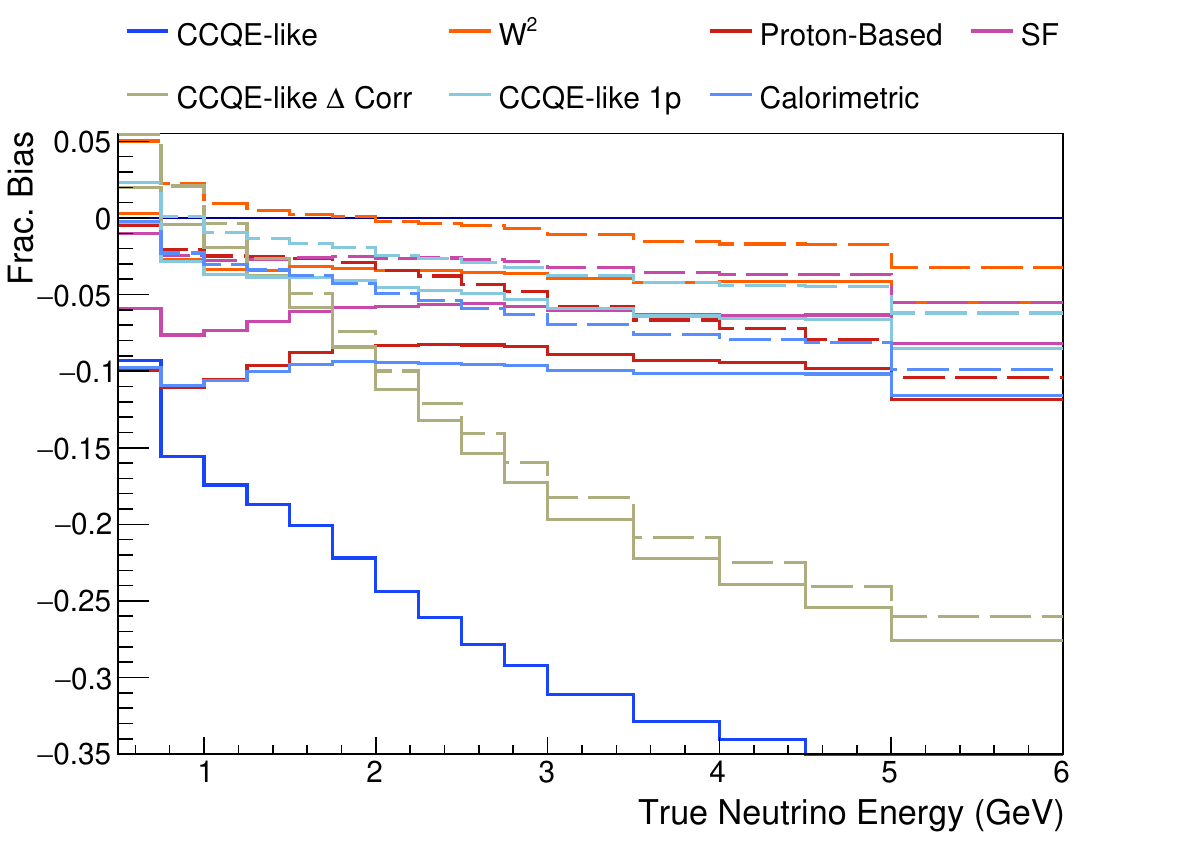}}
    \subfigure[Change in bias from GENIE from applying FSI.]{\label{fig:FSIBiasChangeGENIE}\includegraphics[width=0.47\linewidth]{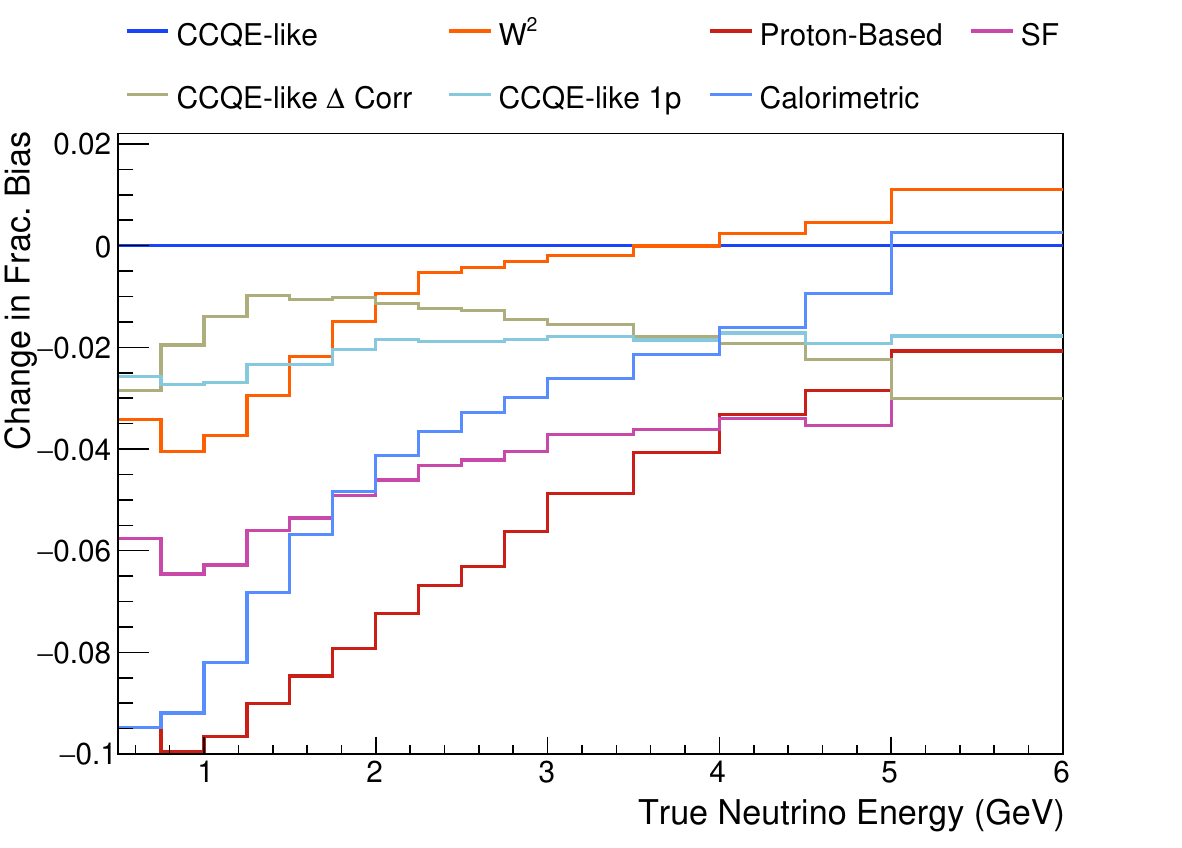}}%
    \hfill
    \subfigure[Change in bias from NuWro from applying FSI.]{\label{fig:FSIBiasChangeNuWro}\includegraphics[width=0.47\linewidth]{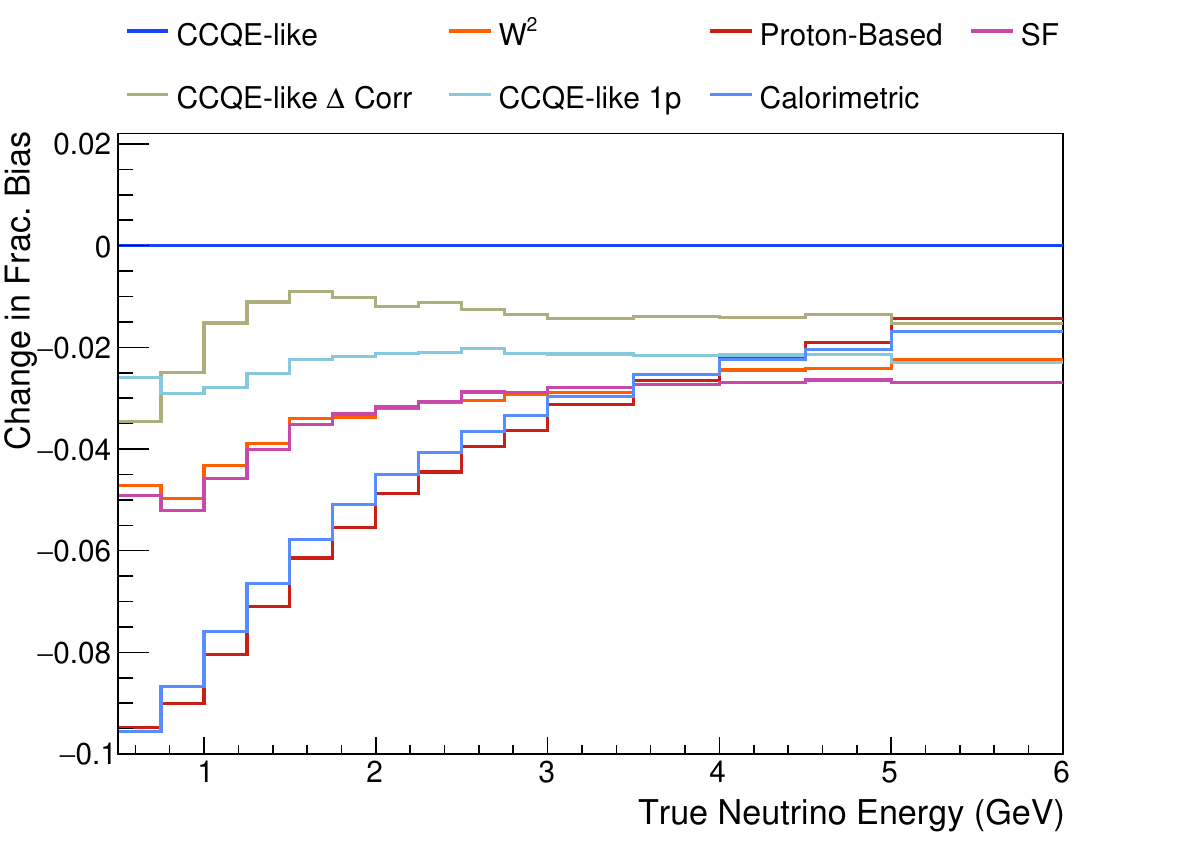}}
    \caption{Fractional bias as a function of true neutrino energy, calculated before and after simulating final state interactions. Dashed lines indicate the calculation without simulating FSI. Panels (c) and (d) show the change in the bias from switching on FSI.}
    \label{fig:BiasFSI}
\end{figure*}

Final state interactions are notoriously challenging to model in neutrino interactions and can lead to substantial uncertainties in cross sections, especially for exclusive processes. We study their impact on energy estimation by comparing the bias and variance of the different energy estimation methods, calculated before and after FSI has taken place. Fig.~\ref{fig:BiasFSI} shows the bias for each estimator, calculated using both the GENIE and NuWro MC simulations. GENIE and NuWro were chosen because they provide the capability to toggle FSI on and off, enabling a controlled comparison of their effects.

The CCQE-like method is insensitive to FSI by construction, as it does not rely on information from the hadronic system. However, this conclusion carries the caveat that the factorized treatment separating FSI from the primary interaction is only an approximation of the true nuclear response. The single proton and $\Delta$ correction variations of the CCQE-like estimator do display some sensitivity to FSI as rescattering of particles can cause events to migrate from one channel to another leading these estimators to examine slightly different populations of events before and after FSI. 

FSI impact both the magnitude and shape of the bias for the $W^2$ and calorimetric methods, albeit in different ways. If GENIE is used, FSI has a minimal effect on the $W^2$ estimator for neutrino energies above approximately 2~GeV, and the non-FSI bias curve follows a very similar shape to the FSI one for both generators. The proton-based estimator is the most strongly affected estimator, a trend which can be seen more clearly in Figs.~\ref{fig:FSIBiasChangeGENIE} and~\ref{fig:FSIBiasChangeNuWro}, which show the change in bias when FSI is switched on. The reliance on proton kinematics in this estimator means the visible hadronic energy in the final state is very susceptible to proton-nucleon rescattering inside the nucleus. The change in bias separates the estimators that utilise directional information from those that do not: these shifts are of a very similar shape (and magnitude when calculated with NuWro) for the $W^2$ and SF estimators, and likewise for the proton-based and calorimetric estimators. This calculation also exposes a noteworthy difference between the FSI modeling in NuWro and GENIE, in that the change in energy bias is far larger in NuWro than GENIE for most of the estimation methods. Figures displaying the variance of each estimator before and after the application of final state interactions can be found in the Supplemental Material~\cite{SupMat}.

\begin{figure*}
    \centering
    \subfigure[GENIE.]{\includegraphics[width=0.47\linewidth]{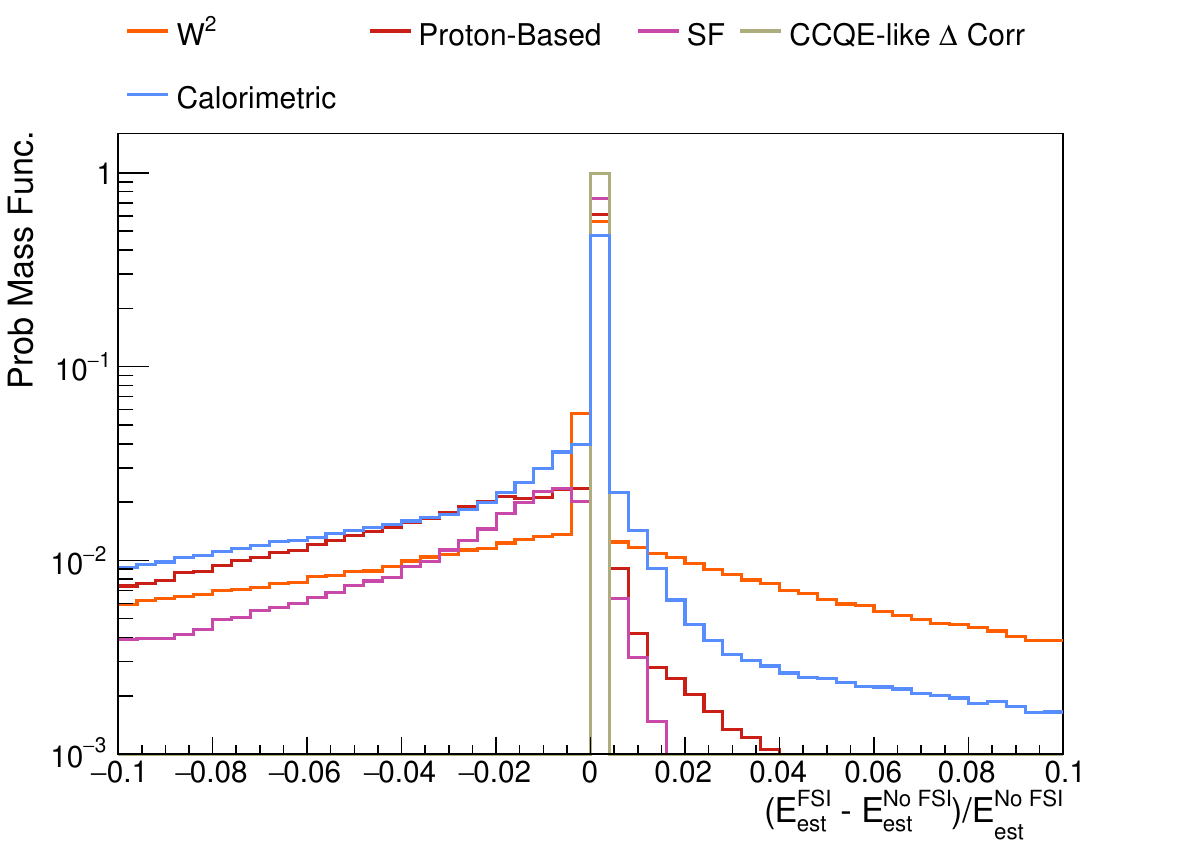}}%
    \hfill
    \subfigure[NuWro.]{\includegraphics[width=0.47\linewidth]{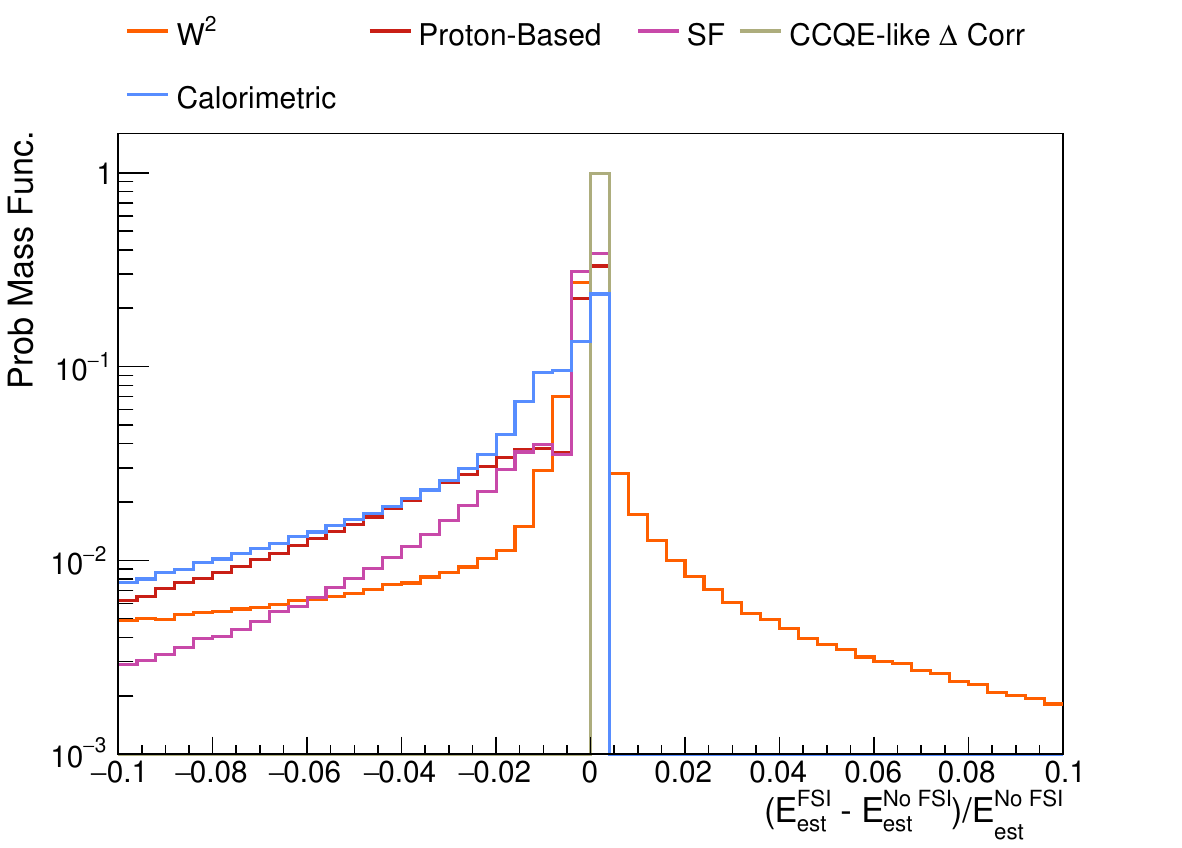}}
    \caption{Fractional change in the estimated neutrino energy due to FSI for each estimation method when only considering events surviving selection criteria before and after FSI. The CCQE-like method is not displayed.}
    \label{fig:ChangeFSI}
\end{figure*}

We further check the susceptibility  of each energy estimation method to FSI by calculating the fractional change in estimated neutrino energy before and after simulating FSI, and comparing the distribution of these values for each energy estimator.  These distributions are shown in Fig.~\ref{fig:ChangeFSI}. For both generators, the SF method produces the most pronounced peak at zero, as expected, since it is explicitly designed to use transverse kinematic information to mitigate the effects of FSI.  However, a shoulder and downward tail are observed when FSI are simulated with NuWro. These features are due to non-CCQE events that populate the 1p$0\pi$ topology through the reabsorption of additional particles during FSI. The proton-based and calorimetric methods are the most affected by FSI, likely due to hadron absorption in the nucleus, to which estimators incorporating directional information—such as the $W^2$ method—are less sensitive.

The similarity between the proton-based and calorimetric estimators displayed in NuWro is attributed to a lower rate of meson production during FSI compared to that of GENIE. The SF and $W^2$ estimators are also very similar, a result of the bias in the $W^2$ being a mixture of events with a single proton in the final state, for which the shift in bias has a similar shape to that of the SF method, and a flat contribution from meson producing events. This relationship is studied further in the Supplemental Material~\cite{SupMat}.

\clearpage

\section{Secondary Variable Dependences}
\label{sec:VariableDependences}

In the previous section, the bias and variance of the different energy estimators was calculated as a function of true neutrino energy. It is also important to understand if any bias and resolution effects have significant dependency on the shape of the neutrino-nucleus cross section in variables that are difficult to measure or poorly modeled. For this study, we consider the visible hadronic invariant mass, missing hadronic energy, and the lepton scattering angle -- deemed as ``secondary variables". We will return to the role of secondary variables in Section~\ref{sec:ImpactOnOsc}, where we evaluate their impact on the oscillation parameters. 

The fractional bias and variance in neutrino energy as a function of some variable of interest $x$ are defined as:
\begin{center}
\begin{minipage}{0.5\textwidth}
\begin{align}
B(x) &= \frac{1}{\mathcal{N}(x)} \iint
\left(\frac{E_{\mathrm{est}} - E_{\mathrm{true}}}{E_{\mathrm{true}}}\right)
N(E_{\mathrm{est}},E_{\mathrm{true}},x) \nonumber \\
&\qquad dE_{\mathrm{true}}\, dE_{\mathrm{est}}, \label{eq:bias} \\[0.6ex]
V(x) &= \frac{1}{\mathcal{N}(x)} \iint
\left(\frac{E_{\mathrm{est}} - \overline{E_{\mathrm{est}}}(E_{\mathrm{true}};x)}
{\overline{E_{\mathrm{est}}}(E_{\mathrm{true}};x)}\right)^{2}
 \nonumber \\
&\qquad N(E_{\mathrm{est}},E_{\mathrm{true}},x) dE_{\mathrm{true}}\, dE_{\mathrm{est}}, \label{eq:variance}
\end{align}
\end{minipage}
\end{center}
where $N(E_{\rm est},E_{\rm true},x)$ is the number of events with the variable of interest at $x$, true neutrino energy $E_{\rm true}$, and reconstructed energy $E_{\rm est}$, and $\mathcal{N}(x) = \iint N(E_{\rm est},E_{\rm true},x) dE_{\rm true} dE_{\rm est}$ is a normalization factor. $\overline{E_{\rm est}}(E_{\rm true};x)$ is the mean estimated energy of events with true energy $E_{\rm true}$ and variable of interest equal to $x$. 

In this section, we use the GENIE generator as our baseline model. For completeness, we repeat the calculations with all other generators, yielding conclusions consistent with GENIE; details are provided in the Supplementary Material~\cite{SupMat}. We find that the estimator least sensitive to the choice of generator is the CCQE-like method; however, though this apparent advantage is offset by its consistently large bias and variance across all variables examined.

\subsection{Visible Hadronic Invariant Mass}
\label{sec:HadronicInvariantMass}
\begin{figure*}[p]
    \centering     
     \subfigure[Distribution.]{\includegraphics[width=0.32\linewidth]{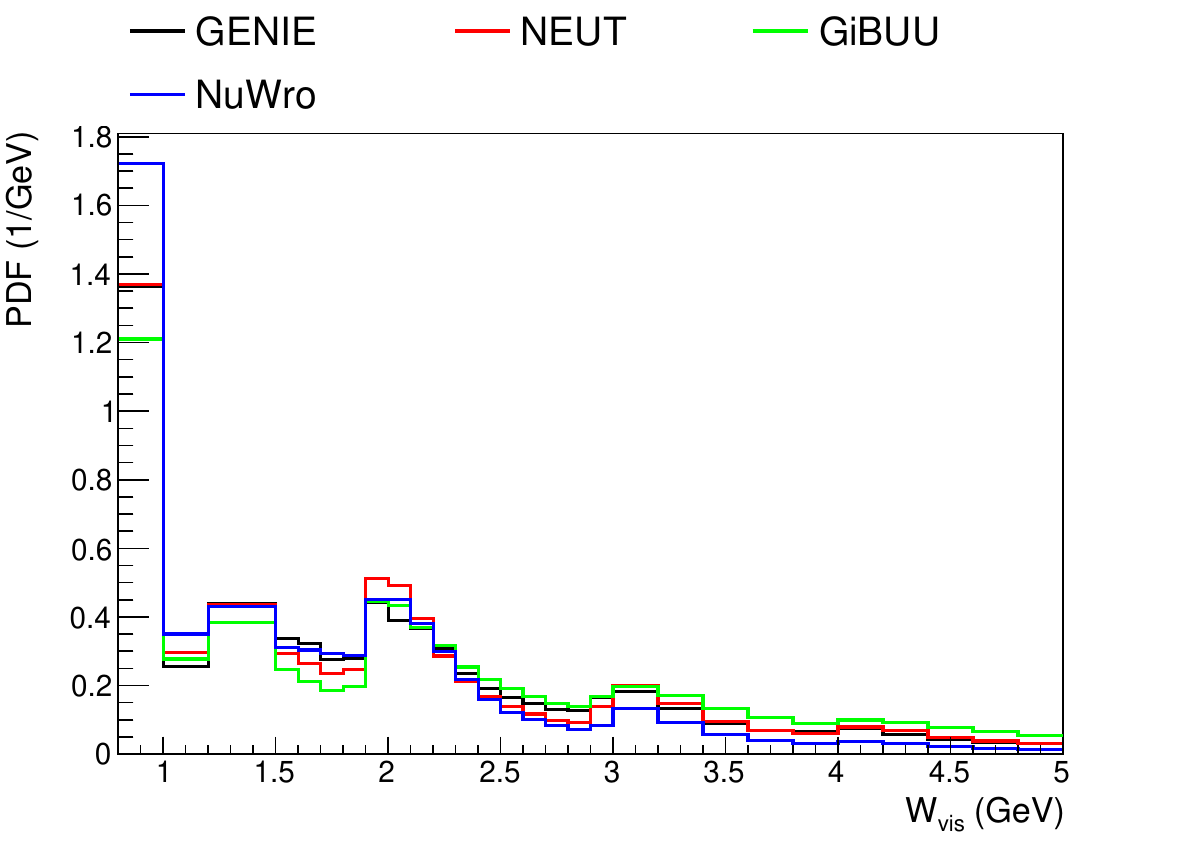}}%
    \hfill
    \subfigure[Fractional bias.]{\includegraphics[width=0.32\linewidth]{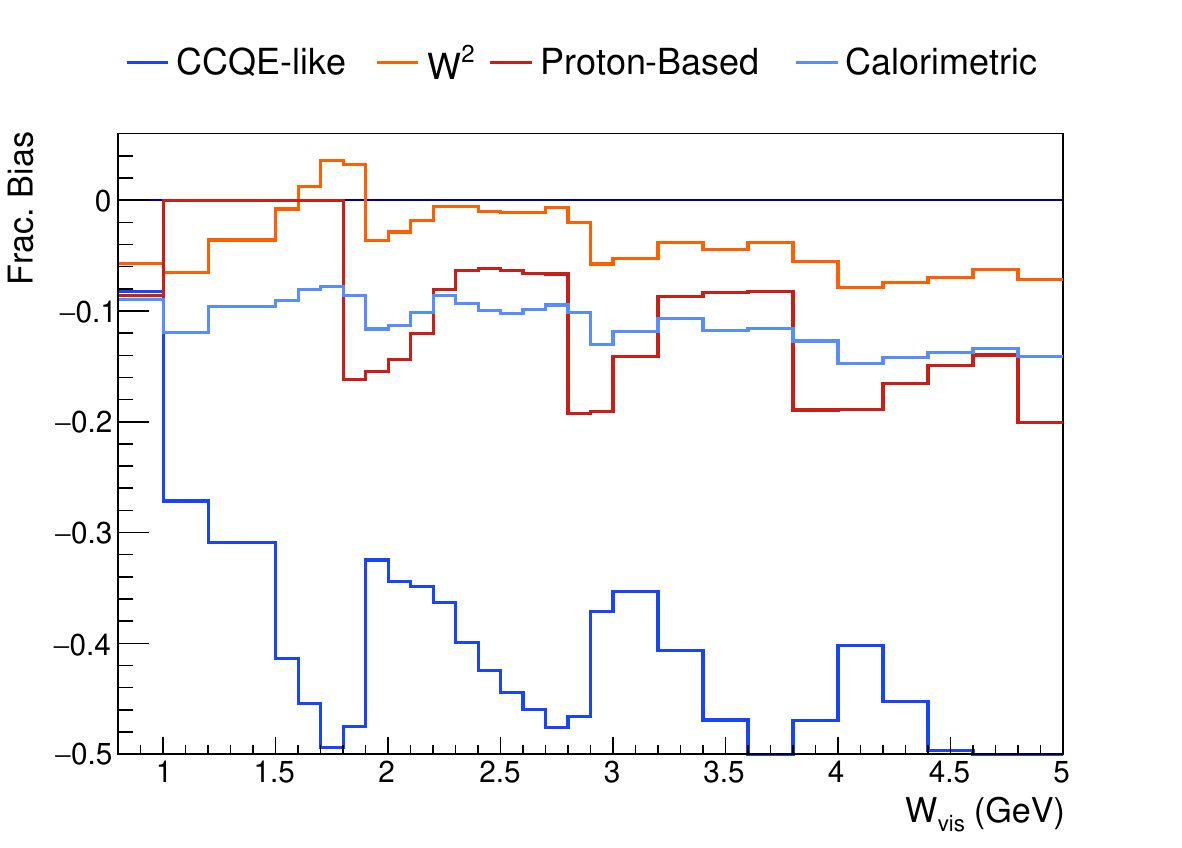}}%
    \hfill
    \subfigure[Fractional variance.]{\includegraphics[width=0.32\linewidth]{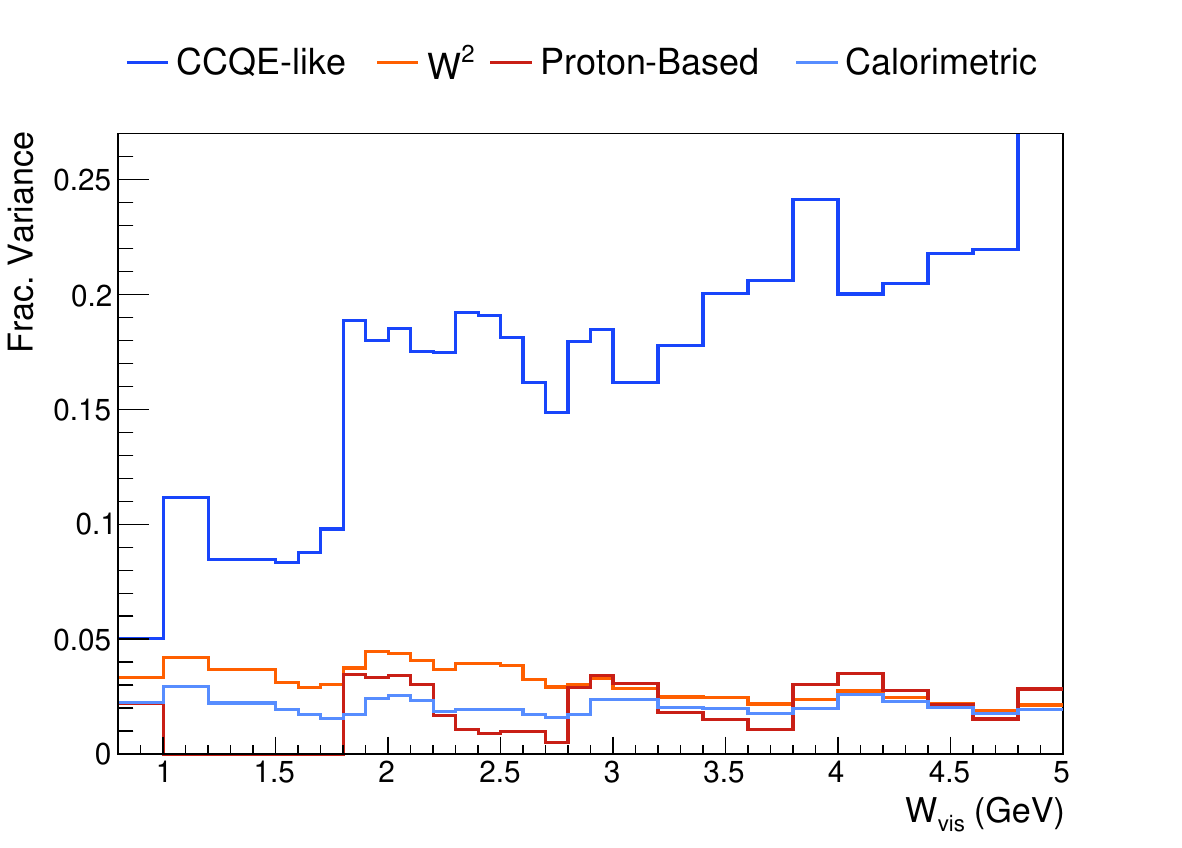}}
    \caption{Fractional bias and variance as a function of visible hadronic invariant mass.}
    \label{fig:BiasPlotsW}    
\end{figure*}

Fig.~\ref{fig:BiasPlotsW} shows the bias as a function of the visible hadronic invariant mass~\footnote{We do not draw the SF method or the modified forms of the CCQE-like estimator in this comparison as these is only valid for specific values of $W_{\rm vis}$.}. The first notable feature is the periodic behaviour of the bias observed across all methods, most prominently in the CCQE-like estimator and, to a lesser extent, in the proton-based method. For the CCQE-like, the bias raises and falls as the true hadronic invariant mass approaches values corresponding to integer multiples of the proton mass. This pattern can be interpreted as a larger fraction of the energy being transferred to pions when $W_{\rm vis}$ lies between the peaks in the $W$ spectrum associated with specific proton multiplicities, resulting in greater bias for estimators that neglect pion information. The CCQE-like estimator exhibits a sharp increase in variance once $W > 2M_{p}$, marking the boundary beyond which the charged-current quasi-elastic (CCQE) hypothesis underlying this method is no longer valid. All other estimators offer similar performance, with the proton based method fluctuating depending on whether the final state invariant mass is close to an integer number of protons or not. The same periodic behavior in the variance with a period roughly equal to $M_p$ is also present, although with a smaller effect.

\subsection{Missing Hadronic Energy}
\label{sec:MissingHadronicEnergy}

\begin{figure*}[p]
    \centering     
    \subfigure[Distribution.]{\includegraphics[width=0.32\linewidth]{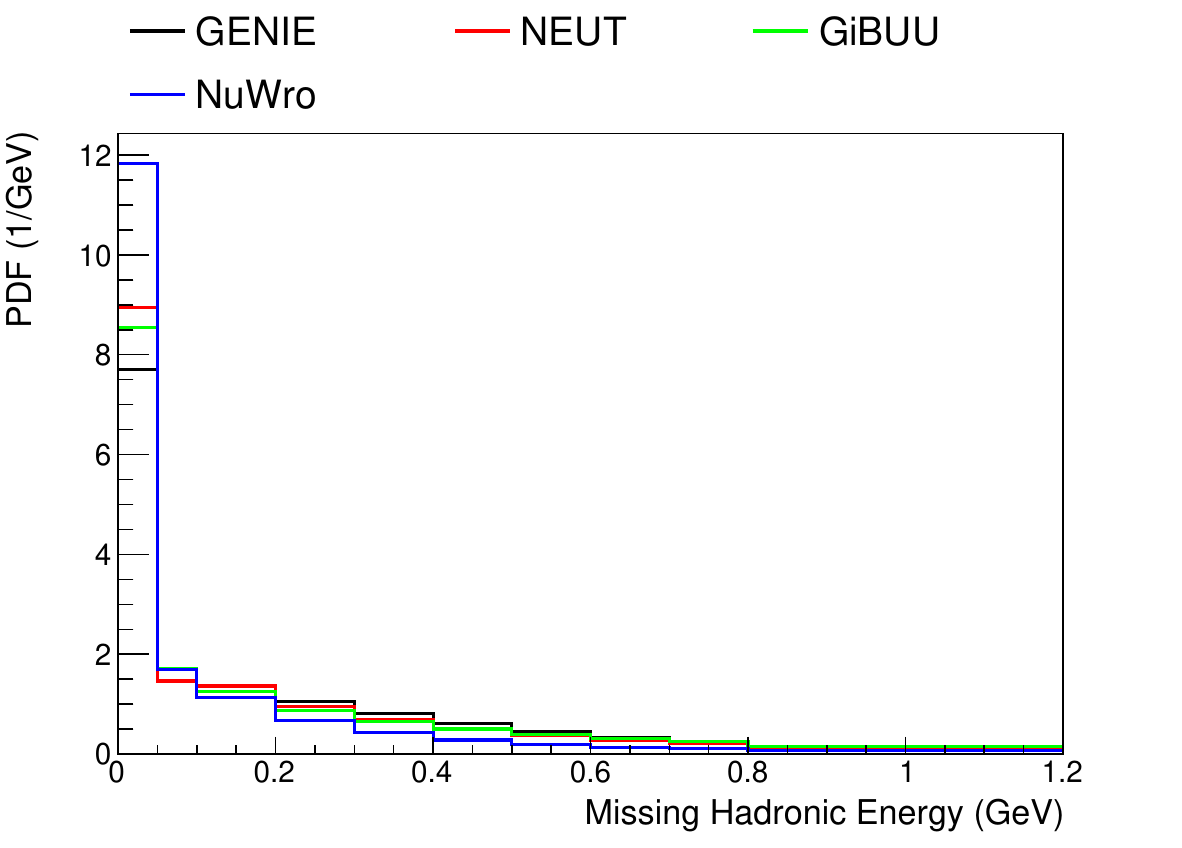}}%
    \hfill
    \subfigure[Fractional bias.]{\includegraphics[width=0.32\linewidth]{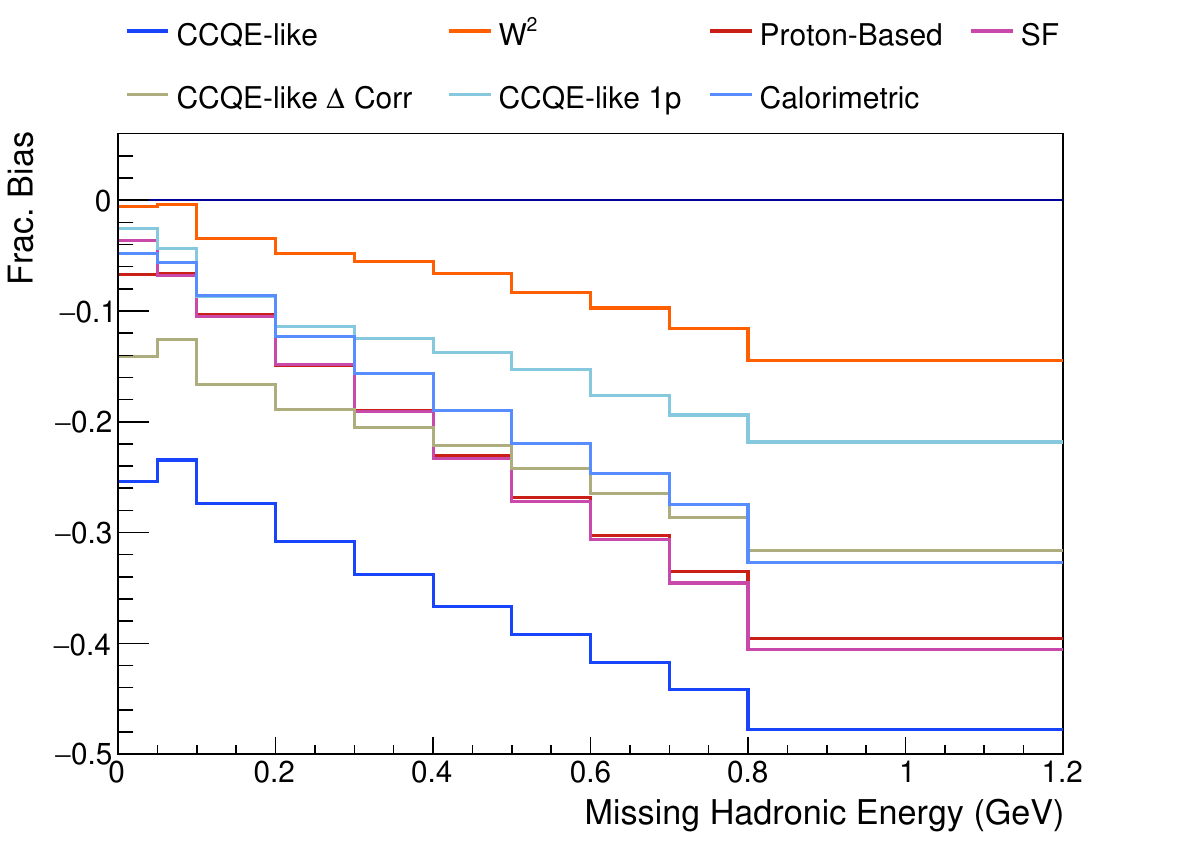}}%
    \hfill
    \subfigure[Fractional variance.]{\label{fig:VarianceMissingE}\includegraphics[width=0.32\linewidth]{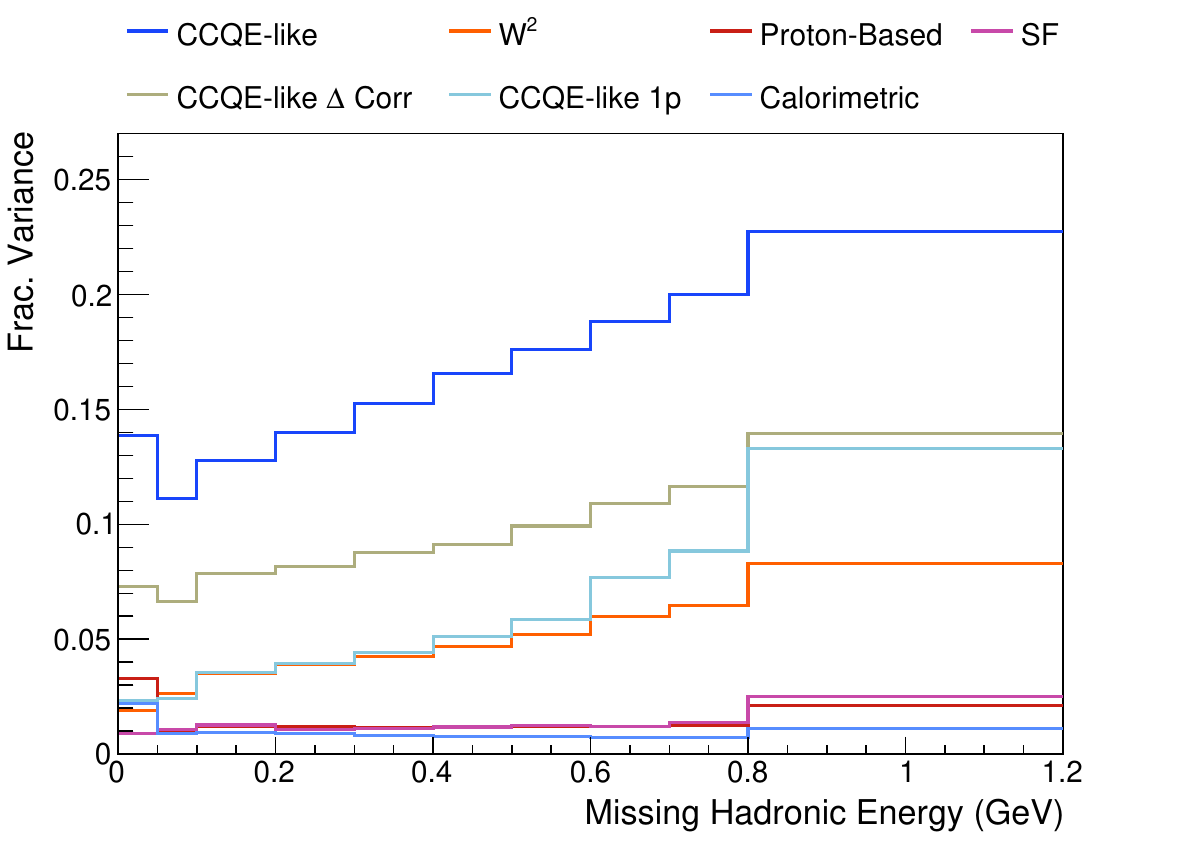}}
    \caption{Fractional bias and variance as a function of missing hadronic energy.}
    \label{fig:BiasPlotsMissingE}
\end{figure*}

We define the missing hadronic energy as the sum of the kinetic energies of all neutrons (as these are assumed to be invisible), the kinetic energies of all protons below detection threshold, and total energies of pions below the detection threshold. Fig.~\ref{fig:BiasPlotsMissingE} shows the distributions of bias and variance in neutrino energy as a function of missing hadronic energy. The distribution of bias produced by the $W^2$ method and CCQE-like methods are flatter than the other estimators, implying greater stability with respect to the variation of missing energy, though this benefit is partially offset by the greater variance shown in Fig.~\ref{fig:VarianceMissingE}. It is interesting to note the SF method is still susceptible to the presence of missing energy despite being constructed to infer the its impact through missing transverse momentum. Comparison of this figure with the corresponding bias distributions from the visible invariant mass and lepton angle shows that the missing hadronic energy has the potential to create the largest bias in neutrino energy.

\subsection{Angular Dependency}
\label{sec:AngularDependences}

\begin{figure*}[p]
    \centering     
     \subfigure[Distribution.]{\includegraphics[width=0.32\linewidth]{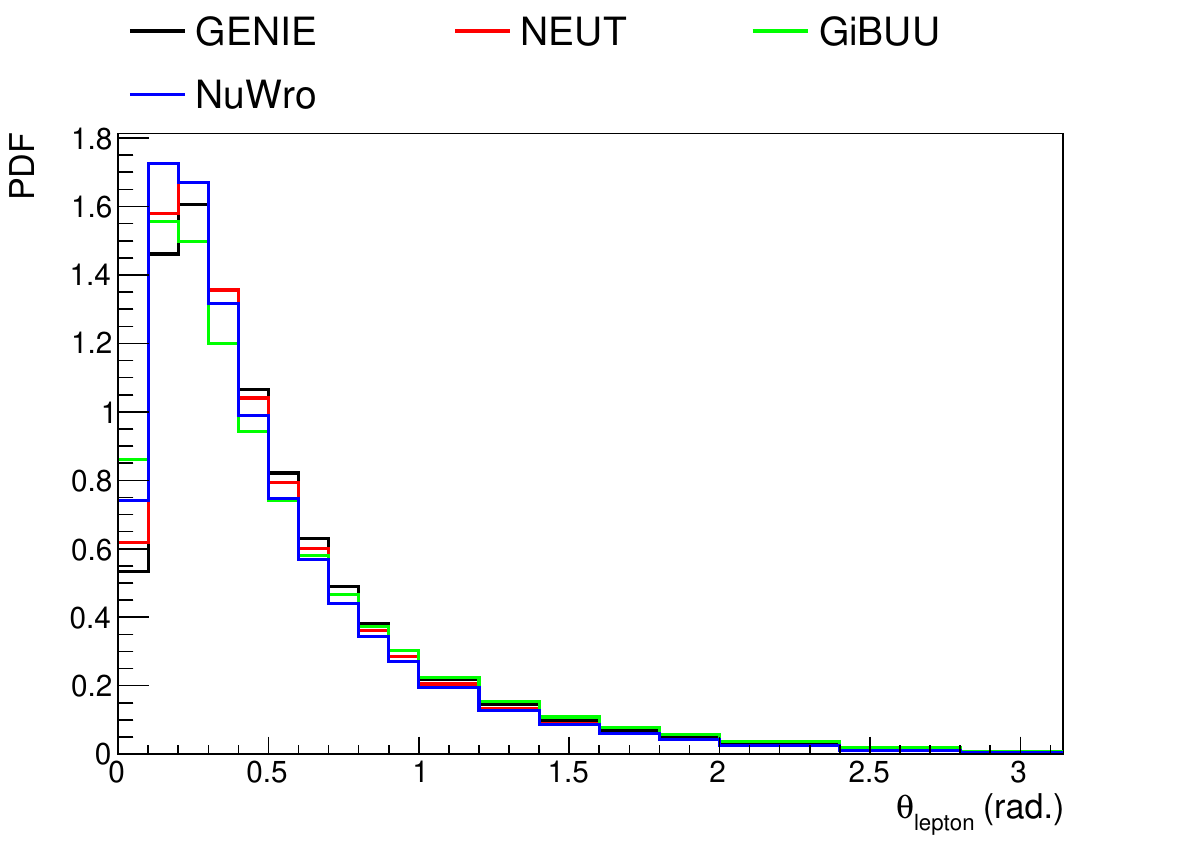}}%
    \hfill
    \subfigure[Fractional bias.]{\includegraphics[width=0.32\linewidth]{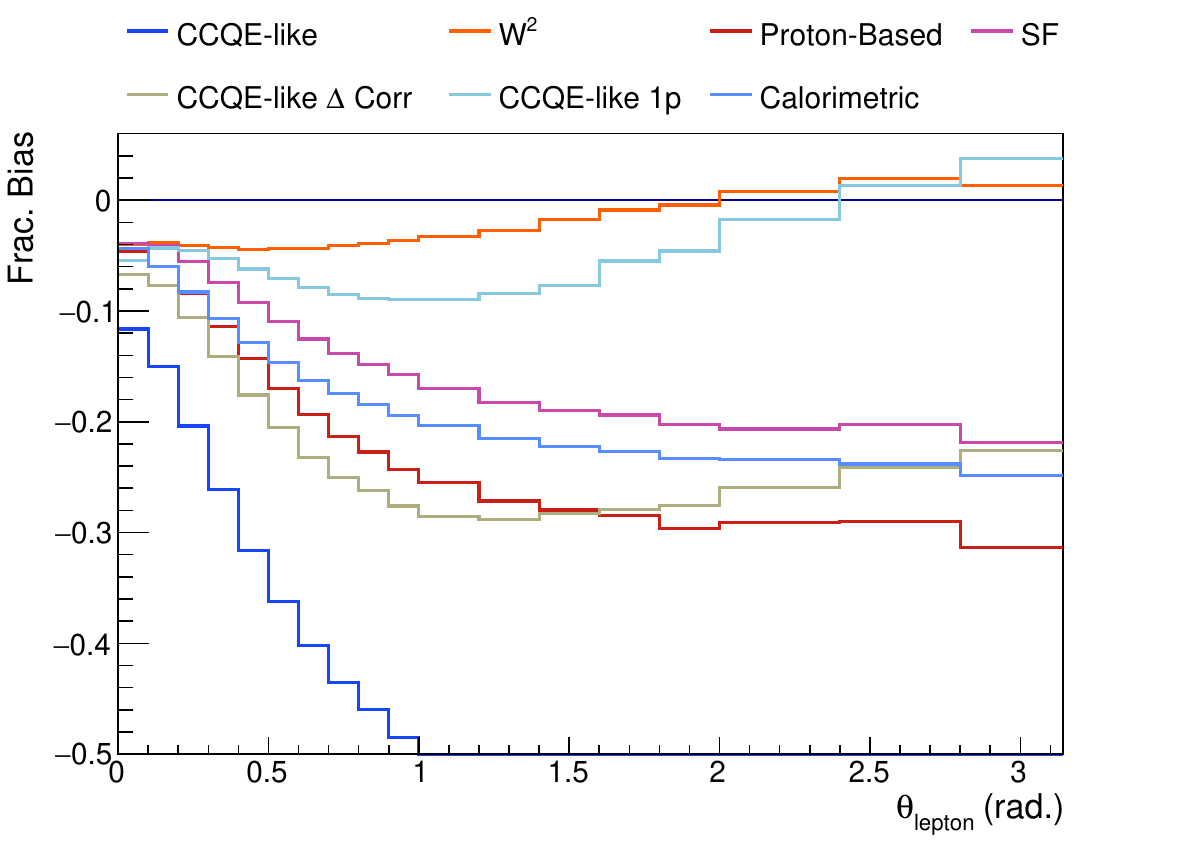}}%
    \hfill
    \subfigure[Fractional variance.]{\includegraphics[width=0.32\linewidth]{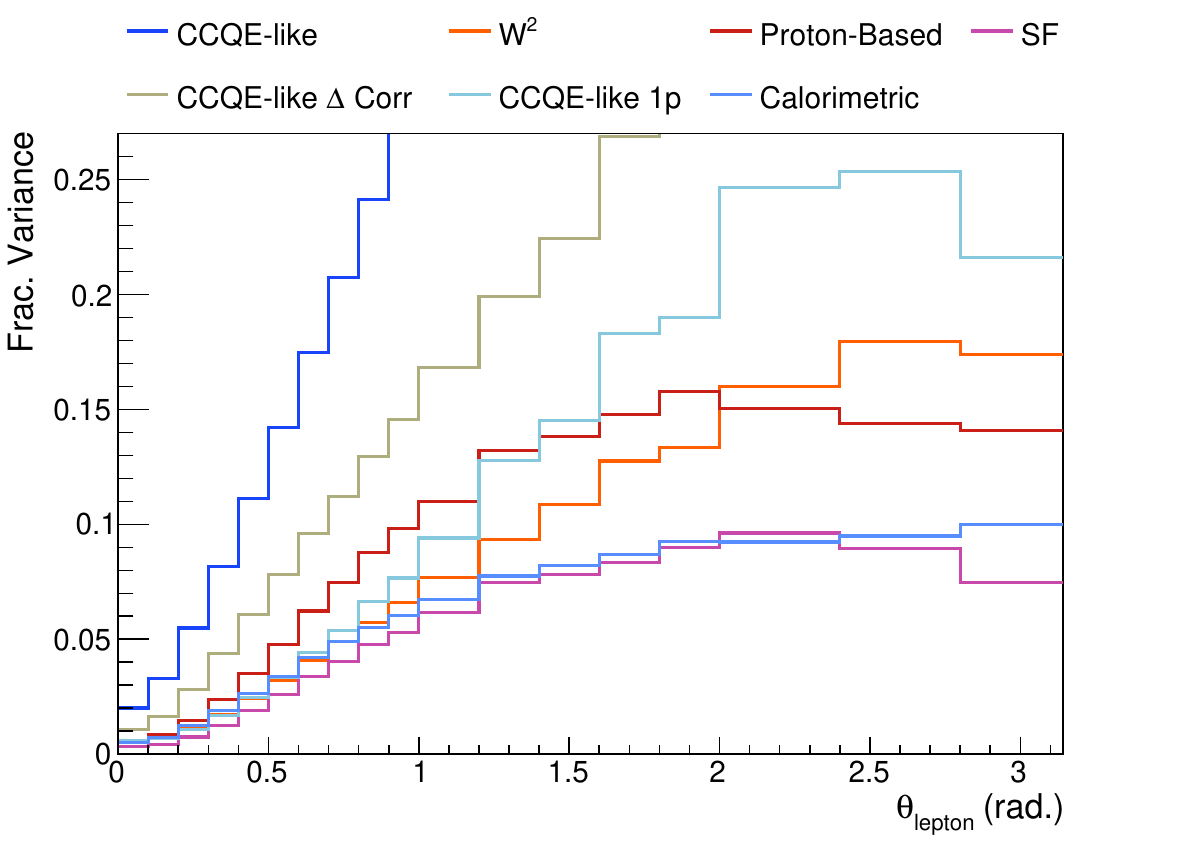}}
    \caption{Fractional bias and variance as a function of the lepton scattering angle.}
    \label{fig:BiasPlotsAngle}
\end{figure*}

One of the features that distinguishes the CCQE-like, $W^2$, and SF estimators from the other methods is that they include angular information in their calculation, in contrast with the proton-based and calorimetric methods which purely utilize the kinetic (or total) energies of final state particles, irrespective of their directions. Fig.~\ref{fig:BiasPlotsAngle} shows the fractional bias and variance as a function of lepton scattering angle $\theta$. The bias calculation reveals the improvement made to the purely muon kinematics method by accounting for the invariant mass of the full final state, as the muon kinematics increasingly underestimates the neutrino energy as the scattering angle increases. At very large angles the interaction will be highly inelastic making the quasielastic approximation inaccurate, leading to large bias. We also see a degradation of performance of the estimators according to the variance calculation at large angles. The $W^2$ method exhibits a flat bias, while all other estimators trend downward, though this may be offset by the worse resolution than the calorimetric and SF methods displayed by the variance calculation.

\subsection{Lepton Momentum}
\label{sec:LeptonMomemtum}

\begin{figure*}[p]
    \centering     
     \subfigure[Distribution.]{\includegraphics[width=0.32\linewidth]{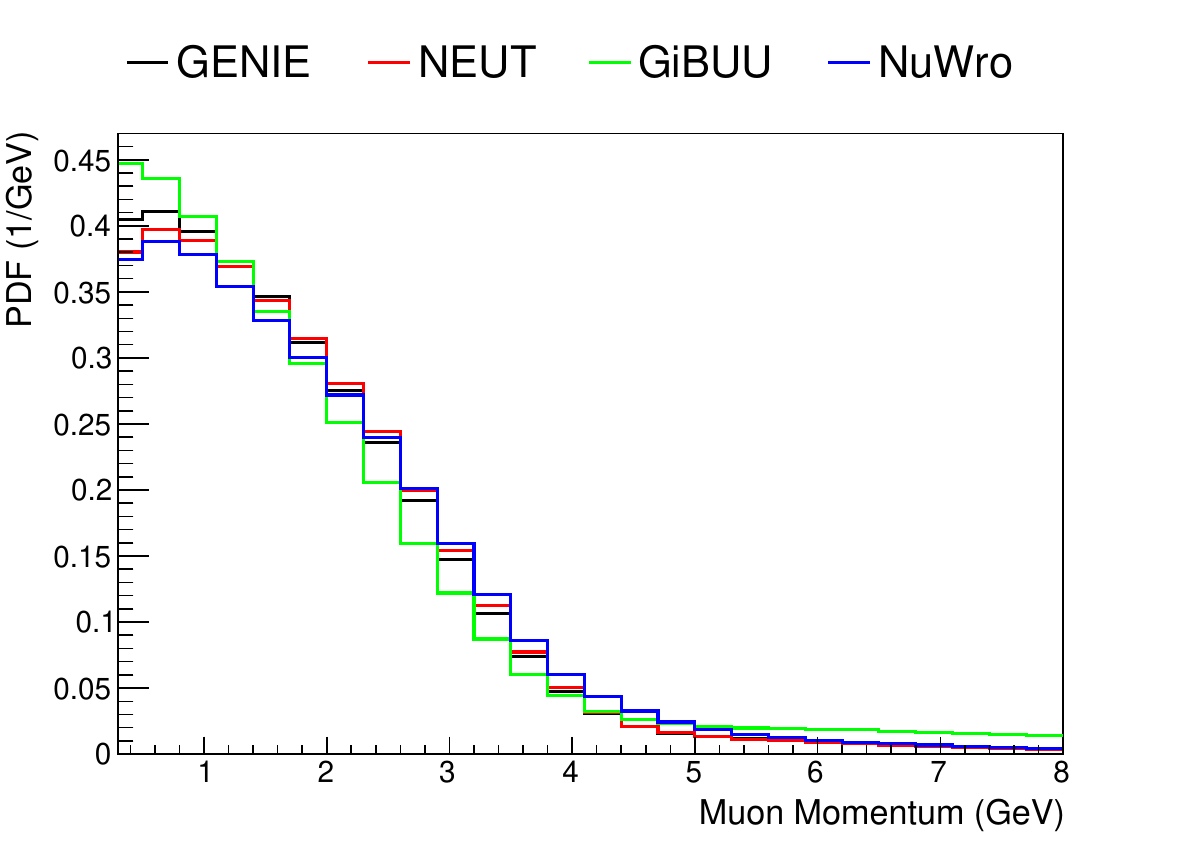}}%
    \hfill
    \subfigure[Fractional bias.]{\includegraphics[width=0.32\linewidth]{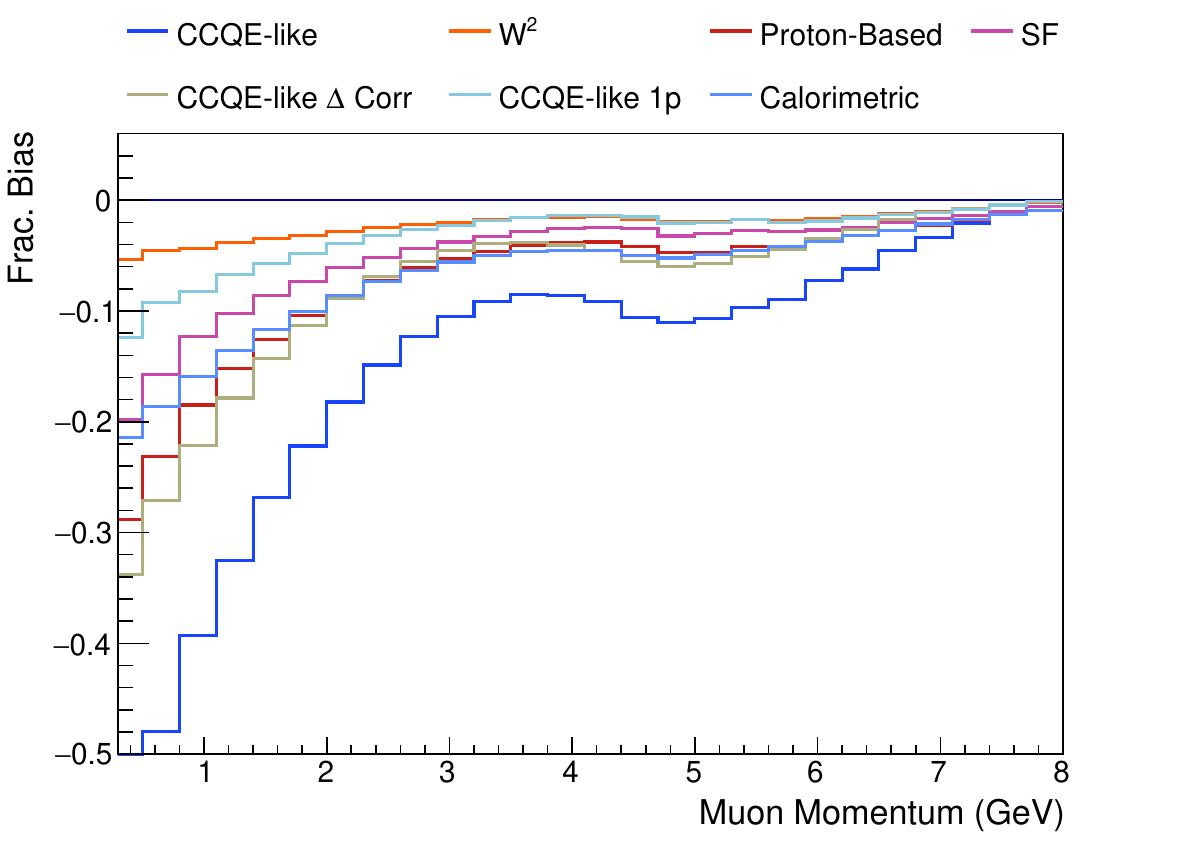}}%
    \hfill
    \subfigure[Fractional variance.]{\includegraphics[width=0.32\linewidth]{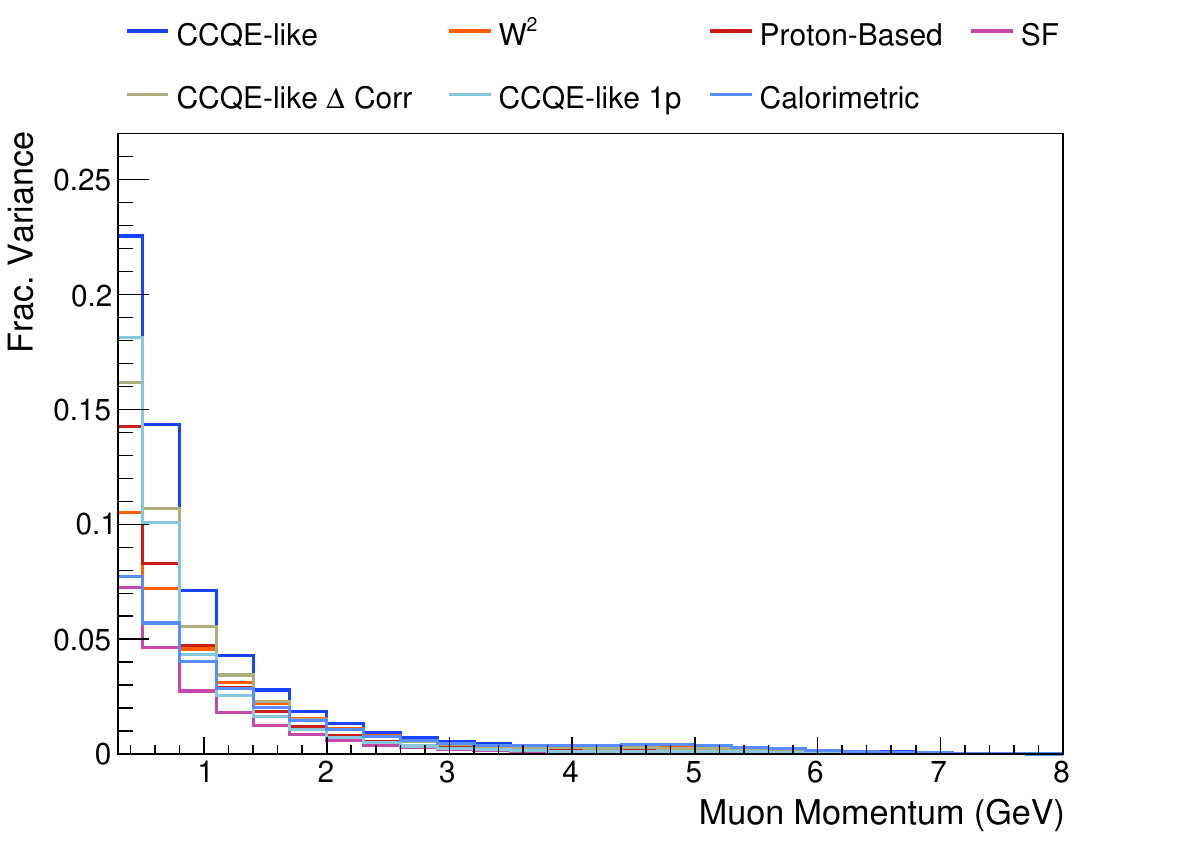}}
    \caption{Fractional bias and variance as a function of the scattered muon's momentum.}
    \label{fig:BiasPlotsLeptonMom}
\end{figure*}

Another quantity that is used by several estimators is the energy and/or momentum of the scattered lepton,; a variable that correlates strongly with neutrino energy. In Fig.~\ref{fig:BiasPlotsLeptonMom} we study the bias and variance in the estimated neutrino energy as a function of the momentum of the scattered lepton. This shows the $W^2$ method has a small, flat bias as a function of this variable, with the next best performing method being the single proton version of the CCQE-like estimator. The SF method offers the lowest variance among the methods compared. The bias and variance both increase for all methods at low muon momentum, as with the lepton scattering angle this is partly due to a larger portion of the energy carried by the neutrino being transferred to the hadronic system in this region of the phase space leading to more complex hadronic processes that are more challenging to simulate and analyze.

\clearpage
\section{Impact of Estimator Choice on Oscillation Parameters}
\label{sec:ImpactOnOsc}
In this section, we examine how the choice of energy estimator influences oscillation measurements, with the aim of identifying which estimator provides the most robust inference of oscillation parameters. Here, robustness refers to reduced susceptibility to biases arising from the choice of event generator in general and from final-state interactions (FSI) in particular.
Since the connection between reconstructed neutrino energy and oscillation parameter inference is non-trivial, the CP-violating phase $\delta_{CP}$ is probed using a toy $\nu_e$-appearance analysis at the far detector, while the mass-splitting $\Delta m^2_{23}$ is constrained through ``shape-only'' fits of $\nu_\mu$ disappearance spectra.  We make no assumption regarding a near detector constraint. 

\subsection{Estimator Comparison in Parameter Exclusion}
\label{sec:ExclusionStudy}
We compare the performance of different energy estimators in excluding benchmark oscillation hypotheses when specific alternate parameters are assumed. We consider two tests: excluding the global central value of $\Delta m^2_{23}=2.451\times 10^{-3}$~eV$^2$~\cite{ParticleDataGroup:2024cfk} using $\nu_{\mu}$ disappearance, and excluding $\delta_{CP}=0$ with $\nu_e$ appearance, both assuming DUNE's far detector flux and baseline. 

For each estimator/generator combination, we compute the estimated energy spectra from $\nu_{\mu}$ assuming the central value and the central value $\pm 3\sigma$, where $\sigma = 0.026\times10^{-3}$~eV$^2$~\cite{ParticleDataGroup:2024cfk}. All distributions are normalized to the same arbitrary detector exposure. We then compute the following metric to quantify the strength of the exclusion of the central value:
\begin{align}
    \sqrt{\chi^2} = \sqrt{\sum_i \left(\frac{N'_i - N_i}{\sqrt{N'_i}} \right)^2},
\end{align}
where $N_{i}$ is the predicted rate in the $i$th estimated energy bin assuming the central value, and $N'_i$ is the rate assuming the central value $\pm 3\sigma$. Similarly, in the $\nu_e$ appearance case, we treat $\delta_{CP}=0$ as the central value and compute $N'_i$ using $\delta_{CP}=\pm \frac{\pi}{2}$. We compute exclusions varying the oscillation parameters in both directions, though the qualitative conclusions are largely the same.

\begin{figure*}
    \centering
    \subfigure[$\Delta m^2_{23}$: central vs.\ $+3\sigma$.]{\includegraphics[width=0.48\linewidth]{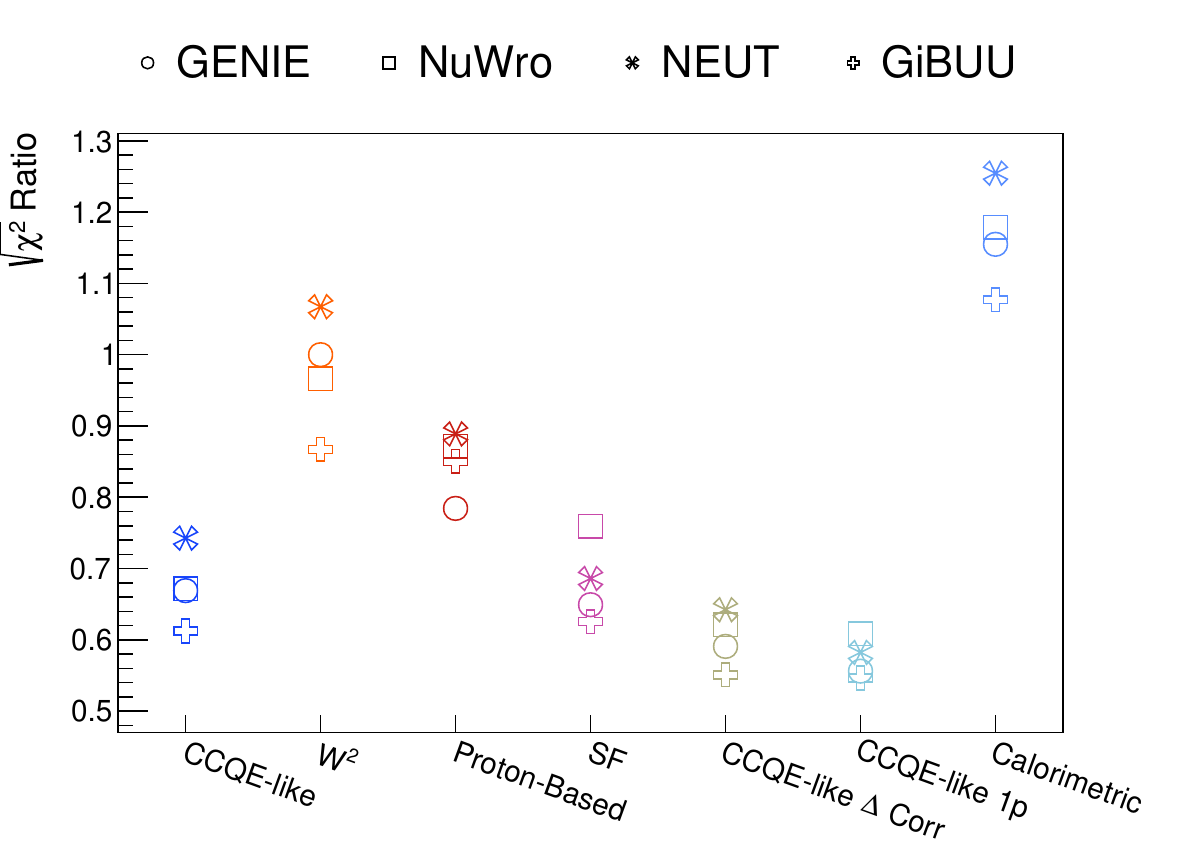}}%
    \hfill
    \subfigure[$\Delta m^2_{23}$: central vs.\ $-3\sigma$.]{\includegraphics[width=0.48\linewidth]{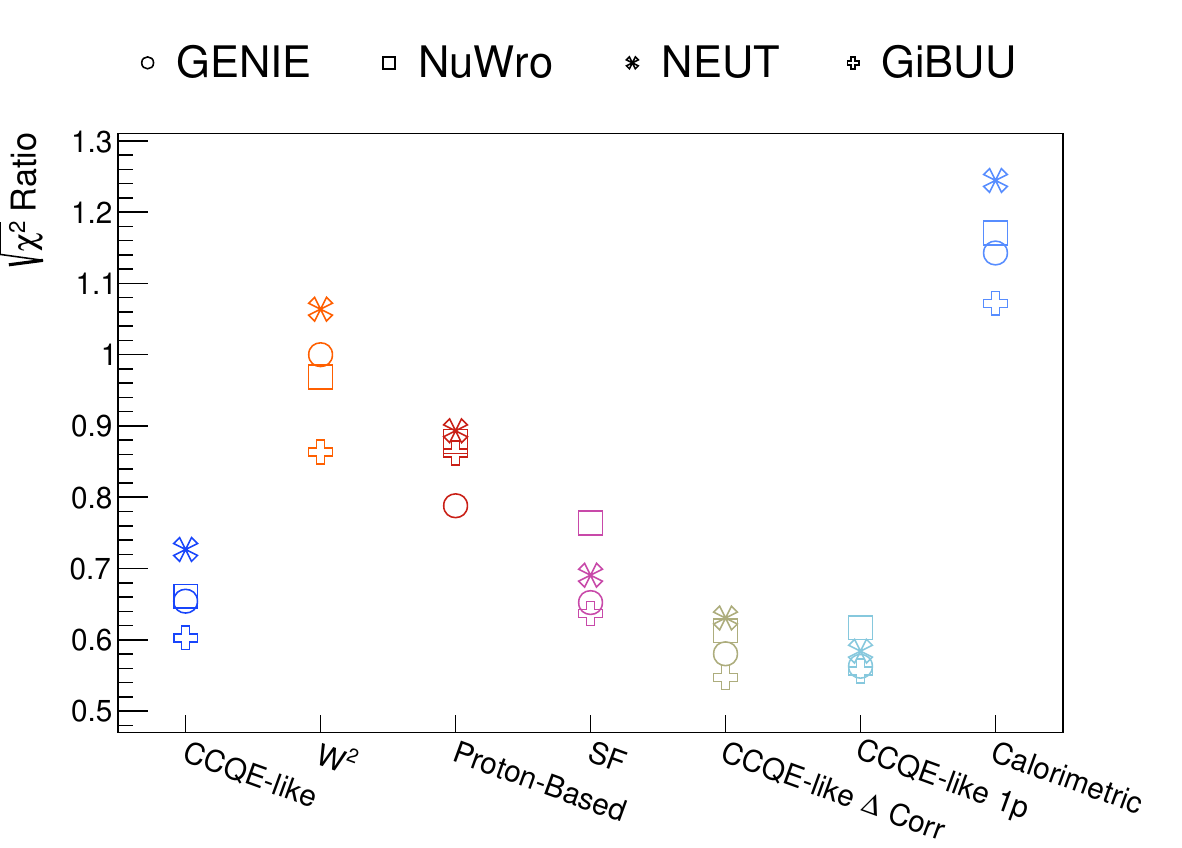}}
    \subfigure[$\delta_{CP}=0$ vs.\ $\tfrac{\pi}{2}$.]{\includegraphics[width=0.48\linewidth]{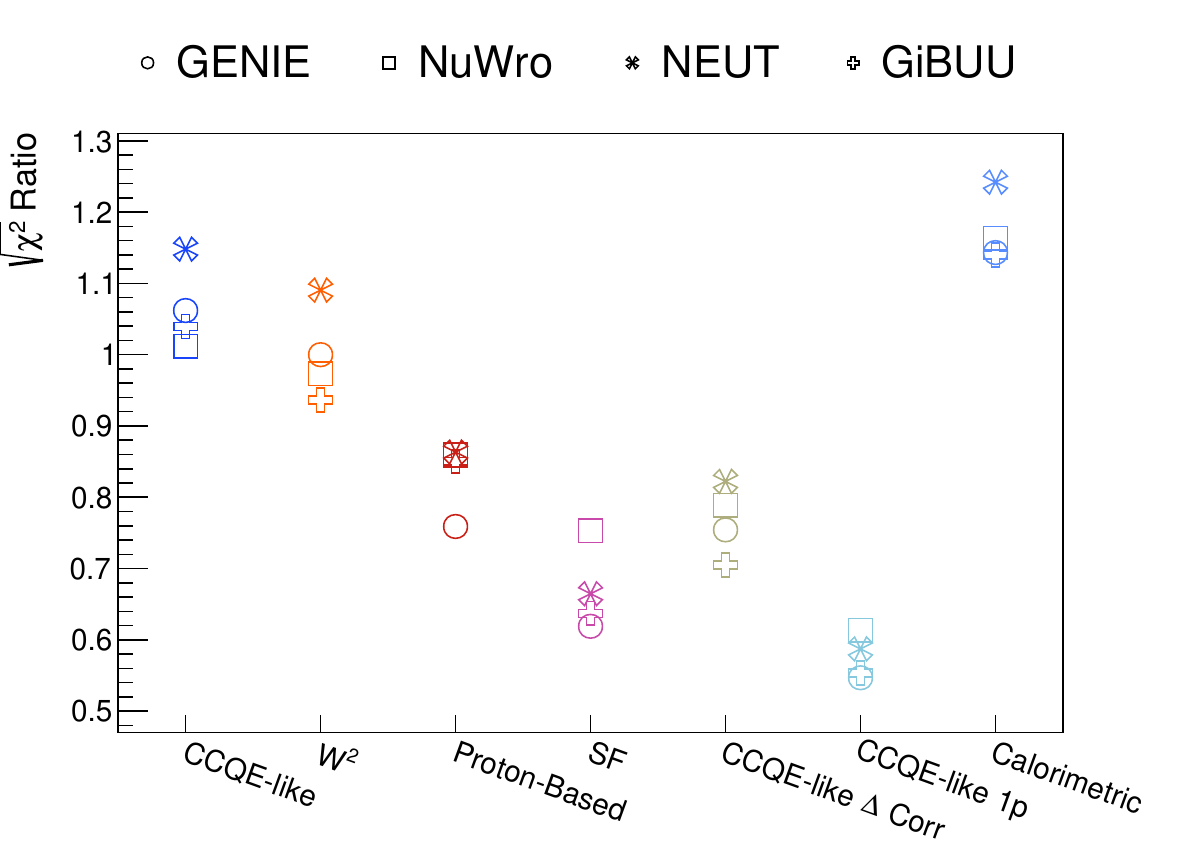}}%
    \hfill
    \subfigure[$\delta_{CP}=0$ vs.\ $-\tfrac{\pi}{2}$.]{\includegraphics[width=0.48\linewidth]{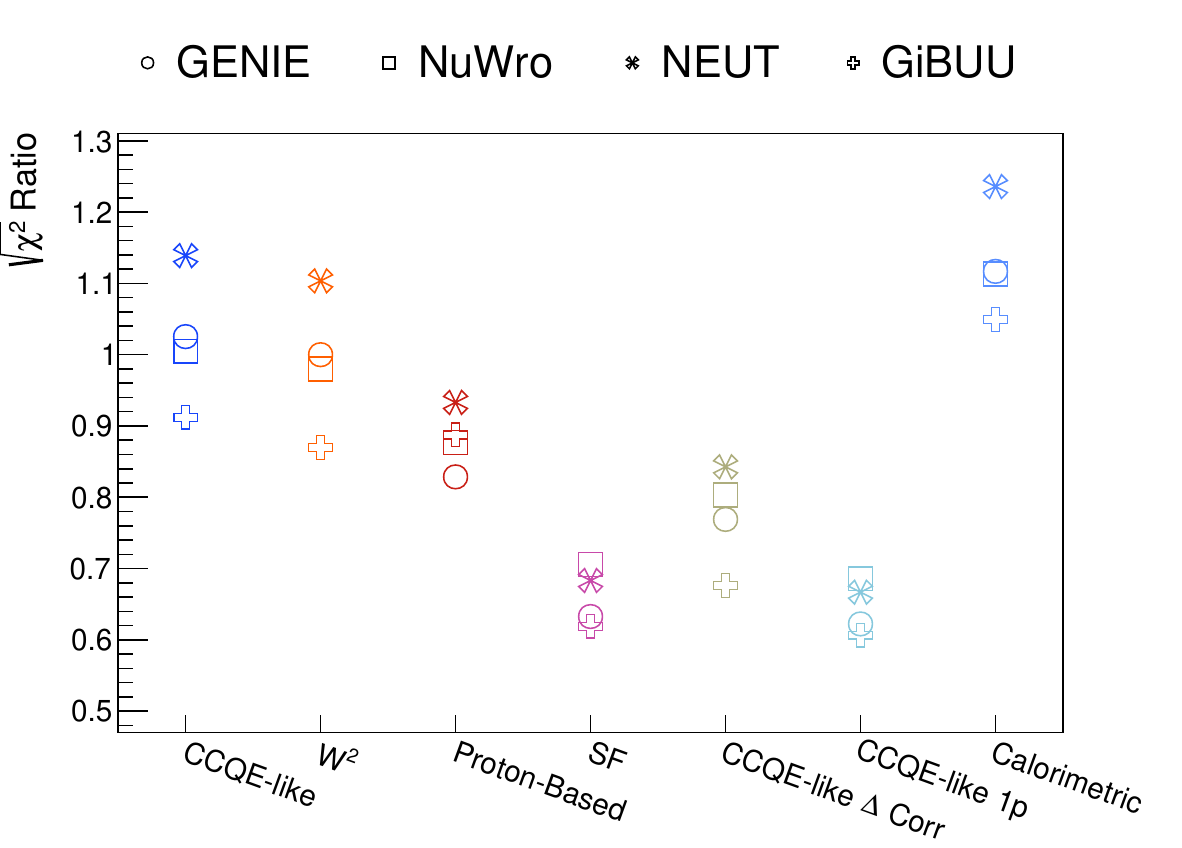}}
    \caption{Relative exclusion power of different estimators, expressed as the ratio of $\sqrt{\chi^2}$ to the GENIE $W^2$ baseline.}
    \label{fig:ExclusionPlots}
\end{figure*}

Figure~\ref{fig:ExclusionPlots} compares the performance of the different estimators relative to a benchmark case: GENIE using the $W^2$ estimator. The ratio of the $\sqrt{\chi^2}$ values obtained for two estimator-generator combinations is independent of the overall exposure the distributions are normalized to.

For a fixed estimator, there is a substantial spread in performance across generators. For a fixed generator, the relative ordering of estimators is fairly consistent, with the calorimetric and $W^2$ estimators providing the best separation power when examining the $\nu_{\mu}$ spectrum. 

We find in both tests the more exclusive methods offer lower sensitivity due to their loss in statistics, with the single proton version of the CCQE-like estimator consistently offering the worse performance. The default CCQE-like and calorimetric estimators perform the best when $\nu_e$ appearance is measured, the benefit of their inclusive approach across all energies, followed by the $W^2$ method which is the slightly less inclusive given the requirement there is a single proton in the final state. The performance of the default CCQE-like estimator changes substantially when switching from $\nu_{\mu}$ to $\nu_e$ spectra, examination of the estimated energy spectra with this estimator shows this is due to worse resolution in the regions sensitive to $\Delta m^2_{23}$. The spectra employed to calculate these exclusions along with the contributions to the $\chi^2$ statistic bin by bin are shown in the Supplemental Material~\cite{SupMat}.


\subsection{Final State Interactions}
\label{sec:FSIFits}
We define the estimated energy spectrum for a given estimator as
\begin{align}
    N(E_{\rm est};\alpha) = \int R(E_{\rm est},E_{\rm true})\, P(E_{\rm true};\alpha)\, \Phi(E_{\rm true})\, dE_{\rm true},
    \label{eq:Spectrum}
\end{align}
where $R(E_{\rm est},E_{\rm true})$ is the response matrix, $P(E_{\rm true};\alpha)$ the oscillation probability, and $\Phi(E_{\rm true})$ the incident flux. The parameter set $\alpha$ denotes the oscillation parameter of interest, i.e.\ $\delta_{CP}$ in the $\nu_e$ appearance channel and $\Delta m^2_{23}$ in the $\nu_\mu$ disappearance channel. When analyzing the FSI impact, we will only consider $\Delta m^2_{23}$ and $\nu_{\mu}$ disappearance~\footnote{Disabling FSI often has a large effect on the shape of the spectrum and this change cannot be offset by adjusting the $\delta_{CP}$ parameter, thus we do not perform this study on $\nu_e$ appearance.}; we will explore the impact of mismodelling secondary variables on $\delta_{CP}$ measurements in Section~\ref{sec:ImpactOnOsc2}. We inject different assumptions (e.g. FSI on or off)  in the estimated energy spectra by replacing the response matrix in equation~\ref{eq:Spectrum}.

To study the effect of FSI, we build two estimated spectra, one using the response matrix with the full simulation and one made without simulating FSI. We denote the two estimated energy spectra as $N(\alpha)$ and $N'(\alpha')$ respectively, in which different sets of oscillation parameters, $\alpha$ and $\alpha'$ have been used in equation~\ref{eq:Spectrum}. We restrict the fit range between true neutrino energies of 0.8~GeV and 5.0~GeV, fitting only the two largest oscillation minima/maxima. We then perform a closure test in which the spectrum constructed without simulating FSI in the response matrix is fitted to spectrum made with the full simulation by varying $\alpha'$ to minimize:
\begin{align}
    \chi^2(\alpha') = \sum_i \left( \frac{N'_i(\alpha') - N_i(\alpha)}{\sqrt{N_i(\alpha)}} \right)^2,
    \label{eq:FitMetric}
\end{align}
where again $i$ indexes the bins in the estimated neutrino energy spectrum. The difference between the best fit of $\alpha'$ and the value of $\alpha$ used to construct the original spectrum with the full simulation then quantifies the impact of FSI in an oscillation context.

\begin{figure*}
    \centering
    \subfigure[GENIE.]{\includegraphics[width=0.48\linewidth]{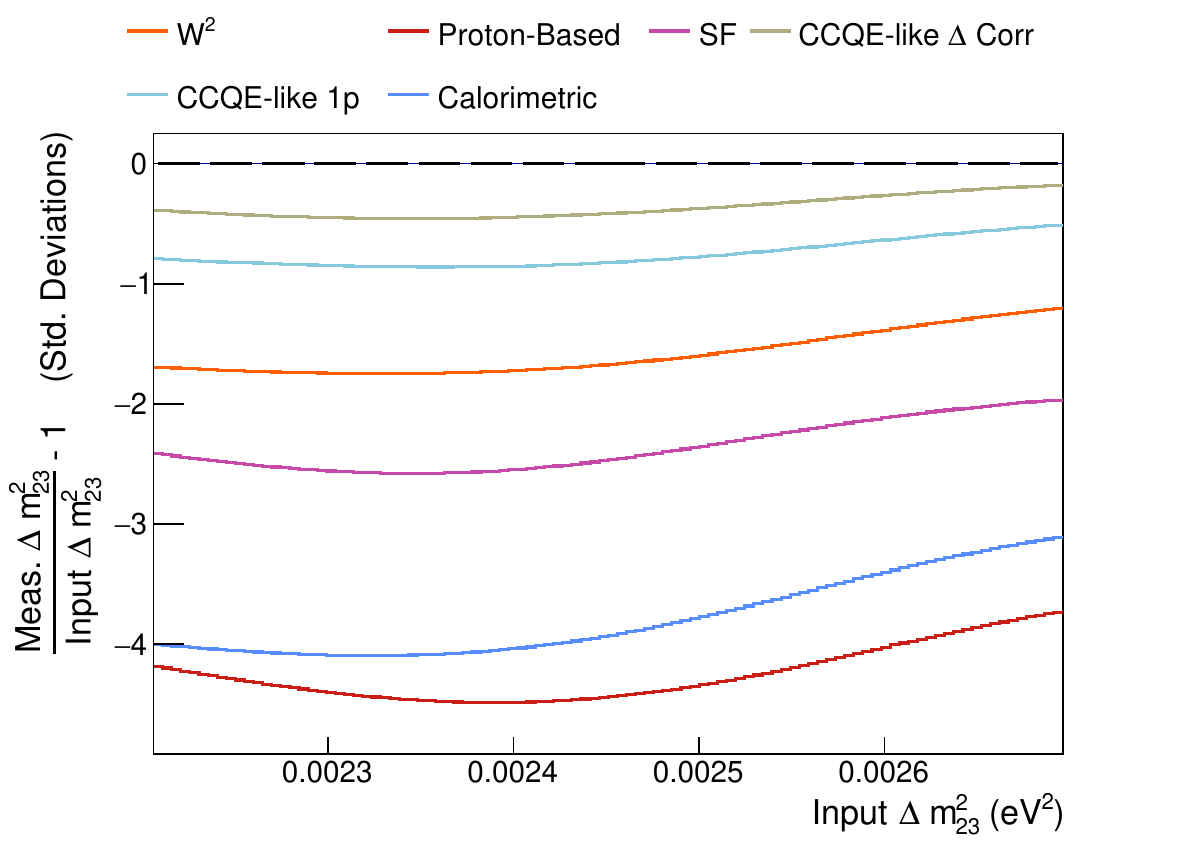}}%
    \hfill
    \subfigure[NuWro.]{\includegraphics[width=0.48\linewidth]{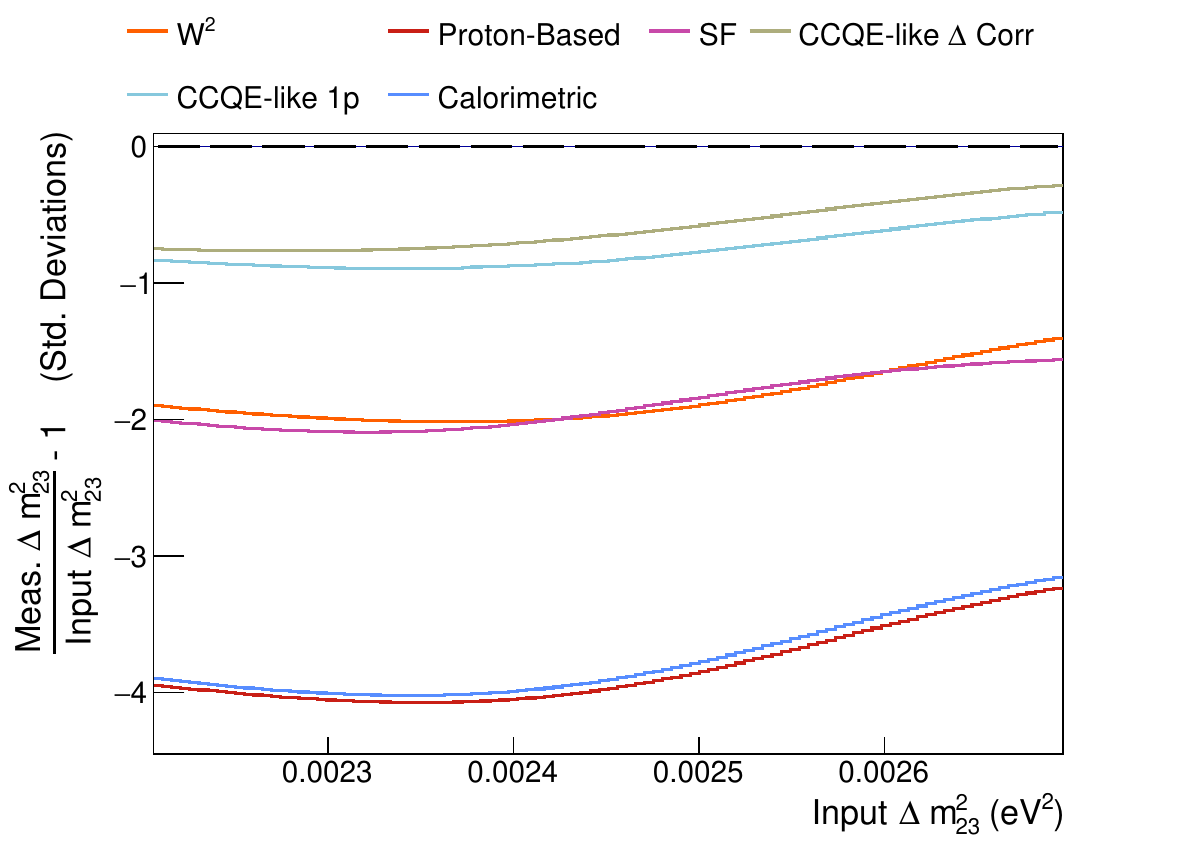}}
    \caption{Bias in extracted $\Delta m^2_{23}$ when fitting spectra generated with and without FSI. We express the bias in standard deviations, where 1 standard deviation is $\frac{0.026\times10^{-3}\textrm{eV}^2}{2.451\times 10^{-3}\textrm{eV}^2} = 0.011$~\cite{ParticleDataGroup:2024cfk}.}
    \label{fig:FSIDeltaM2Fits}
\end{figure*}

Fig.~\ref{fig:FSIDeltaM2Fits} shows the ratio between the extracted value of $\Delta m^2_{23}$ obtained from fitting the non-FSI spectrum with equation~\ref{eq:FitMetric} and the value of $\Delta m^2_{23}$ used to generate the full spectrum with FSI. The estimators least sensitive to FSI in this test are the two modified versions of the CCQE-like estimator, consistent with the study on bias presented in Section~\ref{sec:FSI_Focus}, followed by the $W^2$ and SF methods. The calorimetric and proton based methods show the largest biases. The conclusions of this test are consistent across GENIE and NuWro. 


\clearpage

\section{Impact of secondary variables dependency}
\label{sec:ImpactOnOsc2}
We perform a variation of the closure test described in Section~\ref{sec:FSIFits} to explore the impact of secondary variables (defined in Section~\ref{sec:VariableDependences}) on the oscillation parameter extraction. Here, instead of switching off final state interactions to yield a modified response matrix, we construct a series of alternative response matrices by changing the event selection on one of the secondary variables. We then investigate how the mismodelling of these variables affects the inferred oscillation parameters by fitting the alternative spectrum obtained from the modified response matrix in Eq.~\ref{eq:Spectrum}, and re-minimizing the function in Eq.~\ref{eq:FitMetric} to extract the corresponding parameter $\alpha'$. We do not perform this test with the lepton momentum as this variable strongly correlates with neutrino energy and this test often fails to close as a result.

We perform this test considering both $\nu_{\mu}$ disappearance and $\Delta m^2_{23}$, and $\nu_e$ appearance and $\delta_{CP}$. Contrary to our FSI study, we only consider the global central value of $\Delta m^2_{23}$, and $\delta_{CP} = \{0,\frac{\pi}{2},-\frac{\pi}{2}\}$ to study how varying the selection on the hidden variable influences inference. 

\begin{figure}[p]
    \centering
    \includegraphics[width=\linewidth]{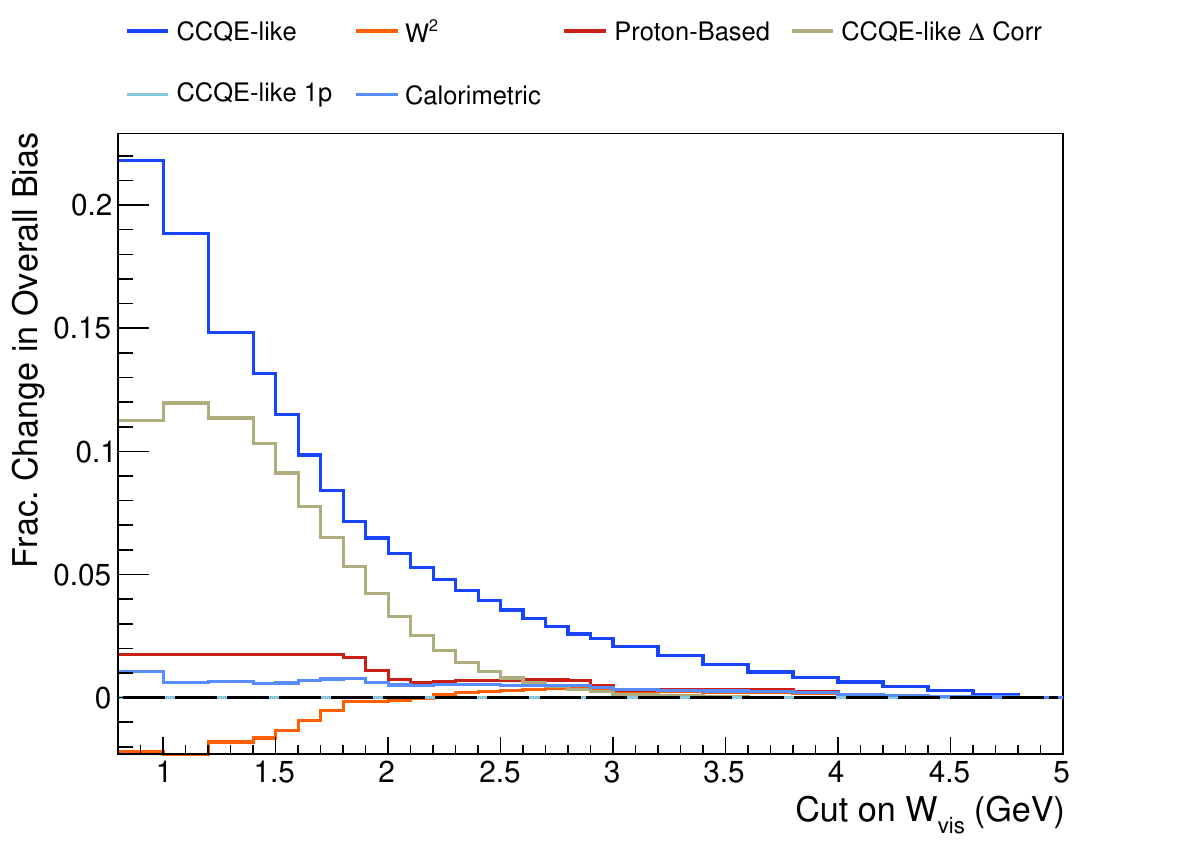}
    \caption{Dependence the overall bias in neutrino energy as a function of the cut visible hadronic invariant mass. Sharp structures occur near integer multiples of the nucleon mass, reflecting sensitivity to resonance and multi-nucleon production.}
    \label{fig:BiasCutW}
\end{figure}

\begin{figure}
    \centering
    \includegraphics[width=\linewidth]{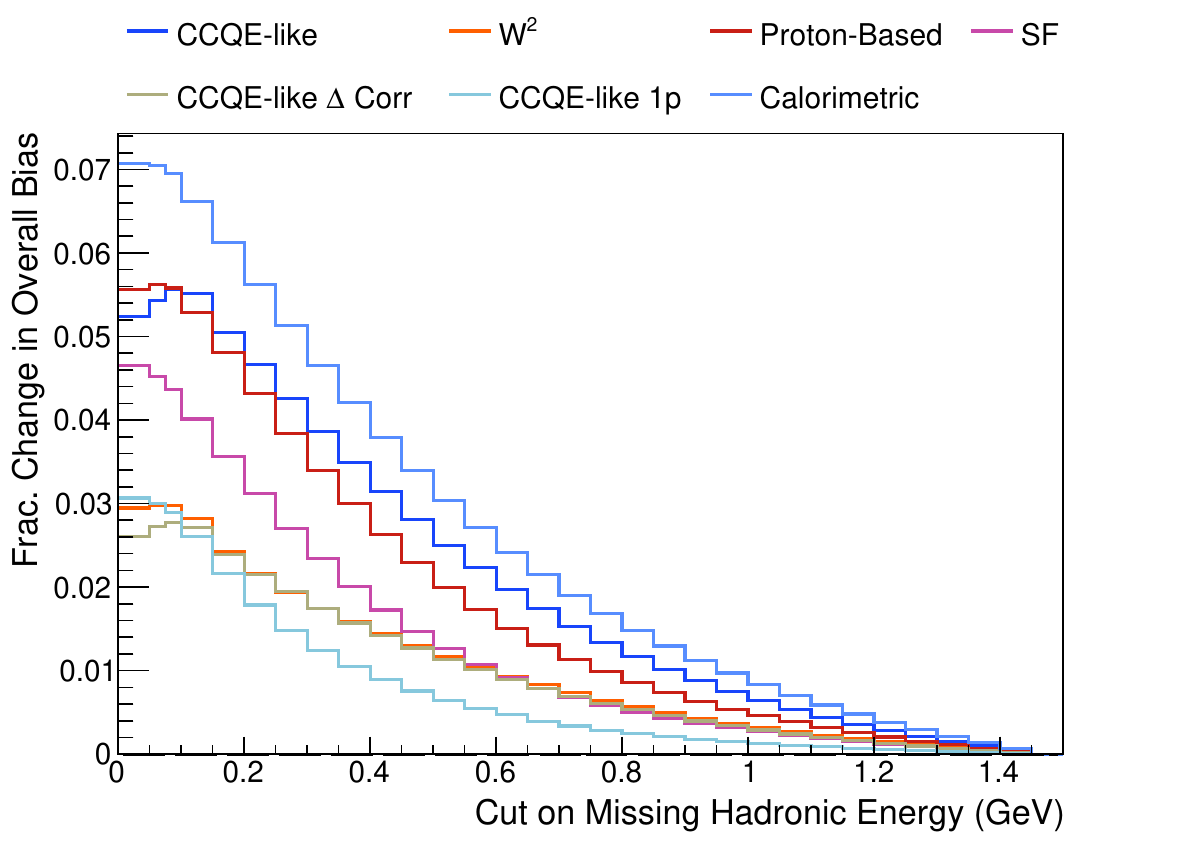}
    \caption{Change in overall bias in neutrino energy as a function of the cut on missing hadronic energy. Methods that incorporate transverse kinematic balance (SF, $W$-based) remain stable, whereas calorimetric and muon-kinematic methods degrade more strongly.}
    \label{fig:BiasCutMissingE}
\end{figure}

\begin{figure}
    \centering
    \includegraphics[width=\linewidth]{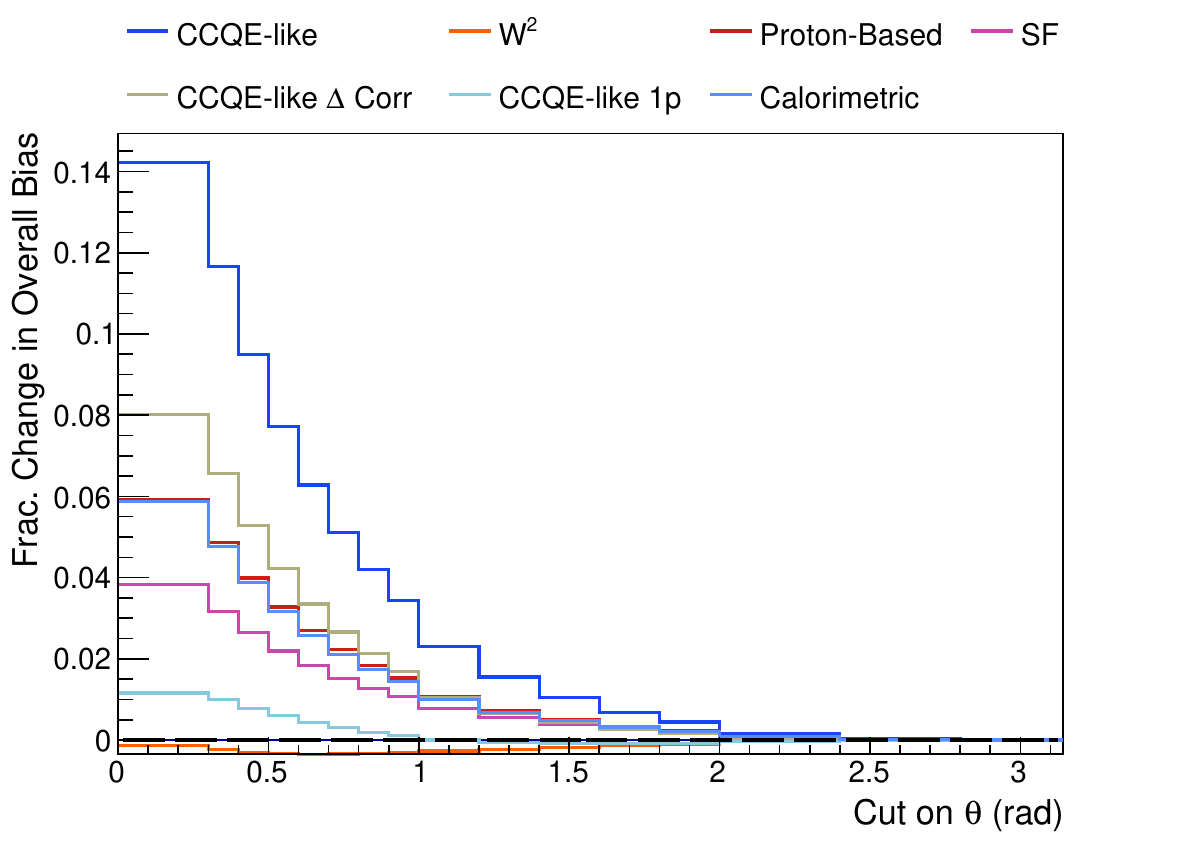}
    \caption{Change in overall bias in neutrino energy as a function of a cut on lepton scattering angle, comparing all estimators. Forward-going selections reduce the impact of angular mismodelling, with the $W$-based method showing the most stable behavior.}
    \label{fig:BiasCutAngle}
\end{figure}

We find it is also informative to calculate the shift in the overall bias in neutrino energy across the entire MC simulation sample, $B'-B$, as a function of the selection on the secondary variable. These calculations are shown in Figures~\ref{fig:BiasCutW}, ~\ref{fig:BiasCutMissingE}, and ~\ref{fig:BiasCutAngle}, and often mimic the bias in the extracted oscillation parameters resulting from the cut on the hidden variable (up to the sign of the bias, as this depends on the oscillation parameter in question). 

\subsection{Hadronic Invariant Mass}
\begin{figure}
    \centering
    \subfigure[$\delta_{CP}=0$.]{\label{fig:OscWCPZero}\includegraphics[width=\linewidth]{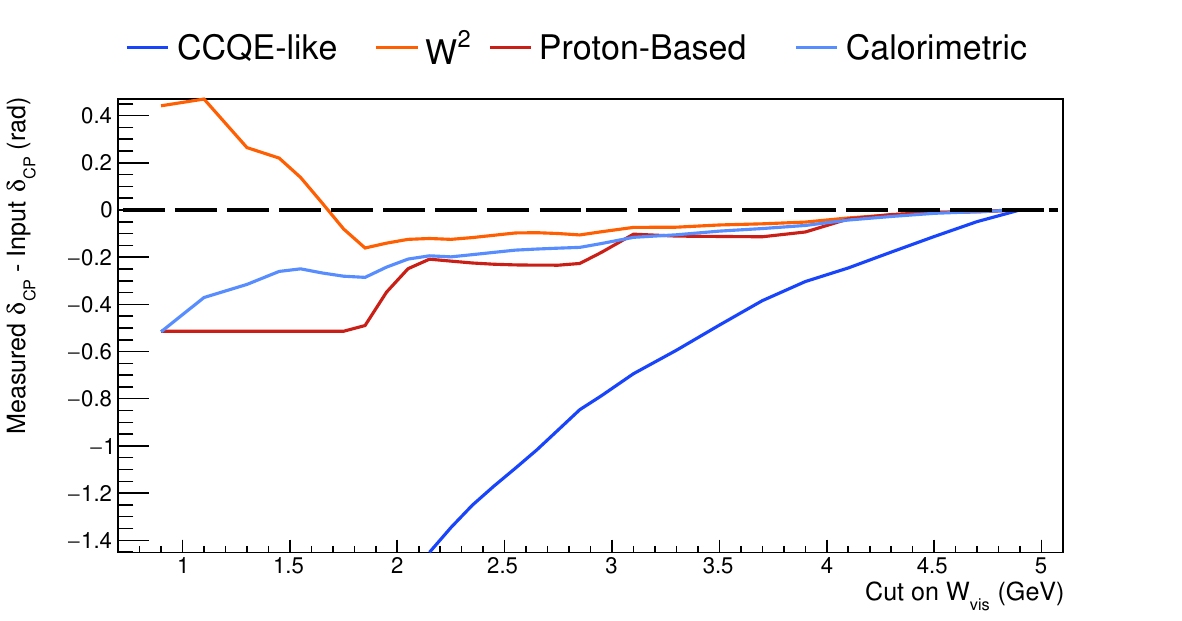}}
    \subfigure[$\delta_{CP}=\tfrac{\pi}{2}$.]{\label{fig:OscWCPPlus}\includegraphics[width=\linewidth]{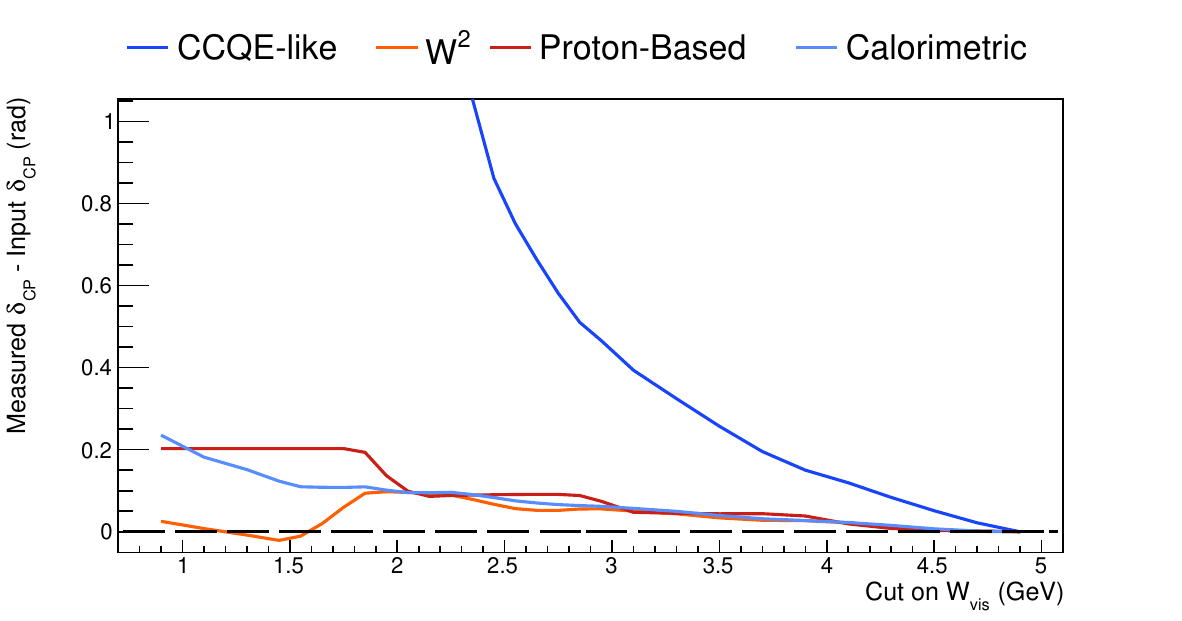}}
    \subfigure[$\delta_{CP}=-\tfrac{\pi}{2}$.]{\label{fig:OscWCPMinus}\includegraphics[width=\linewidth]{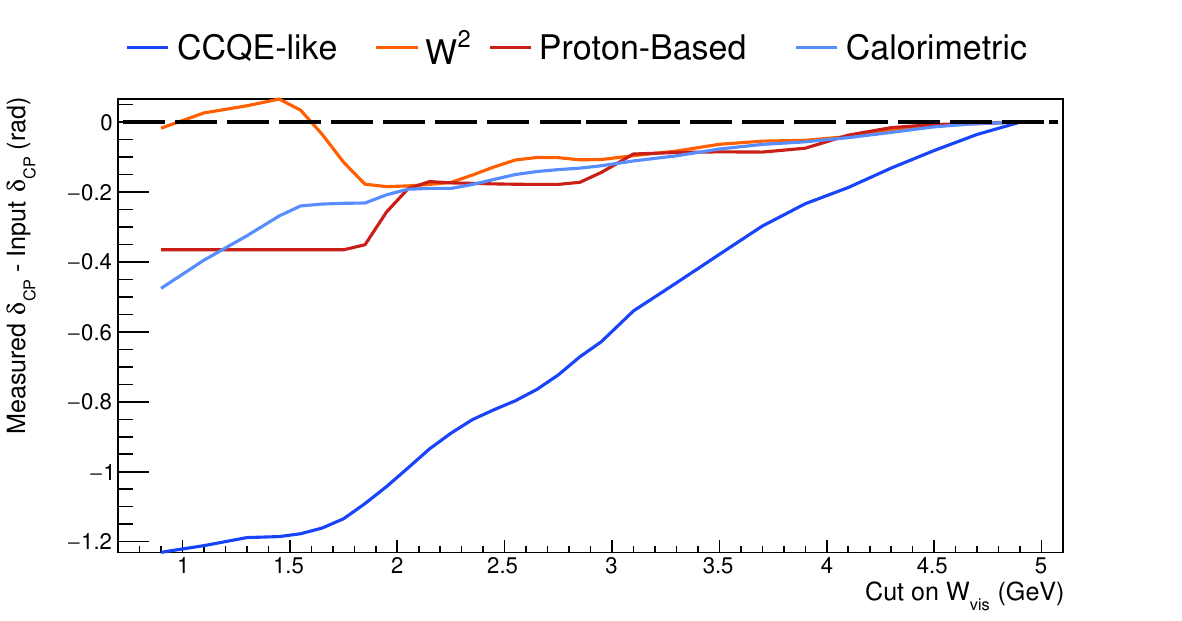}}
    \subfigure[$\Delta m^2_{23}$.]{\label{fig:OscWM2}\includegraphics[width=\linewidth]{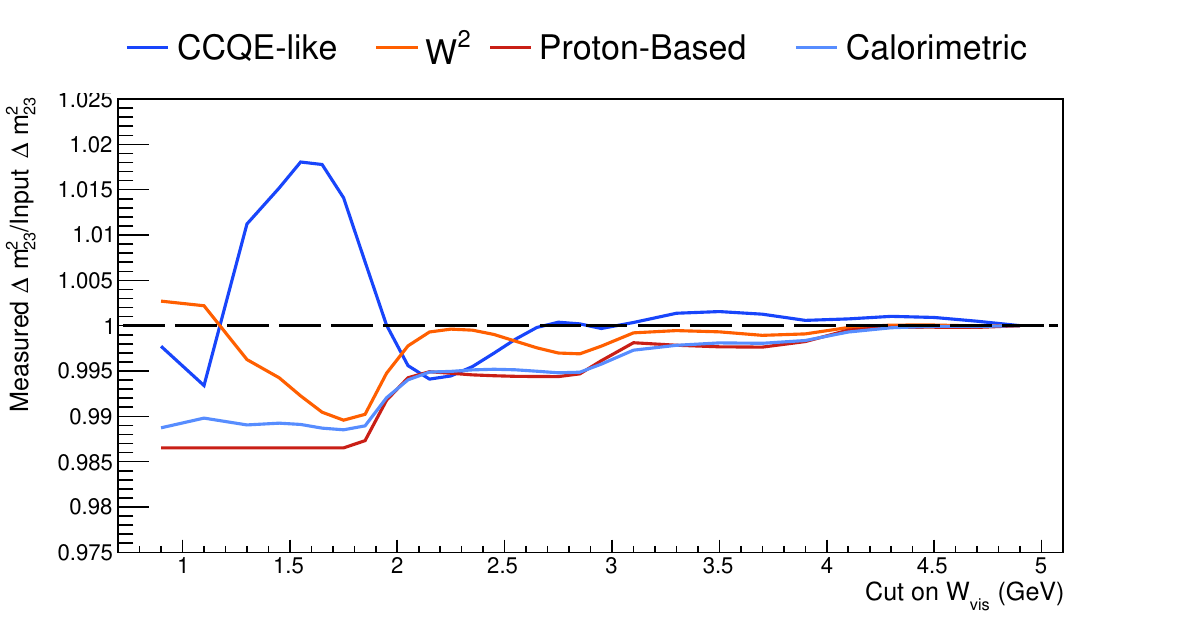}}
    \caption{Oscillation fits when events are selected by visible hadronic invariant mass. Periodic structures in the extracted $\delta_{CP}$ and $\Delta m^2_{23}$ reflect the resonance-like features in Fig.~\ref{fig:OscW}.}
    \label{fig:OscW}
\end{figure}

The oscillation closure tests are produced from cutting on the visible hadronic invariant mass are shown in Fig.~\ref{fig:OscW}. We build the alternative response matrix by placing an upper limit on $W_{\rm vis}$, starting at the single proton mass and gradually allowing more complex final states. We see periodic ``bends'' in the bias in extracted oscillation parameters, similar to the features in the bias in neutrino energy previously seen in Fig.~\ref{fig:BiasPlotsW}, particularly when $W$ approaches integer multiples of the nucleon mass. These underscore the importance of accurate multi-nucleon and resonance modeling in neutrino generators. Comparison of the biases across different estimators indicates the $W^2$ method closes for largest portion of the parameter space, followed closely by the calorimetric method. We do not draw the CCQE-like result for portions of the parameter space in the $\delta_{CP}$ fits as the bias in this case cannot be offset by modifying the value of $\delta_{CP}$ alone. 

\subsection{Missing Hadronic Energy}

\begin{figure}
    \centering
    \subfigure[$\delta_{CP}=0$.]{\label{fig:OscMissingECPZero}\includegraphics[width=\linewidth]{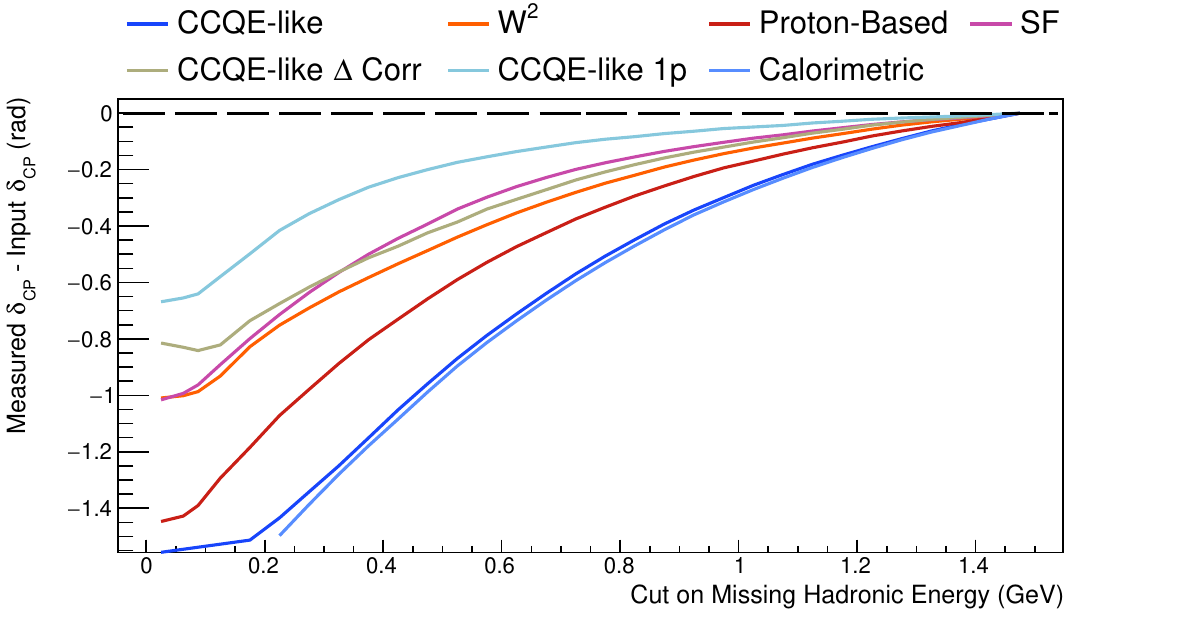}}
    \subfigure[$\delta_{CP}=\tfrac{\pi}{2}$.]{\label{fig:OscMissingECPPlus}\includegraphics[width=\linewidth]{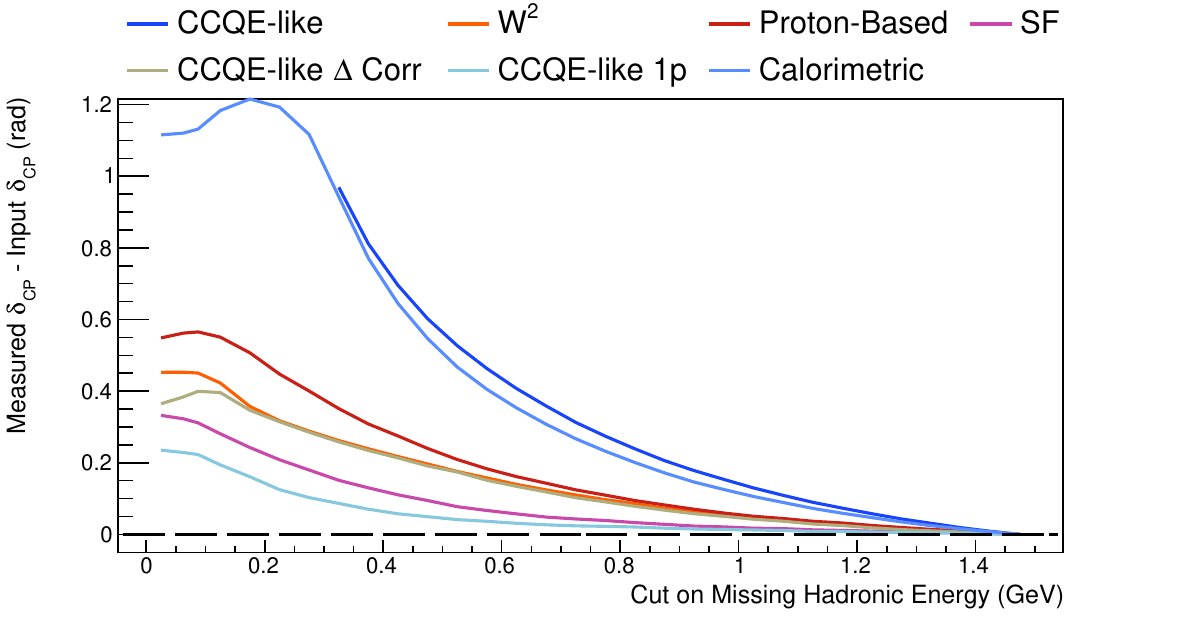}}
    \subfigure[$\delta_{CP}=-\tfrac{\pi}{2}$.]{\label{fig:OscMissingECPMinus}\includegraphics[width=\linewidth]{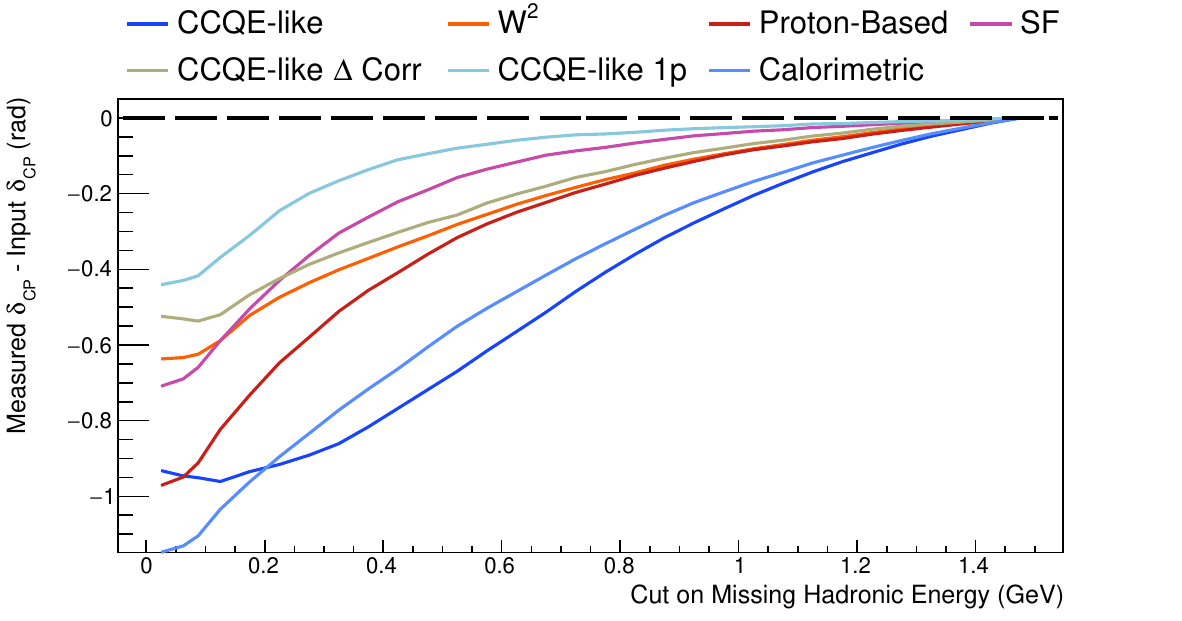}}
    \subfigure[$\Delta m^2_{23}$.]{\label{fig:OscMissingEM2}\includegraphics[width=\linewidth]{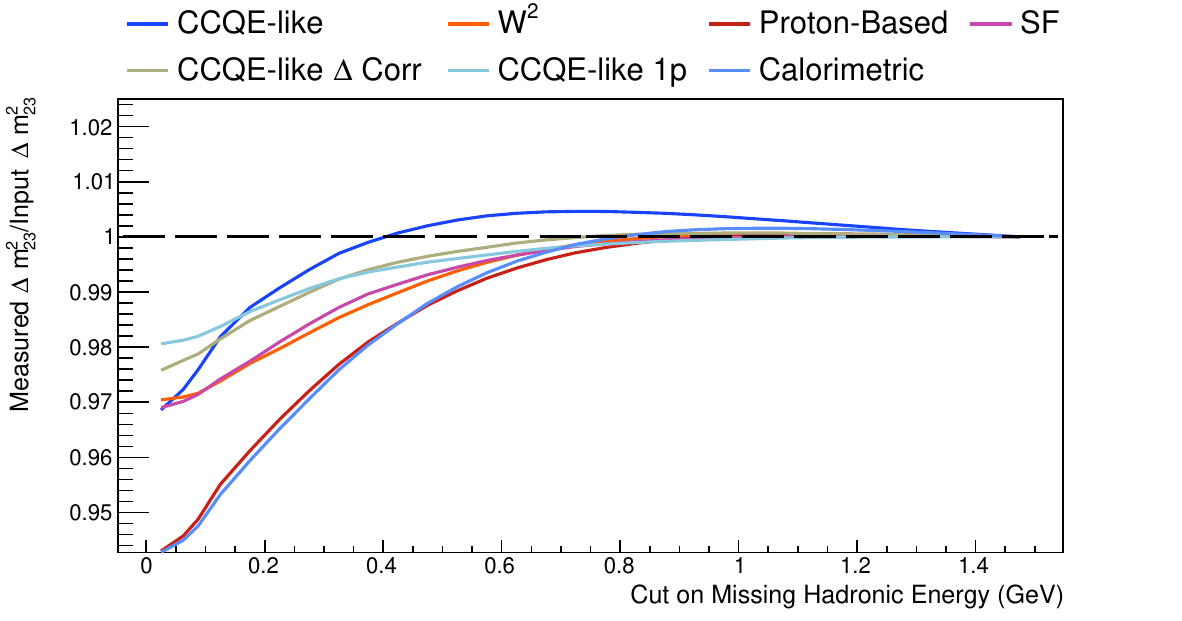}}
    \caption{Oscillation fits under restrictions on missing hadronic energy. The SF method yields the most consistent $\delta_{CP}$ and $\Delta m^2_{23}$ extraction, though with reduced statistics.}
    \label{fig:OscMissingE}
\end{figure}

We explore the impact of missing hadronic energy through constructing the alternative response matrix with an upper limit on this quantity, starting with only events in which all hadronic energy is visible. The results of this closure test are presented in Fig.~\ref{fig:OscMissingE}. The methods displaying the most resilience are the single proton version of the CCQE-like estimator and the SF method, followed by the $W$-based and proton-based estimators. The calorimetric and CCQE-like methods are more sensitive to unobserved hadronic components.

\subsection{Angular Dependence}

\begin{figure}
    \centering
    \subfigure[$\delta_{CP}=0$.]{\label{fig:OscAngleCPZero}\includegraphics[width=\linewidth]{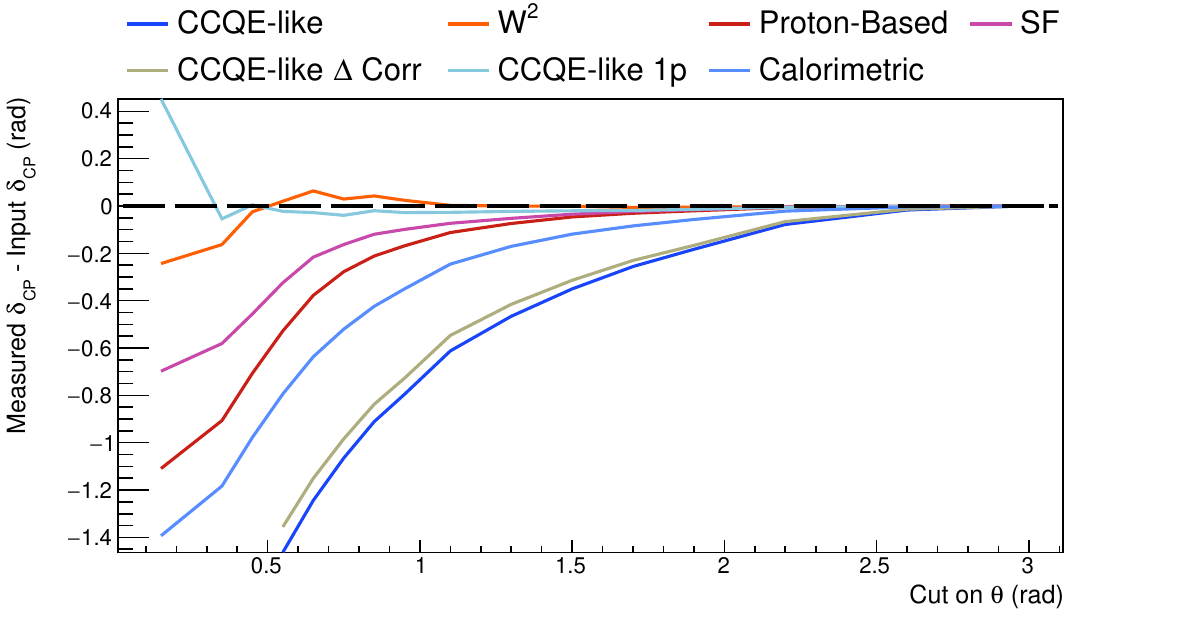}}
    \subfigure[$\delta_{CP}=\tfrac{\pi}{2}$.]{\label{fig:OscAngleCPPlus}\includegraphics[width=\linewidth]{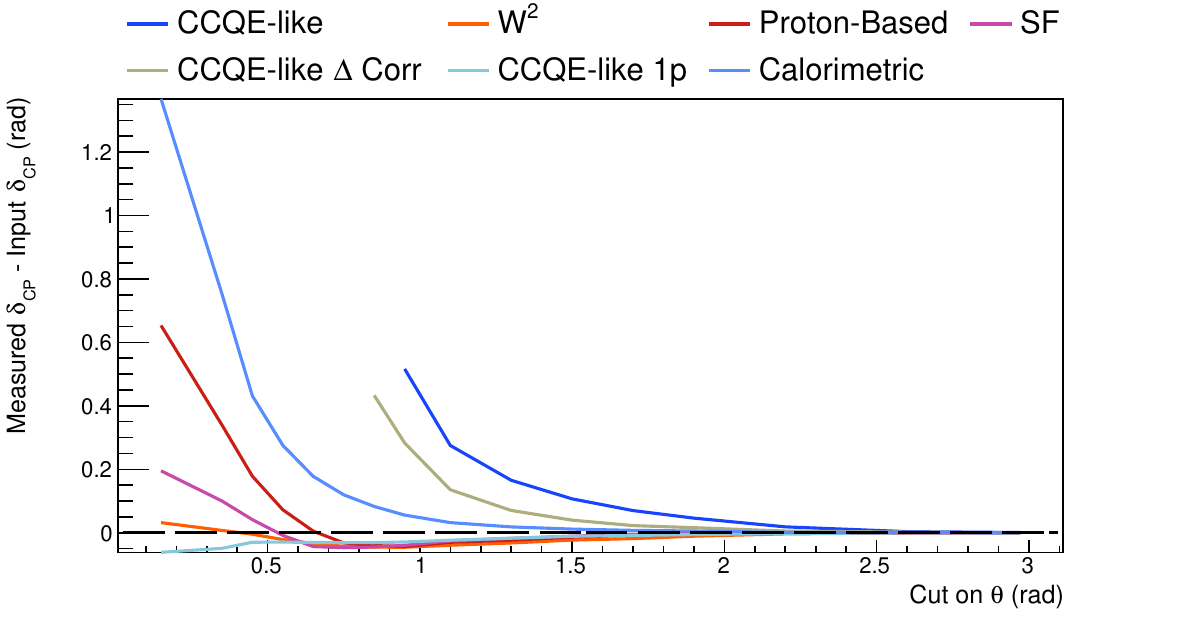}}
    \subfigure[$\delta_{CP}=-\tfrac{\pi}{2}$.]{\label{fig:OscAngleCPMinus}\includegraphics[width=\linewidth]{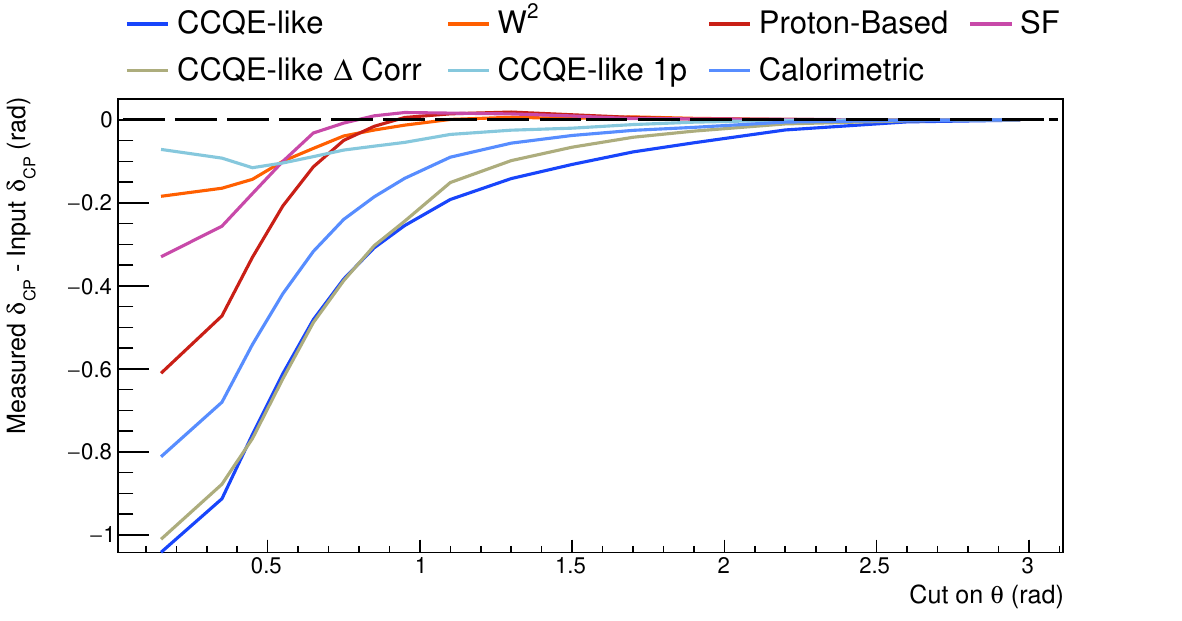}}
    \subfigure[$\Delta m^2_{23}$.]{\label{fig:OscAngleM2}\includegraphics[width=\linewidth]{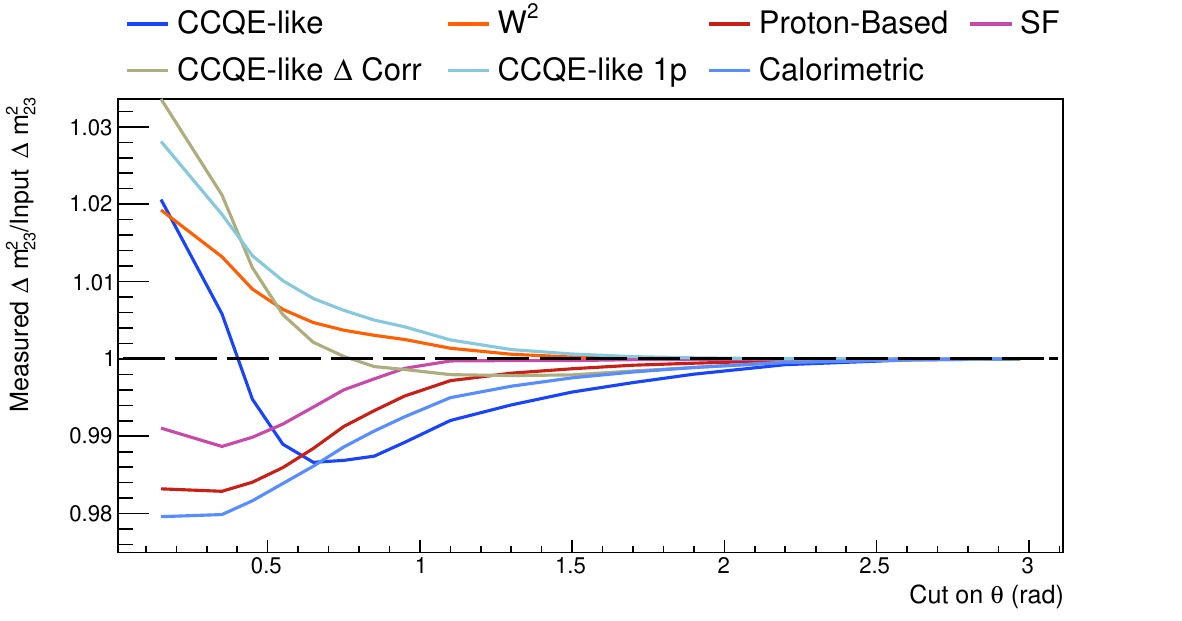}}
    \caption{Extracted oscillation parameters when restricting events by lepton scattering angle. The $W$-based estimator converges rapidly to the correct $\delta_{CP}$ and $\Delta m^2_{23}$ values, while the muon-only approach shows strong dependence on $\theta$.}
    \label{fig:OscAngle}
\end{figure}

We also construct the alternative response matrix by cutting on the lepton scattering angle, starting with only including forward going leptons before loosening the cut to include larger angles, corresponding to more inelastic interactions. Fits of the oscillated spectra are shown in Fig.~\ref{fig:OscAngle}, revealing that the $W^2$-based estimator is comparatively stable, while the purely CCQE-like approach is highly susceptible to angle mismodelling. The SF method offers the second best performance, likely because  this method is applied to selected events that still resemble quasielastic interactions even at large scattering angles. 

\clearpage
\section{Summary}
In this work, we investigate a new method to estimate neutrino energy in broadband beams based on the measurement of the final-state hadronic invariant mass, $W$.  This method is compared against several other commonly used estimators: the CCQE kinematics formula, summation of the energies of final state protons, a calorimetry based approach, and the Sobczyk-Furmanski method. 

The $W^2$ method consistently exhibits the smallest overall bias as functions of the true energy, with a variance comparable to other inclusive estimators, particularly once resolution effects are considered. Additionally, the $W^2$ method shows the weakest dependence on FSI effects, most notably when using the GENIE neutrino event generator. When probing the stability of each estimator against the visible hadronic invariant mass, missing hadronic energy, and the angle and momentum of the scattered lepton, we find that the $W^2$ method often shows the least variation in bias with respect to these variables, especially in the case of the lepton's scattering angle.  

These findings are mirrored in an oscillation analysis context. The most powerful methods in the exclusion tests are the calorimetric, $W^2$, and CCQE-like estimators. While the Sobczyk–Furmanski method achieves the lowest variance, its sensitivity in the toy oscillation analyses is reduced due to the substantial loss of statistics. In the closure tests, the $W^2$ method emerges as the most stable among the inclusive estimators, often exhibiting the smallest bias in the extracted oscillation parameters across the widest regions of parameter space.

In this work, we examine each method individually, highlighting the contexts and underlying motivations for which each estimator performs best. This thorough examination provides a detailed guidance for future experiments on how to combine estimators to enhance overall performance. Certain considerations when combining estimators have not explored in this work, for example migration of events between channels due to reconstruction failures possibly leading to a decline in performance. 

 In this analysis we calculate the estimated energy event by event using each method assuming the detection thresholds established by existing LArTPCs coupled with perfect efficiency and particle separation performance. Future studies of this subject should attempt to use more complete detector response and reconstruction models to further explore the interplay between the cross section modeling effects explored here and these other aspects of oscillation measurements. 

\begin{acknowledgments}
We would like to thank Afroditi Papadopoulou, Justin Evans, Stefan S\"oldner-Rembold for useful conversations. This work was directly supported by the Science and Technology Facilities Council (STFC) grant number ST/W003945/1. 
\end{acknowledgments}


\bibliography{bib}

\end{document}